\documentclass[10pt, journal]{IEEEtran}
\IEEEoverridecommandlockouts
\usepackage{lineno}
\usepackage{hyperref}
\usepackage{cite}
\usepackage{amsmath,amssymb,amsfonts}
\usepackage{amsmath}
\usepackage{amsthm}

\usepackage{graphicx}
\usepackage{tikz}
\usetikzlibrary{shapes,shapes.geometric,arrows.meta,positioning,patterns,calc,decorations.pathreplacing}
\usepackage{textcomp}
\usepackage[table]{xcolor}
\usepackage{graphicx}
\usepackage{float}
\usepackage{subfigure}
\usepackage{amsmath}
\usepackage{amsfonts,amssymb}
\usepackage{mathrsfs}
\usepackage{mathtools}
\usepackage{algorithm}
\usepackage{algorithmicx}
\usepackage{algpseudocode}
\usepackage{bm}
\usepackage{multirow}
\usepackage{array}
\usepackage{amssymb}
\usepackage{amsmath}
\usepackage{cite}
\usepackage{url}
\usepackage{xcolor}
\usepackage{cite,graphicx,amsmath,amssymb}
\usepackage{subfigure}
\usepackage{fancyhdr}
\usepackage{mdwmath}
\usepackage{mdwtab}
\usepackage{caption}
\usepackage{amsthm}
\usepackage{setspace}
\usepackage{bm}
\usepackage{algorithm}
\usepackage{algpseudocode}
\usepackage{mathtools}
\usepackage{dsfont}
\usepackage{bbm}
\usepackage{esint}
\usepackage{scalerel}
\usepackage{balance}

\usepackage[font=small]{caption}

\hypersetup{colorlinks=true,
linkcolor=blue,
citecolor=blue,      
urlcolor=black,
}

\renewcommand{\normalsize}{\fontsize{9.5}{11.4}\selectfont}

\expandafter\def\expandafter\normalsize\expandafter{%
    \normalsize%
    \setlength\abovedisplayskip{5pt}%
    \setlength\belowdisplayskip{5pt}%
    \setlength\abovedisplayshortskip{2pt}%
    \setlength\belowdisplayshortskip{2pt}%
}
\makeatletter
\def\section{\@startsection{section}{1}{\z@}{1.5ex plus 0.4ex minus 0.3ex}%
{0.7ex plus 1ex minus 0ex}{\normalfont\normalsize\centering\scshape}}%
\def\subsection{\@startsection{subsection}{2}{\z@}{1.5ex plus 0.4ex minus 0.3ex}%
{0.7ex plus .5ex minus 0ex}{\normalfont\normalsize\itshape}}%
\makeatother

\definecolor{TableHeader}{RGB}{44,62,92}
\definecolor{TableStripe}{RGB}{246,248,252}
\definecolor{TableGroup}{RGB}{229,236,246}
\definecolor{TableRule}{RGB}{186,196,210}

\newcommand{\rev}[1]{{\color{black}#1}}

\newtheorem{remark}{Remark}
\newtheorem{theorem}{Theorem}

\newtheorem{lemma}{Lemma}

\newtheorem{corollary}{Corollary}

\makeatletter

\newcommand*{\paral}{\stretchrel*{\parallel}{\perp}}

\newcommand{\biggg}{\bBigg@{3}}
\newcommand{\Biggg}{\bBigg@{3.5}}
\makeatother
\def\BibTeX{{\rm B\kern-.05em{\sc i\kern-.025em b}\kern-.08em
    T\kern-.1667em\lower.7ex\hbox{E}\kern-.125emX}}
\begin{document}
\title{Electromagnetic Signal and Information Theory: \\ A Continuous-Aperture Array Perspective}

\author{Zhaolin Wang,~\IEEEmembership{Member, IEEE}, Chongjun Ouyang,~\IEEEmembership{Member, IEEE}, \\
Kuranage Roche Rayan Ranasinghe,~\IEEEmembership{Graduate Student Member, IEEE}, \\
Shuai S. A. Yuan,~\IEEEmembership{Member, IEEE}, Giuseppe Thadeu Freitas de Abreu,~\IEEEmembership{Senior Member, IEEE}, \\
Emil Björnson,~\IEEEmembership{Fellow, IEEE}, and Yuanwei Liu,~\IEEEmembership{Fellow, IEEE}
\thanks{Zhaolin Wang and Yuanwei Liu are with the Department of Electrical and Computer Engineering, The University of Hong Kong, Hong Kong (e-mail: \{zhaolin.wang, yuanwei\}@hku.hk).}
\thanks{Chongjun Ouyang is with the School of Electronic Engineering and Computer Science, Queen Mary University of London, London, E1 4NS, U.K. (e-mail: c.ouyang@qmul.ac.uk).}
\thanks{Kuranage Roche Rayan Ranasinghe and Giuseppe Thadeu Freitas de Abreu are with the School of Computer Science and Engineering, Constructor University
(previously Jacobs University Bremen), Campus Ring 1, 28759 Bremen,
Germany (e-mail: \{kranasinghe, gabreu\}@constructor.university).}
\thanks{Shuai S. A. Yuan is with the Department of Electronics and Nano Engineering, School of Electrical Engineering, Aalto University, 02150 Espoo, Finland (e-mail: shuai.yuan@aalto.fi).}
\thanks{Emil Björnson is with the School
of Electrical Engineering and Computer Science, KTH Royal Institute of Technology, 100 44 Stockholm, Sweden (e-mail: emilbjo@kth.se).}
\vspace{-0.8cm}
}

\maketitle
\begin{abstract}
    Emerging wireless systems are evolving toward larger, denser, higher-frequency, and more reconfigurable apertures, which motivates the study of continuous-aperture arrays (CAPAs). Unlike conventional spatially discrete arrays (SPDAs), CAPAs are more naturally modeled as spatially continuous electromagnetic apertures and therefore call for a fundamental shift in both signal processing and information-theoretic analysis. In particular, the underlying channels, signals, and beamformers are no longer finite-dimensional vectors and matrices, but continuous fields and operators governed by Maxwell’s equations. This paper provides a tutorial overview of CAPA systems from the perspective of electromagnetic signal and information theory (ESIT), with an emphasis on the transition from discrete array models to physics-consistent continuous-aperture formulations. We review the electromagnetic foundations of CAPAs, practical hardware implementations, line-of-sight and multipath channel modeling, continuous-space beamforming and channel estimation, and the fundamental degrees of freedom and capacity limits of CAPA systems. We also highlight how tools such as wavenumber-domain methods, functional analysis, and compressive sensing can transform challenging infinite-dimensional problems into tractable finite-dimensional ones while preserving the essential physical structure of the channel. Overall, this tutorial aims to clarify the key principles, analytical tools, and open challenges that shape CAPA-enabled wireless communications.
\end{abstract}

\begin{IEEEkeywords}
    Continuous-aperture array, electromagnetic signal and information theory, channel modeling, beamforming design, channel estimation, fundamental limits.
\end{IEEEkeywords}

\section{Introduction}
\IEEEPARstart{S}INCE Hertz verified the existence of radio waves with dipole antennas in 1887, electromagnetic (EM) waves have served as the physical carrier of wireless information transmission. Yet the two main theories underpinning wireless communications, namely EM theory and information theory, have historically evolved along largely separate tracks. Since Shannon’s seminal work in 1948 \cite{shannon1948mathematical}, information theory has mainly described channels through finite-dimensional mathematical abstractions, enabling elegant characterizations of reliability and capacity. In contrast, EM theory is built on continuous fields, boundary conditions, and Maxwell’s equations. Although some early works, such as the study by Bucci \emph{et al.} on the spatial degrees of freedom (DoFs) of scattered fields \cite{bucci1989spatial}, attempted to connect these two viewpoints, modern wireless system design still often treats them separately.

This separation is becoming increasingly inadequate. As wireless networks evolve toward sixth-generation (6G) systems and beyond, they are expected to support extreme data rates, massive connectivity, and highly reliable links. Meeting these demands requires a more faithful account of the physical behavior of EM waves than what is offered by purely finite-dimensional channel models. In particular, the spatial DoFs of wireless channels need to be revisited from a physics-based perspective. This need has motivated the development of \emph{Electromagnetic Signal and Information Theory (ESIT)} \cite{di2023electromagnetic}, which seeks to bridge continuous-space EM propagation formulations and Shannon’s information-theoretic framework. By doing so, ESIT provides a principled way to study information generation, transmission, and recovery under the fundamental laws of EM fields.

One important driver behind this convergence is the historical evolution of multiple-antenna technology. In the early 1990s, initial studies explored the use of multiple antennas to distinguish multiple communication users in the spatial domain \cite{Swales1990a,Anderson1991a}. The emergence of multiple-input multiple-output (MIMO) systems in the late 1990s further showed that spatial multiplexing can increase channel capacity linearly with the number of antennas without consuming extra bandwidth \cite{telatar1999capacity, foschini1998on}. This idea later developed into massive MIMO \cite{marzetta2010noncooperative, lu2014overview}, where very large antenna arrays deliver substantial gains in spectral and energy efficiency. More recently, to address the stringent requirements of 6G, a variety of large-scale antenna architectures have been proposed \cite{liu2024near}, including extremely large antenna arrays (ELAAs) \cite{Bjornson2019c}, reconfigurable intelligent surfaces (RISs) \cite{liu2021reconfigurable}, fluid/movable antennas (FAs/MAs) \cite{new2024tutorial}, reconfigurable holographic surfaces (RHSs) \cite{deng2021reconfigurable}, and dynamic metasurface antennas (DMAs) \cite{shlezinger2021dynamic}. Despite their hardware differences, these technologies share a common evolutionary direction: the aperture is becoming \emph{larger} in physical size, \emph{denser} in spatial control, \emph{higher} in operating frequency, and \emph{more flexible} in geometry and reconfigurability. This trend naturally leads to the concept of the \emph{continuous-aperture array (CAPA)} \cite{11095329}.

In contrast to a conventional spatially discrete array (SPDA), which consists of a finite set of separated antenna elements, a CAPA is modeled as a spatially continuous EM aperture. Conceptually, it can be interpreted as the limiting form of an SPDA with increasingly dense spatial sampling and increasingly fine aperture controllability \cite{11095329,liaskos2018new}. From an EM perspective, the physically available spatial DoFs are determined by the aperture size, operating wavelength, and propagation environment, rather than by the sampling density alone. Under standard assumptions, a sufficiently sampled discrete array, such as a half-wavelength-spaced array, can already represent the propagating field without spatial aliasing. Therefore, the significance of CAPA lies in providing a continuous-aperture abstraction that more directly characterizes the physically available spatial modes, improves aperture utilization, and enables finer field synthesis within the fundamental EM limits. More fundamentally, the continuous nature of CAPA allows the synthesis of continuous source-current distributions over the aperture, making it a natural physical platform for studying communication limits under the ESIT framework~\cite{11095329,10417101,bjornson2024towards}.


However, this shift from SPDA to CAPA also calls for a substantial rethinking of signal processing and information theory. In SPDA-based systems, the spatial response is usually represented by finite-dimensional vectors and matrices. In CAPA-based systems, by contrast, the spatial response is inherently continuous and is more naturally described by EM-based integral operators. Under this perspective, transmitted and received signals are no longer discrete vectors, but continuous spatial random fields. Consequently, many standard tools developed for SPDAs, including channel modeling, capacity analysis, performance evaluation, and beamforming design, are no longer directly suitable. To fully exploit CAPA, new analytical and algorithmic frameworks are needed. In particular, an ESIT-based treatment enables channel models to be derived directly from Maxwell’s equations \cite{1386525, 9139337}, extends capacity analysis from matrices to infinite-dimensional operators through functional analysis \cite{10807262, 11045763}, and supports continuous beamforming design using tools such as the calculus of variations \cite{10910020, 10938678}. A systematic study of CAPA from this viewpoint is therefore valuable both for revealing the ultimate performance limits of physically realizable apertures and for guiding practical transceiver design in a physics-consistent manner.

Since 2020, several overview, survey, and tutorial papers have discussed holographic MIMO and related \rev{aperture architectures}. Early overview papers mainly focused on hardware structures, application scenarios, and high-level opportunities \cite{huang2020holographic}. Subsequent tutorial works emphasized communication-theoretic issues such as near-field channel modeling, covariance-based beamforming, channel estimation, and performance analysis \cite{an2023tutorial1,an2023tutorial2,an2023tutorial3}. More recent studies have moved closer to the EM layer by incorporating spatial correlation, mutual coupling, circuit-level effects, and over-the-air EM signal processing \rev{\cite{gong2024holographic,wei2024electromagnetic,bjornson2024towards,dardari2026over,di2023electromagnetic}}. \rev{These works are summarized and compared with this tutorial in Table~\ref{tab:survey_comparison}.}

\begin{table*}[!t]
\centering
\setlength{\abovecaptionskip}{0pt}
\caption{\rev{Comparison of This Work With Existing Magazine, Survey, and Tutorial Papers.}}
\vspace{0.1cm}
\label{tab:survey_comparison}
\footnotesize
\setlength{\tabcolsep}{2pt}
\renewcommand{\arraystretch}{1.15}
\rev{
\begin{tabular}{|>{\centering\arraybackslash}m{1.5cm}|>{\centering\arraybackslash}m{0.75cm}|>{\centering\arraybackslash}m{1.3cm}|*{6}{>{\centering\arraybackslash}m{\dimexpr(\textwidth-3.55cm-18\tabcolsep-10\arrayrulewidth)/6\relax}|}}
\hline
\textbf{Ref.} & \textbf{Year} & \textbf{Type} & \textbf{Continuous aperture} & \textbf{EM/\allowbreak circuit modeling} & \textbf{Channel modeling} & \textbf{Beamforming} & \textbf{Channel estimation} & \textbf{DoF \& capacity} \\ \hline
\cite{huang2020holographic} & 2020 & Magazine & $\triangle$ & -- & -- & $\triangle$ & $\triangle$ & $\triangle$ \\ \hline
\cite{an2023tutorial1,an2023tutorial2,an2023tutorial3} & 2023 & Tutorial & $\triangle$ & $\triangle$ & \checkmark & $\triangle$ & \checkmark & \checkmark \\ \hline
\cite{gong2024holographic} & 2024 & Survey & $\triangle$ & $\triangle$ & \checkmark & $\triangle$ & $\triangle$ & $\triangle$ \\ \hline
\cite{bjornson2024towards} & 2024 & Tutorial & $\triangle$ & \checkmark & \checkmark & $\triangle$ & \checkmark & \checkmark \\ \hline
\cite{di2023electromagnetic} & 2024 & Magazine & $\triangle$ & \checkmark & $\triangle$ & -- & -- & \checkmark \\ \hline
\cite{11095329} & 2025 & Magazine & \checkmark & -- & -- & $\triangle$ & -- & $\triangle$ \\ \hline
\cite{wei2024electromagnetic} & 2026 & Survey & \checkmark & \checkmark & \checkmark & $\triangle$ & -- & \checkmark \\ \hline
\cite{dardari2026over} & 2026 & Tutorial & \checkmark & \checkmark & \checkmark & \checkmark & -- & \checkmark \\ \hline
This work & -- & Tutorial & \checkmark & \checkmark & \checkmark & \checkmark & \checkmark & \checkmark \\ \hline
\multicolumn{9}{|l|}{Continuous aperture: \checkmark: primary; $\triangle$: partial.} \\
\multicolumn{9}{|l|}{Other topics: \checkmark: substantive mathematical treatment; $\triangle$: brief treatment or literature overview; --: not substantively covered.} \\ \hline
\end{tabular}}
\vspace{-10pt}
\end{table*}

\rev{Building on these contributions,} this article provides a comprehensive tutorial on CAPA systems from the ESIT perspective. We cover the full research chain, from physical radiation principles and continuous-space signal modeling to beamforming design, channel estimation, and fundamental limits. Our objective is to clarify the major conceptual shifts that arise when moving from spatially discrete arrays to truly continuous apertures, and to present the key ideas needed to exploit the unique advantages of CAPA while addressing the associated theoretical and practical challenges.

The remainder of this tutorial is organized as follows. Section \ref{sec:theory} reviews the EM foundations needed for CAPA analysis, including radiation, near- and far-field regions, polarization, power, and mutual coupling. Section \ref{sec:fundamentals} introduces the fundamentals of CAPA systems, covering the continuous-space signal model, practical implementations, the circuit-to-field representation, and power constraints. Section \ref{sec:channel_model} presents CAPA channel models, including line-of-sight, physics-based multipath, correlation-based multipath, and time-domain doubly dispersive formulations. Section \ref{sec:design_optimization} discusses CAPA design and optimization, with emphasis on continuous beamforming and channel estimation. Section \ref{sec:fundamental_limits} studies the fundamental limits of CAPA systems, including spatial DoFs, spatial multiplexing, Shannon capacity, Kolmogorov capacity, and the role of physical power and colored noise. Section \ref{sec:conclusion} summarizes the tutorial and concludes with open research challenges.

\emph{Notations:} Boldface symbols denote vectors or matrices, while calligraphic symbols denote sets, surfaces, domains, and function spaces. The superscripts $(\cdot)^{\mathsf{T}}$, $(\cdot)^*$, and $(\cdot)^{\mathsf{H}}$ denote transpose, complex conjugate, and Hermitian transpose, respectively. The operators $|\cdot|$, $\|\cdot\|$, $\Re\{\cdot\}$, $\mathbb{E}\{\cdot\}$, $\mathrm{tr}(\cdot)$, and $\det(\cdot)$ denote absolute value, Euclidean norm, real part, expectation, trace, and determinant, respectively. The sets of real numbers, complex numbers, and integers are denoted by $\mathbb{R}$, $\mathbb{C}$, and $\mathbb{Z}$, respectively. The constants $\mathrm{j}=\sqrt{-1}$ and $\mathrm{e}$ denote the imaginary unit and Euler's number, respectively. $\mathcal{CN}(0,\mathbf{C})$ denotes a circularly-symmetric complex Gaussian distribution with covariance $\mathbf{C}$, and $\delta(\cdot)$ denotes the Dirac delta function. Spatial points on the transmit and receive apertures are denoted by $\mathbf{s}$ and $\mathbf{r}$, respectively, with Cartesian components indicated by sans-serif subscripts such as $s_{\mathsf{x}}$ and $r_{\mathsf{z}}$. Unless otherwise stated, integrals over $\mathcal{S}_{\mathrm{t}}$ and $\mathcal{S}_{\mathrm{r}}$ are surface integrals over the transmitter (Tx) and receiver (Rx) apertures. For a Euclidean domain $\mathcal{S}$, its Lebesgue measure and associated Hilbert space are denoted by $|\mathcal{S}|$ and $L^2(\mathcal{S})$, respectively.

\section{EM Theory in Wireless Systems}  \label{sec:theory}

In conventional communication models, the EM details of the antenna and propagation environment are often absorbed into an array response or a channel coefficient. For CAPAs, however, this abstraction must be revisited, since their behavior is fundamentally governed by EM theory and constrained by Maxwell’s equations. Therefore, before developing CAPA-specific signal and channel models, it is helpful to first review the EM principles that govern aperture radiation, propagation, reception, and power transfer. This section reviews these foundations, which will be used throughout the rest of the tutorial.

\subsection{How Are EM Waves Radiated?}

In wireless communications, EM waves propagating through space are the physical carriers of information. A wireless transmitter must therefore generate a time-varying EM field, and the device that performs this function is typically an antenna. Like all EM devices, antennas are governed by Maxwell's equations. We begin by recalling their differential form for a point $\mathbf{r} \in \mathbb{R}^{3}$ in free space:
\begin{subequations}
    \begin{align}
        & \nabla \cdot \boldsymbol{\mathcal{E}}(\mathbf{r}, t) = \frac{\rho(\mathbf{r}, t)}{\epsilon_0}, \\
        & \nabla \cdot \boldsymbol{\mathcal{H}}(\mathbf{r}, t) = 0, \\
        \label{MW_3}
        & \nabla \times \boldsymbol{\mathcal{E}}(\mathbf{r}, t) = - \mu_0\frac{\partial \boldsymbol{\mathcal{H}}(\mathbf{r}, t)}{\partial t}, \\
        \label{MW_4}
        & \nabla \times \boldsymbol{\mathcal{H}}(\mathbf{r}, t) = \boldsymbol{\mathcal{J}}(\mathbf{r}, t) +  \epsilon_0 \frac{\partial \boldsymbol{\mathcal{E}}(\mathbf{r}, t)}{\partial t},
    \end{align}
\end{subequations}
where $\epsilon_0$ and $\mu_0$ denote free-space permittivity and permeability, respectively, and $t$ denotes the time. The field variables are defined as the electric field $\boldsymbol{\mathcal{E}}(\mathbf{r}, t) \in \mathbb{R}^{3 \times 1}$, the magnetic field $\boldsymbol{\mathcal{H}}(\mathbf{r}, t) \in \mathbb{R}^{3 \times 1}$, the electric charge density $\rho(\mathbf{r}, t) \in \mathbb{R}$, and the current density $\boldsymbol{\mathcal{J}}(\mathbf{r}, t) \in \mathbb{R}^{3 \times 1}$. Additionally, $\nabla \triangleq [\frac{\partial}{\partial \mathsf{x}}, \frac{\partial}{\partial \mathsf{y}}, \frac{\partial}{\partial \mathsf{z}}]^{\mathsf{T}}$ represents the spatial differential operator, while $\cdot$ and $\times$ denote the dot and cross products, respectively.

EM radiation is characterized by \emph{spatiotemporal} field variations governed by the last two equations, \eqref{MW_3} and \eqref{MW_4}, namely Faraday's law and the Ampère-Maxwell law. 
These laws give the foundations of how an antenna can initiate the EM wave by generating a time-varying source current and through the following sequence:
\begin{enumerate}
    \item \emph{Source generation:} The antenna supports a time-varying source current $\boldsymbol{\mathcal{J}}(\mathbf{r}, t)$ within a finite source region $\mathcal{V}$.
    \item \emph{Magnetic-field excitation:} This current excites a time-varying magnetic field $\boldsymbol{\mathcal{H}}(\mathbf{r}, t)$ near the antenna according to the Ampère-Maxwell law \eqref{MW_4}.
    \item \emph{Electric-field excitation:} The time variation of the magnetic field induces spatial variation of the electric field $\boldsymbol{\mathcal{E}}(\mathbf{r}, t)$ according to Faraday's law \eqref{MW_3}.
    \item \emph{Self-sustaining propagation:} The time-varying electric field further induces the magnetic field, creating a coupled field evolution that transports energy away from the source as an EM wave.
\end{enumerate} 

For wireless communication modeling, the key task is to establish the relationship between the controllable source excitation and the field observed at the receiver. In the EM domain, the source current $\boldsymbol{\mathcal{J}}(\mathbf{r}, t)$ serves as the transmit signal, while the received signal is typically represented by the propagating electric field $\boldsymbol{\mathcal{E}}(\mathbf{r}, t)$ that induces a circuit voltage at the receiver \cite{1386525, 9139337, 10938994}. To characterize their relationship,  we adopt frequency-domain representations: $\boldsymbol{\mathcal{J}}_\omega(\mathbf{r}) \triangleq \mathcal{F}\{\boldsymbol{\mathcal{J}}(\mathbf{r},t)\}$, $\boldsymbol{\mathcal{E}}_\omega(\mathbf{r}) \triangleq \mathcal{F}\{\boldsymbol{\mathcal{E}}(\mathbf{r}, t)\}$, and $\boldsymbol{\mathcal{H}}_\omega(\mathbf{r}) \triangleq \mathcal{F}\{\boldsymbol{\mathcal{H}}(\mathbf{r},t)\}$, where $\mathcal{F}\{\cdot\}$ denotes the Fourier transform operator and $\omega = 2 \pi f$ is the angular frequency and $f$ is the frequency. Assuming a $e^{\mathrm{j} \omega t}$ time factor, the Fourier transform $\mathcal{F}\{\cdot\}$ can be explicitly defined as $\mathcal{F}\{x(t)\} = \int_{-\infty}^{\infty} x(t) e^{-\mathrm{j} \omega t} \mathrm{d}t$. In the frequency domain, the time derivatives in \eqref{MW_3} and \eqref{MW_4} become multiplications by $\mathrm{j} \omega$, leading to the following equations:
\begin{subequations}
    \begin{align}
        \label{Fourier_MW}
        & \nabla \times \boldsymbol{\mathcal{E}}_\omega(\mathbf{r}) = -\mathrm{j} \omega \mu_0 \boldsymbol{\mathcal{H}}_\omega(\mathbf{r}), \\
        \label{Fourier_MW_2}
        & \nabla \times \boldsymbol{\mathcal{H}}_\omega(\mathbf{r}) = \boldsymbol{\mathcal{J}}_\omega(\mathbf{r}) + \mathrm{j} \omega \epsilon_0 \boldsymbol{\mathcal{E}}_\omega(\mathbf{r}).
    \end{align}
\end{subequations}
By eliminating the magnetic field $\boldsymbol{\mathcal{H}}_\omega$, we obtain the differential equation relating the current to the electric field:
\begin{equation} \label{wave_equation}
    \nabla \times \nabla \times \boldsymbol{\mathcal{E}}_\omega(\mathbf{r}) - k_0^2 \boldsymbol{\mathcal{E}}_\omega(\mathbf{r}) = -\mathrm{j} \omega \mu_0 \boldsymbol{\mathcal{J}}_\omega(\mathbf{r}),
\end{equation}
where $k_0 = \omega/c$ is the wavenumber and $c = 1/\sqrt{\mu_0 \epsilon_0}$ is the speed of light. This equation is referred to as the \emph{inhomogeneous Helmholtz wave equation}. Solving this equation for a source current confined within a region $\mathcal{V}$ yields the following transmit-receive relationship in the EM domain:
\begin{equation}
    \label{EM_model_green}
    \boldsymbol{\mathcal{E}}_\omega(\mathbf{r}) = \iiint_{\mathcal{V}} \, \boldsymbol{\mathcal{G}}_{\omega}(\mathbf{r}, \mathbf{s}) \, \boldsymbol{\mathcal{J}}_\omega(\mathbf{s}) \, \mathrm{d} \mathbf{s},
\end{equation}
where $\boldsymbol{\mathcal{G}}_{\omega}(\mathbf{r}, \mathbf{s})$ denotes the dyadic Green's function. For the detailed derivation and solution of \eqref{wave_equation}, we refer the reader to \cite[Appendix I]{poon2005degrees}. Here, we directly provide the explicit expression for the free-space dyadic Green's function:
\begin{align}
    \label{dyadic_Green}
    \boldsymbol{\mathcal{G}}_{\omega} (\mathbf{r}, \mathbf{s}) & = -\mathrm{j} k_0 \eta_0  \left(\mathbf{I} + \frac{1}{k_0^2} \nabla \nabla^{\mathsf{T}}\right) g \left(\|\mathbf{r} - \mathbf{s}\|\right), \\
    g(R) & = \frac{\mathrm{e}^{-\mathrm{j} k_0 R}}{4 \pi R},
\end{align}
where $\eta_0 = \sqrt{\mu_0/\epsilon_0}$ is the intrinsic impedance of free space and $g(\cdot)$ is the scalar Green's function.  

\begin{figure}[!t]
 \centering
\includegraphics[width=0.5\textwidth]{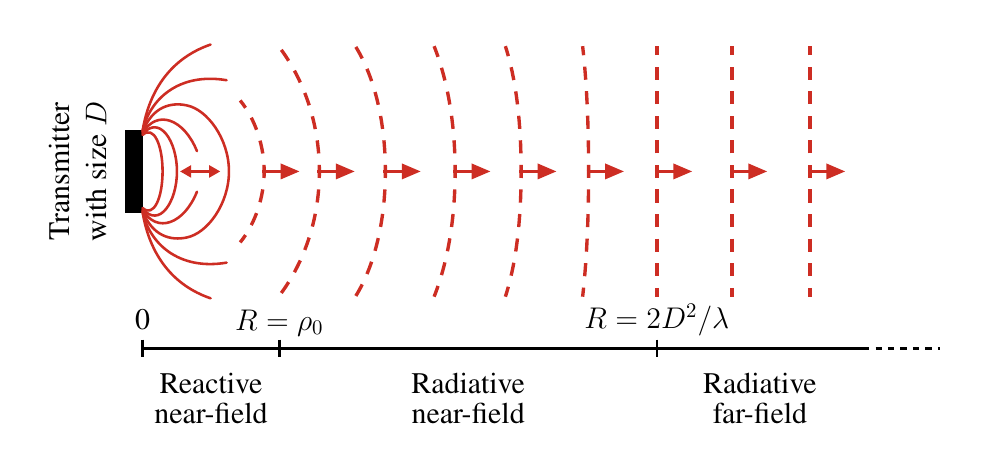}
\caption{Illustration of the spatial evolution of EM waves from the reactive near-field to the radiative far-field. The boundary $R=\rho_0$ marks the transition from the reactive near-field to the radiative near-field, where $\rho_0$ is criterion-dependent, e.g., $\rho_0=0.62\sqrt{D^3/\lambda}$ under the classical antenna criterion or $\rho_0=\lambda$ when the reactive terms are negligible for the channel gain metric \cite{10614327}. The boundary to the far-field is the Fraunhofer distance $R=2D^2/\lambda$.}
\label{fig:spatial_regions}
\end{figure}


This dyadic Green’s-function model serves as the foundation for the remainder of this tutorial. It explicitly reveals how distance, direction, polarization, power flow, and coupling are incorporated into the continuous-space signal model. We next review these properties.

\subsection{Near Field and Far Field} \label{sec:near_far}

The propagation characteristics of EM waves depend fundamentally on the distance between the observation point and the source region. For a finite radiating aperture, classical antenna theory therefore partitions the surrounding space into the reactive near-field region, the radiative near-field (or Fresnel) region, and the far-field (or Fraunhofer) region, with practical boundaries typically expressed in terms of the aperture size and wavelength \cite{balanis2016antenna,selvan2017fraunhofer,bjornson2021primer,liu2023near}. However, the term ``near field'' is often used ambiguously. In some contexts, it refers only to the reactive region dominated by stored energy, whereas in others it also includes the radiative region, where the wavefront remains spherical and varies across the aperture. To clarify this distinction, we explicitly expand the dyadic Green's function in \eqref{dyadic_Green}. Evaluating the operator $\nabla\nabla^{\mathsf{T}}$ yields
\begin{align}
    \boldsymbol{\mathcal{G}}_{\omega}(\mathbf{r}, \mathbf{s}) = -\mathrm{j} k_0 \eta_0 & \Bigg[ (\mathbf{I} - \hat{\mathbf{u}} \hat{\mathbf{u}}^{\mathsf{T}} ) - \frac{\mathrm{j}}{k_0 R}\left( \mathbf{I} - 3 \hat{\mathbf{u}} \hat{\mathbf{u}}^{\mathsf{T}} \right) \nonumber \\
    & - \frac{1}{(k_0 R)^2}\left( \mathbf{I} - 3 \hat{\mathbf{u}} \hat{\mathbf{u}}^{\mathsf{T}} \right) \Bigg] \, \frac{\mathrm{e}^{-\mathrm{j} k_0 R}}{4 \pi R}, \label{eq:dyadicGreen}
\end{align}
where $R = \|\mathbf{r}-\mathbf{s}\|$ denotes the distance between the observation point $\mathbf{r}$ and the source point $\mathbf{s}$, and $\hat{\mathbf{u}} = (\mathbf{r}-\mathbf{s})/R$ is the corresponding unit direction vector. This expression contains three groups of terms that scale as $1/R$, $1/R^2$, and $1/R^3$, respectively. The $1/R$ term represents the radiative component, whereas the $1/R^2$ and $1/R^3$ terms correspond to the induction and quasi-static components. 

In the existing communication literature, the near-field and far-field regions are often defined \rev{primarily} in terms of wavefront phase variation \cite{balanis2016antenna,selvan2017fraunhofer,bjornson2021primer,liu2023near}, overlooking important physical insights such as the roles of the $1/R$, $1/R^2$, and $1/R^3$ terms, as well as the relationship between the electric and magnetic fields. In the following, we present a more physics-consistent classification of the spatial regions based on the relative magnitudes of these terms and the resulting field characteristics, as illustrated in Fig.~\ref{fig:spatial_regions}.

\subsubsection{Reactive Near-Field Region} This region is governed by the $1/R^2$ and $1/R^3$ terms, which contribute significantly when $k_0 R \ll 1$. The Green's function can thus be approximated by 
\begin{equation}
    \boldsymbol{\mathcal{G}}_{\omega}(\mathbf{r}, \mathbf{s}) \approx \mathrm{j} \eta_0 \left( \frac{\mathrm{j}}{R} + \frac{1}{k_0 R^2} \right) \left( \mathbf{I} - 3 \hat{\mathbf{u}} \hat{\mathbf{u}}^{\mathsf{T}} \right) \frac{\mathrm{e}^{-\mathrm{j} k_0 R}}{4 \pi R}.
\end{equation}
In this region, the fields are not purely transverse, i.e., they are not orthogonal to the propagation direction. This is because $(\mathbf{I} - \hat{\mathbf{u}}\hat{\mathbf{u}}^{\mathsf{T}})$ is the orthogonal projector onto the plane perpendicular to $\hat{\mathbf{u}}$, whereas $(\mathbf{I} - 3\hat{\mathbf{u}}\hat{\mathbf{u}}^{\mathsf{T}})$ is not a projection operator and, in general, couples transverse and radial components. As a result, the induction and quasi-static terms produce non-negligible radial field components, and the energy exchange is dominated by reactive storage rather than radiation.
    
\subsubsection{Radiative Near-Field Region} As the distance increases, the $1/R$ term in \eqref{eq:dyadicGreen} becomes dominant, and the field begins to exhibit radiative behavior. \rev{Comparing the terms in \eqref{eq:dyadicGreen} shows that the $1/R^2$ and $1/R^3$ reactive terms become negligible relative to the $1/R$ radiative term once $k_0 R \gg 1$, i.e., when the observation point is several wavelengths away from every source point. For a finite aperture, classical antenna theory adopts the engineering boundary} $R \ge 0.62 \sqrt{D^3/\lambda}$, where $D$ denotes the maximum dimension of the cross section of $\mathcal{V}$ orthogonal to the propagation direction, and $\lambda = c/f$ is the wavelength \cite{balanis2016antenna}. However, it is important to note that this boundary should not be interpreted as a sharp physical demarcation, but rather as a practical guideline, particularly because it is originally derived from a phase-error criterion based on approximating $e^{-\mathrm{j}k_0 R}$ through a binomial expansion of $R$ \cite{walter1965traveling, selvan2017fraunhofer}, instead of directly evaluating the relative magnitudes of the $1/R$, $1/R^2$, and $1/R^3$ terms. Under some performance metrics, such as channel gain, the contributions of the $1/R^2$ and $1/R^3$ terms have been shown to be negligible whenever $R \ge \lambda$ \cite{10614327}. \rev{This distinction becomes particularly pronounced for electrically large CAPAs. For an aperture of $D=1$~m operating at $30$~GHz ($\lambda=1$~cm), the classical boundary yields $0.62\sqrt{D^3/\lambda}\approx 6.2$~m, whereas the actual reactive fields remain confined to within a few wavelengths of the aperture surface \cite{10614327}. Consequently, the reactive near-field region of a CAPA forms a thin layer clinging to the aperture, rather than a vast volume whose extent scales with $D^{3/2}$.}

In this region, the wavefront remains spherical, but the fields are predominantly transverse, and the energy is mainly radiated outward. Accordingly, the Green's function can be approximated by retaining only the radiative component \cite{poon2005degrees, 9139337}:
\begin{equation}
    \boldsymbol{\mathcal{G}}_{\omega}(\mathbf{r}, \mathbf{s}) \approx -\mathrm{j} k_0 \eta_0 \left(\mathbf{I} - \hat{\mathbf{u}} \hat{\mathbf{u}}^{\mathsf{T}} \right) \frac{\mathrm{e}^{-\mathrm{j} k_0 R}}{4 \pi R}.
\end{equation}  
To characterize the corresponding magnetic field, we evaluate the curl of the Green's function according to \eqref{Fourier_MW} as follows:
\begin{align} \label{m_green_function}
    \nabla \times \boldsymbol{\mathcal{G}}_{\omega}(\mathbf{r}, \mathbf{s}) = &\mathrm{j} k_0 \eta_0 \left( \hat{\mathbf{u}} \times \mathbf{I} \right) \frac{\mathrm{e}^{-\mathrm{j}k_0 R}}{4\pi} \left( \frac{1}{R^2} + \frac{\mathrm{j} k_0}{R} \right) \nonumber \\
    \approx & - k_0^2 \eta_0 \left( \hat{\mathbf{u}} \times \mathbf{I} \right) \frac{\mathrm{e}^{-\mathrm{j}k_0 R}}{4\pi R},
\end{align}
where the $1/R^2$ term is neglected in the last step. Let us now consider a specific point current source located at $\mathbf{s}_0 \in \mathcal{V}$ with current density $\boldsymbol{\mathcal{J}}_\omega(\mathbf{s}_0)$. The resulting electric field is
\begin{align}
    \bar{\boldsymbol{\mathcal{E}}}_\omega(\mathbf{r}) &= \boldsymbol{\mathcal{G}}_{\omega}(\mathbf{r}, \mathbf{s}_0) \, \boldsymbol{\mathcal{J}}_\omega(\mathbf{s}_0) \nonumber \\
    & = -\mathrm{j} k_0 \eta_0 \frac{\mathrm{e}^{-\mathrm{j} k_0 R_0}}{4 \pi R_0} \left(\mathbf{I} - \hat{\mathbf{u}}_0 \hat{\mathbf{u}}_0^{\mathsf{T}} \right) \boldsymbol{\mathcal{J}}_\omega(\mathbf{s}_0),
\end{align} 
where $R_0 = \|\mathbf{r} - \mathbf{s}_0\|$ and $\hat{\mathbf{u}}_0 = (\mathbf{r} - \mathbf{s}_0)/R_0$. Then, we have $\hat{\mathbf{u}}_0 \cdot \boldsymbol{\mathcal{G}}_{\omega}(\mathbf{r}, \mathbf{s}_0) = \mathbf{0}$, implying the electric field is orthogonal to the propagation direction. Furthermore, using the result in \eqref{m_green_function}, the magnetic field is given by
\begin{align}
    \bar{\boldsymbol{\mathcal{H}}}_\omega(\mathbf{r}) &= -\frac{1}{\mathrm{j} \omega \mu_0} \nabla \times \left(\boldsymbol{\mathcal{G}}_{\omega}(\mathbf{r}, \mathbf{s}_0) \, \boldsymbol{\mathcal{J}}_\omega(\mathbf{s}_0)\right) \nonumber \\
    & \overset{(a)}{=} -\frac{1}{\mathrm{j} \omega \mu_0} \left( \nabla \times \boldsymbol{\mathcal{G}}_{\omega}(\mathbf{r}, \mathbf{s}_0)\right) \, \boldsymbol{\mathcal{J}}_\omega(\mathbf{s}_0) \nonumber \\
    & = -\mathrm{j} k_0 \frac{\mathrm{e}^{-\mathrm{j} k_0 R_0}}{4 \pi R_0} \left(\hat{\mathbf{u}}_0 \times  \boldsymbol{\mathcal{J}}_\omega(\mathbf{s}_0)\right) \nonumber \\
    & \overset{(b)}{=} -\mathrm{j} k_0 \frac{\mathrm{e}^{-\mathrm{j} k_0 R_0}}{4 \pi R_0} \Big[\hat{\mathbf{u}}_0 \times \left( (\mathbf{I} - \hat{\mathbf{u}}_0 \hat{\mathbf{u}}_0^{\mathsf{T}}) \boldsymbol{\mathcal{J}}_\omega(\mathbf{s}_0)\right)\Big] \nonumber \\
    & = \frac{1}{\eta_0} \hat{\mathbf{u}}_0 \times \bar{\boldsymbol{\mathcal{E}}}_\omega(\mathbf{r}),
\end{align}
where $(a)$ holds because $\boldsymbol{\mathcal{J}}_\omega(\mathbf{s}_0)$ is independent of the observation point $\mathbf{r}$, and $(b)$ uses the identity $\hat{\mathbf{u}}_0 \times \hat{\mathbf{u}}_0 = \mathbf{0}$. These expressions show that in the radiative region the fields exhibit local plane-wave behavior, where the electric field, magnetic field, and propagation direction form an orthogonal triplet, i.e.,
\begin{equation} \label{NF_ortho_condition}
\bar{\boldsymbol{\mathcal{E}}}_\omega(\mathbf{r})\,\bot \, \bar{\boldsymbol{\mathcal{H}}}_\omega(\mathbf{r}) \, \bot \, \hat{\mathbf{u}}_0.   
\end{equation} 

However, this region is still classified as the near field because this plane-wave property is only local. The total field is the superposition of contributions from the entire finite source region $\mathcal{V}$, cf. \eqref{EM_model_green}, and the direction vector $\hat{\mathbf{u}}$ varies across the aperture relative to the observation point. As a result, the aggregate wavefront is still spherical, and therefore the planar orthogonality condition in \eqref{NF_ortho_condition} does not hold globally. This spherical-wave behavior enables many near-field effects of practical interest; see \cite{liu2023near} and \cite{10934790} for further discussion.

\subsubsection{Radiative Far-Field Region} 
When the distance exceeds the \emph{Fraunhofer distance}, i.e., $R \ge 2D^2/\lambda$, the local plane-wave behavior becomes global, and the radiation pattern becomes distance-independent. This region is referred to as the far field or Fraunhofer region. This threshold follows from bounding the maximum phase error in $e^{-\mathrm{j} k_0 R}$ introduced by \eqref{far_field_approx} to not exceed $\pi/8$ \cite{liu2023near}, which is an arbitrarily selected limit. 

To illustrate the far-field behavior, define a reference point $\mathbf{s}_{\mathrm{ref}} \in \mathcal{V}$. The far-field condition requires $\hat{\mathbf{u}} \approx \hat{\boldsymbol{\kappa}}$ for all $\mathbf{s} \in \mathcal{V}$, where the global direction vector is $\hat{\boldsymbol{\kappa}} \triangleq (\mathbf{r} - \mathbf{s}_{\mathrm{ref}})/r$ and $r \triangleq \|\mathbf{r} - \mathbf{s}_{\mathrm{ref}}\|$ \cite{poon2005degrees, liu2023near}. Under this assumption, the distance $R = \|\mathbf{r} - \mathbf{s}\| = \| r \hat{\boldsymbol{\kappa}} + \mathbf{s}_{\mathrm{ref}} - \mathbf{s} \|$ can be approximated as
\begin{equation}
    \label{far_field_approx}
    R \approx r - \hat{\boldsymbol{\kappa}}^{\mathsf{T}} \left(\mathbf{s} - \mathbf{s}_{\mathrm{ref}} \right).
\end{equation}  
With this approximation, the electric field in \eqref{EM_model_green} simplifies to~\cite{poon2005degrees, liu2023near}
\begin{equation} 
    \label{far_field_E_field}
    \boldsymbol{\mathcal{E}}_{\omega}(\mathbf{r}) \approx -\mathrm{j} k_0 \eta_0 \frac{\mathrm{e}^{-\mathrm{j} k_0 r }}{4 \pi r} (\mathbf{I} - \hat{\boldsymbol{\kappa}} \hat{\boldsymbol{\kappa}}^{\mathsf{T}}) \boldsymbol{\mathcal{A}}(\hat{\boldsymbol{\kappa}}), 
\end{equation}
where $\boldsymbol{\mathcal{A}}(\hat{\boldsymbol{\kappa}}) = \iiint_{\mathcal{V}} \mathrm{e}^{\mathrm{j} k_0 \hat{\boldsymbol{\kappa}}^{\mathsf{T}}(\mathbf{s} - \mathbf{s}_{\mathrm{ref}})} \boldsymbol{\mathcal{J}}_\omega(\mathbf{s}) \, \mathrm{d} \mathbf{s}$ is the distance-independent radiation pattern. Another useful far-field approximation is $\nabla \approx -\mathrm{j} k_0 \hat{\boldsymbol{\kappa}}$, which leads to the magnetic field expression
\begin{equation}
    \label{far_field_H_field}
    \boldsymbol{\mathcal{H}}_{\omega}(\mathbf{r}) = -\frac{1}{\mathrm{j} \omega \mu_0} \nabla \times \boldsymbol{\mathcal{E}}_{\omega}(\mathbf{r}) \approx \frac{1}{\eta_0} \hat{\boldsymbol{\kappa}} \times \boldsymbol{\mathcal{E}}_{\omega}(\mathbf{r}).
\end{equation}
In this case, the plane-wave orthogonality relation is no longer merely local but holds globally:
\begin{equation}
    \boldsymbol{\mathcal{E}}_\omega(\mathbf{r})\,\bot \, \boldsymbol{\mathcal{H}}_\omega(\mathbf{r}) \, \bot \, \hat{\boldsymbol{\kappa}}. 
\end{equation}

Based on the above analysis, near-field and far-field behavior can be characterized more precisely from the perspective of the orthogonality among the electric field, magnetic field, and propagation direction. More specifically, 1) in the reactive near-field region, these quantities are not mutually orthogonal; 2) in the radiative near-field region, they are locally orthogonal, while the overall wavefront remains spherical; and 3) in the radiative far-field region, they are globally orthogonal, and the wavefront becomes effectively planar. This perspective provides a more fundamental understanding of the spatial evolution of EM waves.

\subsection{Polarization} \label{sec:polarization}
Polarization describes the directional structure of an EM field. Since EM fields are vector fields, the source current, electric field, and magnetic field can each be decomposed into three orthogonal spatial components, which describe oscillations along different spatial directions. This directional property is referred to as polarization. To illustrate this, consider decomposing the source current density into orthogonal $\mathsf{x}$-, $\mathsf{y}$-, and $\mathsf{z}$-components:
\begin{equation}
    \label{tri_pol_current}
    \boldsymbol{\mathcal{J}}_\omega(\mathbf{s}) = \mathcal{J}_{\mathsf{x}}(\mathbf{s})\, \hat{\bm{\mathsf{x}}} + \mathcal{J}_{\mathsf{y}}(\mathbf{s})\, \hat{\bm{\mathsf{y}}} + \mathcal{J}_{\mathsf{z}}(\mathbf{s})\, \hat{\bm{\mathsf{z}}},
\end{equation}   
where $\mathcal{J}_{\mathsf{x}}(\mathbf{s})$, $\mathcal{J}_{\mathsf{y}}(\mathbf{s})$, and $\mathcal{J}_{\mathsf{z}}(\mathbf{s})$ are the current components along the $\mathsf{x}$-, $\mathsf{y}$-, and $\mathsf{z}$-directions, respectively, and $\hat{\bm{\mathsf{x}}}$, $\hat{\bm{\mathsf{y}}}$, and $\hat{\bm{\mathsf{z}}}$ are the corresponding unit vectors. An analogous decomposition applies to the electric field $\boldsymbol{\mathcal{E}}_{\omega}(\mathbf{r})$ and the magnetic field $\boldsymbol{\mathcal{H}}_{\omega}(\mathbf{r})$. Based on which components are excited at the Tx and captured at the Rx, several polarization configurations can be distinguished as follows.

\subsubsection{Uni-Polarization} In a uni-polarized system, the source current is excited along only one polarization direction at the Tx and/or only one polarization component of the EM field is measured at the Rx. In particular, for a uni-polarized Tx, the source current can be written as
\begin{equation}
    \boldsymbol{\mathcal{J}}_{\omega} (\mathbf{s}) = \mathcal{J}_{\mathrm{p}}(\mathbf{s}) \, \hat{\mathbf{p}}_{\mathrm{t}},
\end{equation}
where $\hat{\mathbf{p}}_{\mathrm{t}} \in \mathbb{R}^{3 \times 1}$ is a unit direction vector. Thus, the source current oscillates only along $\hat{\mathbf{p}}_{\mathrm{t}}$. Nevertheless, even a uni-polarized source generally produces a vector EM field with multiple components:
\begin{equation}
\boldsymbol{\mathcal{E}}_\omega(\mathbf{r}) = \iiint_{\mathcal{V}}\, \boldsymbol{\mathcal{G}}_{\omega}(\mathbf{r}, \mathbf{s}) \, \mathcal{J}_{\mathrm{p}}(\mathbf{s}) \, \hat{\mathbf{p}}_{\mathrm{t}} \, \mathrm{d} \mathbf{s},
\end{equation}
If the Rx is sensitive to only a single polarization direction $\hat{\mathbf{p}}_{\mathrm{r}} \in \mathbb{R}^{3 \times 1}$, then the effective measured field reduces to
\begin{equation}
    \mathcal{E}_{\mathrm{p}}(\mathbf{r}) = \hat{\mathbf{p}}_{\mathrm{r}}^{\mathsf{T}} \boldsymbol{\mathcal{E}}_\omega(\mathbf{r})
    = \iiint_{\mathcal{V}} \Big[\hat{\mathbf{p}}_{\mathrm{r}}^{\mathsf{T}} \boldsymbol{\mathcal{G}}_{\omega}(\mathbf{r}, \mathbf{s}) \hat{\mathbf{p}}_{\mathrm{t}} \Big] \mathcal{J}_{\mathrm{p}}(\mathbf{s}) \, \mathrm{d} \mathbf{s}.
\end{equation}

The above expression reveals the issue of \emph{polarization mismatch}. Beyond the reactive near-field region, the kernel term $\hat{\mathbf{p}}_{\mathrm{r}}^{\mathsf{T}} \boldsymbol{\mathcal{G}}_{\omega}(\mathbf{r}, \mathbf{s}) \hat{\mathbf{p}}_{\mathrm{t}}$ is proportional to $\hat{\mathbf{p}}_{\mathrm{r}}^{\mathsf{T}} (\mathbf{I} - \hat{\mathbf{u}} \hat{\mathbf{u}}^{\mathsf{T}}) \hat{\mathbf{p}}_{\mathrm{t}}$, which can vanish for certain polarization choices. For example, if $\hat{\mathbf{p}}_{\mathrm{t}} = \hat{\bm{\mathsf{x}}}$, $\hat{\mathbf{p}}_{\mathrm{r}} = \hat{\bm{\mathsf{y}}}$, and $\hat{\mathbf{u}} = \hat{\bm{\mathsf{z}}}$, then the received field is zero. Hence, the polarization directions at the Tx and Rx must be chosen carefully to avoid severe mismatch losses.

\subsubsection{Dual-Polarization} Dual-polarization is the most widely used polarization configuration in modern wireless base stations, and frequently also in user devices. In this case, the source current is expressed as a superposition of two orthogonal components:
\begin{equation}
    \boldsymbol{\mathcal{J}}_{\omega} (\mathbf{s}) = \mathcal{J}_1(\mathbf{s}) \, \hat{\mathbf{p}}_{\mathrm{t}, 1} + \mathcal{J}_2(\mathbf{s}) \, \hat{\mathbf{p}}_{\mathrm{t}, 2},
\end{equation}
where the unit vectors satisfy $\hat{\mathbf{p}}_{\mathrm{t}, 1} \perp \hat{\mathbf{p}}_{\mathrm{t}, 2}$. Introducing a second polarization brings several important benefits. First, it roughly doubles the available field DoFs while using nearly the same physical aperture. Second, it improves robustness to polarization mismatch, since the Tx and/or Rx can exploit two orthogonal components even when the channel rotates the polarization or induces partial depolarization. Third, it enables more flexible polarization synthesis. By adjusting the relative amplitudes and phases of the two components, one can generate a variety of polarization states, including circular polarization when the two components have equal amplitude and a $90^\circ$ phase offset. If the Tx and Rx both exploit two orthogonal polarizations, the link becomes a $2\times2$ polarized coupling channel, allowing the system to better trade off multiplexing and diversity gains.

\subsubsection{Tri-Polarization} In an ideal tri-polarized system, the Tx can excite source currents along all three orthogonal polarization directions, as described by \eqref{tri_pol_current}. At first sight, one may expect that this would always triple the available DoFs. In reality, the achievable gain depends strongly on the propagation environment.

First consider the free-space line-of-sight (LoS) model in \eqref{EM_model_green} with the dyadic Green's function in \eqref{dyadic_Green}. In the far-field region, the electric field generated by a tri-polarized source remains transverse to the propagation direction, cf. \eqref{far_field_E_field}. Therefore, even if the Rx can measure the full three-dimensional electric field vector, it can capture at most two independent electric-field components. As a result, the DoFs offered by a tri-polarized source cannot be fully exploited in far-field LoS scenarios. In contrast, in near-field LoS propagation, the aggregate electric field is not strictly transverse, and a non-negligible radial component may arise. In principle, this allows a full-rank $3 \times 3$ electric-field coupling between the Tx and Rx. Hence, under LoS propagation, the gain of tri-polarization is most significant in the near-field region.

The situation is different in scattering environments. Scattering can rotate and mix the field polarization, so the electric field observed at the Rx is no longer constrained to remain perpendicular to a single LoS propagation direction. Consequently, even in far-field scenarios, the received electric field need not be confined to a two-dimensional subspace associated with one propagation direction. In such environments, using a tri-polarized source is generally more beneficial than in pure far-field LoS propagation, assuming the same signal strength. Moreover, after scattering, the magnetic field also provides three additional vector dimensions that are distinct from the electric field. In principle, this yields up to six field DoFs. Fully exploiting them would require not only tri-polarized electric antennas, which excite currents and sense electric fields, but also their tri-polarized magnetic counterparts. Such a system can, in principle, support a full-rank $6 \times 6$ Tx-Rx coupling in rich scattering environments \cite{andrews2001tripling}, although realizing this architecture in practice remains highly challenging.

\rev{Specifically, exploiting all six dimensions requires co-locating three orthogonal electric dipoles and three orthogonal magnetic dipoles (e.g., small loops) at each point, with each element driven by a dedicated radio-frequency (RF) chain, akin to EM vector sensors \cite{nehorai1994vector}. Accordingly, the noise model, the mutual-coupling model, and the power constraint all need to account for the correlations and cross-couplings between the electric and magnetic components. In rich scattering, compact EM vector antennas can exploit angular and polarization diversity to access six independent local field components. In particular, the authors of \cite{tian2012experimental} experimentally verified six DoFs using six-port electric-magnetic antenna arrays. For a single plane wave, however, the magnetic field is determined by the electric field through \eqref{far_field_H_field}. Thus, for a large planar aperture that resolves the propagation directions in one half-space, each direction still supports only two independent transverse polarizations.}

\subsection{Power} \label{sec:power}

Power is an important factor in EM theory. According to Poynting's theorem, the work done by a time-varying current on the surrounding field is balanced by radiated energy, stored reactive energy, and material dissipation \cite{pozar2021microwave,orfanidis2016electromagnetic}. Thus, the power associated with a radiating structure cannot, in general, be determined from the source current alone. It also depends on the field induced by that current and on the EM properties of the supporting medium. We therefore first review the radiated power of a lossless source region, and then include material dissipation through a surface-current formulation.

\subsubsection{Radiated Power} In the present setting, radiated power can be computed in two standard ways: the \emph{flux method} and the \emph{source method}. The flux method is based on the time-averaged Poynting vector, which describes the directional power flux density of the EM field and is given by
\begin{equation}
    \boldsymbol{\mathcal{S}}_{\omega}(\mathbf{r})
    = \frac{1}{2}\,\boldsymbol{\mathcal{E}}_{\omega}(\mathbf{r}) \times \boldsymbol{\mathcal{H}}^*_{\omega}(\mathbf{r}).
\end{equation}
The total radiated power is obtained by integrating the outward power flux over a closed surface $\partial\mathcal{V}$ enclosing the source region $\mathcal{V}$ as follows:
\begin{equation}\label{power_Poynting_vector}
    P_{\mathrm{rad}}
    = \oiint_{\partial \mathcal{V}}
    \Re\!\left\{\boldsymbol{\mathcal{S}}_{\omega}(\mathbf{r})\right\}
    \cdot \hat{\mathbf{n}}\, \mathrm{d}A,
\end{equation}
where $\hat{\mathbf{n}}$ is the outward unit normal vector on $\partial\mathcal{V}$ and $\mathrm{d}A$ is the surface area element. For analytical convenience, $\partial\mathcal{V}$ is typically chosen as a sphere of radius $r$ centered at a reference point $\mathbf{s}_{\mathrm{ref}}$, and the limit $r\to\infty$ is taken. Assuming $\mathbf{s}_{\mathrm{ref}}=\mathbf{0}$ without loss of generality, the outward normal becomes $\hat{\mathbf{n}}=\hat{\boldsymbol{\kappa}}$, and
\begin{align}\label{power_farfield_sphere}
    P_{\mathrm{rad}} = &\lim_{r\to\infty} \oiint_{\|\mathbf{r}\|=r}
    \! \Re\left\{ \boldsymbol{\mathcal{S}}_{\omega}(\mathbf{r}) \right\}\cdot \hat{\boldsymbol{\kappa}} \, \mathrm{d}A \nonumber \\
    = &\lim_{r\to\infty}\oiint_{\|\mathbf{r}\|=r}
    \frac{1}{2\eta_{0}}\,
    \big\|\boldsymbol{\mathcal{E}}_{\omega}(\mathbf{r})\big\|^{2}\,
    \mathrm{d}A, 
\end{align}
where the last step follows from the far-field relation \eqref{far_field_H_field} together with the orthogonality of the triplet $(\boldsymbol{\mathcal{E}}_{\omega},\boldsymbol{\mathcal{H}}_{\omega},\hat{\boldsymbol{\kappa}})$.

The source method provides an alternative, and often more efficient, route by calculating the power directly from the source current within $\mathcal{V}$. In a lossless system, the radiated power equals the total power delivered by the source current density $\boldsymbol{\mathcal{J}}_{\omega}$ to the field:
\begin{align}
    \label{source_method_radiation}
    P_{\mathrm{rad}} = & - \iiint_{\mathcal{V}} \frac{1}{2} \Re \big\{ \boldsymbol{\mathcal{J}}^*_{\omega}(\mathbf{s}) \cdot \boldsymbol{\mathcal{E}}_{\omega}(\mathbf{s})   \big\} \, \mathrm{d} \mathbf{s} \nonumber \\
    = & - \iiint_{\mathcal{V}} \iiint_{\mathcal{V}} \frac{1}{2}
    \Re \left\{ \boldsymbol{\mathcal{J}}_{\omega}^{\mathsf{H}}(\mathbf{s}) \, \boldsymbol{\mathcal{G}}_{\omega}(\mathbf{s}, \mathbf{s}^{\prime}) \, \boldsymbol{\mathcal{J}}_\omega(\mathbf{s}^{\prime}) \right\}
    \, \mathrm{d} \mathbf{s}^{\prime} \mathrm{d} \mathbf{s} \nonumber \\
    = & - \iiint_{\mathcal{V}} \iiint_{\mathcal{V}} \frac{1}{2}
    \boldsymbol{\mathcal{J}}_{\omega}^{\mathsf{H}}(\mathbf{s}) \, \Re \big\{\boldsymbol{\mathcal{G}}_{\omega}(\mathbf{s}, \mathbf{s}^{\prime}) \big\} \, \boldsymbol{\mathcal{J}}_\omega(\mathbf{s}^{\prime})
    \, \mathrm{d} \mathbf{s}^{\prime} \mathrm{d} \mathbf{s},
\end{align}
where the negative sign follows the convention that $\boldsymbol{\mathcal{J}}_{\omega}^* \cdot \boldsymbol{\mathcal{E}}_{\omega}$ represents absorbed power, so its negation represents radiated power. The last step follows from \cite[Appendix B]{11006094} \rev{and relies on the reciprocity $\boldsymbol{\mathcal{G}}_{\omega}(\mathbf{s},\mathbf{s}^{\prime})=\boldsymbol{\mathcal{G}}_{\omega}^{\mathsf{T}}(\mathbf{s}^{\prime},\mathbf{s})$}. Here, the full dyadic Green's function \eqref{dyadic_Green} \rev{is needed} because $\mathbf{s}$ and $\mathbf{s}^{\prime}$ are both located inside the source region $\mathcal{V}$. The flux method and the source method are equivalent by the Poynting theorem; see \cite[Appendix A]{wang2025mutual} for a proof connecting \eqref{power_farfield_sphere} and \eqref{source_method_radiation}.

\subsubsection{Dissipated Power}
In realistic structures, not all power delivered by the source is converted into radiation. When the material in the region $\mathbf{r}\in\mathcal{V}$ has nonzero conductivity, part of the supplied power is dissipated as heat. In this case, \eqref{Fourier_MW_2} is modified as \cite{pozar2021microwave}
\begin{align}
    \nabla \times \boldsymbol{\mathcal{H}}_\omega(\mathbf{r})
    = & \boldsymbol{\mathcal{J}}_\omega(\mathbf{r}) + \sigma(\mathbf{r}) \boldsymbol{\mathcal{E}}_\omega(\mathbf{r}) + \mathrm{j} \omega \epsilon'(\mathbf{r}) \boldsymbol{\mathcal{E}}_\omega(\mathbf{r}) \nonumber \\
    = & \boldsymbol{\mathcal{J}}_\omega(\mathbf{r}) + \mathrm{j} \omega \tilde{\epsilon}(\mathbf{r}) \boldsymbol{\mathcal{E}}_\omega(\mathbf{r}),
\end{align}
where $\sigma(\mathbf{r}) \boldsymbol{\mathcal{E}}_\omega(\mathbf{r})$ is the conduction current, and $\tilde{\epsilon}(\mathbf{r}) \triangleq \epsilon'(\mathbf{r}) - \mathrm{j}\sigma(\mathbf{r})/\omega$ is the effective complex permittivity. In the free-space background considered above, we have $\epsilon'(\mathbf{r})=\epsilon_0$. The imaginary part of $\tilde{\epsilon}(\mathbf{r})$ accounts for ohmic loss, showing that conduction can be equivalently incorporated into the material response through a complex permittivity. Once such losses are included, however, the dyadic Green's function generally becomes much more difficult to characterize explicitly, and the resulting field model differs fundamentally from the lossless case.

For many thin conducting sheets, apertures, and metasurfaces, solving the full volumetric loss problem is unnecessary and often inconvenient. In such cases, it is more natural to adopt a \emph{surface-current} formulation, in which the material properties are absorbed into a surface impedance \cite{yang2019surface}. Specifically, consider a two-dimensional surface $\mathcal{S}$ with surface impedance $Z_s(\mathbf{s})$ and surface current $\boldsymbol{\mathcal{J}}^{\mathrm{surf}}_{\omega}(\mathbf{s})$. The corresponding surface-impedance boundary condition is
\begin{equation} \label{boundary_condition}
    \boldsymbol{\mathcal{E}}_{\mathrm{src}, \omega}^t(\mathbf{s}) + \boldsymbol{\mathcal{E}}_{\omega}^t(\mathbf{s}) = Z_s(\mathbf{s}) \boldsymbol{\mathcal{J}}^{\mathrm{surf}}_{\omega}(\mathbf{s}), \, \forall \mathbf{s} \in \mathcal{S}.
\end{equation}
Here, $\boldsymbol{\mathcal{E}}_{\mathrm{src}, \omega}^t(\mathbf{s})$ denotes the tangential component of the impressed field that drives the surface current, whereas $\boldsymbol{\mathcal{E}}_{\omega}^t(\mathbf{s})$ is the tangential component of the field reradiated by the induced surface current itself. Their sum therefore represents the total tangential electric field acting on the surface. The reradiated electric field and its tangential component are given by
\begin{align}
    \label{surface_current_Green}
    \boldsymbol{\mathcal{E}}_{\omega}(\mathbf{r}) = & \iint_{\mathcal{S}} \boldsymbol{\mathcal{G}}_{\omega}(\mathbf{r}, \mathbf{s}) \, \boldsymbol{\mathcal{J}}^{\mathrm{surf}}_{\omega}(\mathbf{s}) \, \mathrm{d}\mathbf{s}, \\
    \boldsymbol{\mathcal{E}}_{\omega}^t(\mathbf{s}) = & \big( \mathbf{I} - \hat{\mathbf{q}}\hat{\mathbf{q}}^{\mathsf{T}} \big)\boldsymbol{\mathcal{E}}_{\omega}(\mathbf{s}),
\end{align}
where $\hat{\mathbf{q}}$ is the unit normal vector to $\mathcal{S}$. Since the current is constrained to flow tangentially along the surface, it satisfies $\big( \mathbf{I} - \hat{\mathbf{q}}\hat{\mathbf{q}}^{\mathsf{T}} \big)\boldsymbol{\mathcal{J}}^{\mathrm{surf}}_{\omega}(\mathbf{s}) = \boldsymbol{\mathcal{J}}^{\mathrm{surf}}_{\omega}(\mathbf{s})$. Multiplying \eqref{boundary_condition} by $\frac{1}{2}\boldsymbol{\mathcal{J}}_{\omega}^{\mathrm{surf},*}(\mathbf{s})$, integrating over $\mathcal{S}$, and taking the real part yields the following power relation
\begin{equation}
    P_{\mathrm{src}} = P_{\mathrm{rad}} + P_{\mathrm{loss}},
\end{equation}
where the source power, dissipated power, and radiated power are given by
\begin{align}
    P_{\mathrm{src}} & = \iint_{\mathcal{S}} \frac{1}{2} \Re \Big\{\boldsymbol{\mathcal{J}}^{\mathrm{surf},*}_{\omega}(\mathbf{s}) \cdot \boldsymbol{\mathcal{E}}_{\mathrm{src}, \omega}^t(\mathbf{s}) \Big\} \, \mathrm{d}\mathbf{s}, \\
    P_{\mathrm{loss}} & = \iint_{\mathcal{S}} \frac{1}{2} \Re\{Z_s(\mathbf{s})\} \big\| \boldsymbol{\mathcal{J}}^{\mathrm{surf}}_{\omega}(\mathbf{s}) \big\|^2 \, \mathrm{d}\mathbf{s}, \\
    \label{radiation_power_surface}
    P_{\mathrm{rad}} & = - \iint_{\mathcal{S}} \frac{1}{2} \Re \Big\{ \boldsymbol{\mathcal{J}}^{\mathrm{surf},*}_{\omega}(\mathbf{s}) \cdot \boldsymbol{\mathcal{E}}_{\omega}^t(\mathbf{s}) \Big\} \, \mathrm{d}\mathbf{s} \nonumber \\
    & = - \iint_{\mathcal{S}} \iint_{\mathcal{S}} \frac{1}{2}
    \boldsymbol{\mathcal{J}}^{\mathrm{surf},\mathsf{H}}_{\omega}(\mathbf{s})
    \Re \Big\{ \boldsymbol{\mathcal{G}}_{\omega}(\mathbf{s}, \mathbf{s}^{\prime}) \Big\}
    \boldsymbol{\mathcal{J}}^{\mathrm{surf}}_{\omega}(\mathbf{s}^{\prime})
    \, \mathrm{d}\mathbf{s}^{\prime}\mathrm{d}\mathbf{s}.
\end{align}

In the following analysis, we adopt the surface-current formulation as the default model. This model is especially suitable for thin apertures and metasurfaces, as it captures radiation and dissipation in a compact form without introducing a full volumetric material model. To simplify notation, we henceforth write $\boldsymbol{\mathcal{J}}_{\omega}(\mathbf{s}) \triangleq \boldsymbol{\mathcal{J}}^{\mathrm{surf}}_{\omega}(\mathbf{s})$.

\subsection{Mutual Coupling}  \label{sec:mutual_coupling}

Mutual coupling describes the EM interaction among different points or ports on an aperture. Physically, a current excited at one location generates an EM field that modifies the local voltage-current relationship at other locations. This phenomenon arises in both transmission and reception, and becomes particularly important for electrically large or densely excited apertures such as CAPAs.

To illustrate this effect, we again consider the surface-current model governed by \eqref{boundary_condition}. In this model, the current distribution on the aperture is not determined solely by the external excitation at each point, but also by the fields reradiated by currents over the entire surface. Consequently, mutual coupling should be viewed as an intrinsic field-mediated interaction within the aperture.

\begin{itemize}
    \item \emph{Tx-side coupling:}  
    For a Tx aperture, the surface current $\boldsymbol{\mathcal{J}}_{\omega}(\mathbf{s})$ at point $\mathbf{s}$ is driven by the impressed source field $\boldsymbol{\mathcal{E}}_{\mathrm{src}, \omega}^t(\mathbf{s})$. However, this current is also affected by the reradiated field produced by currents at all other points $\mathbf{s}'$ on the surface, as described by \eqref{surface_current_Green}. Therefore, the actual current distribution is the result of a collective interaction over the entire aperture, rather than an independent pointwise excitation. To realize a desired transmit current profile, the source excitation must compensate for these mutual interactions.

    \item \emph{Rx-side coupling:}  
    For a Rx aperture, $\boldsymbol{\mathcal{E}}_{\mathrm{src}, \omega}^t(\mathbf{s})$ can be interpreted as the incident field radiated by an external Tx. This incident field induces a surface current on the receiving aperture, which in turn reradiates and produces an additional field over the same surface through \eqref{surface_current_Green}. As a result, the total tangential field acting on the aperture is modified, and the final induced current distribution must be determined self-consistently. Hence, even at the receiver, the current at one point is generally coupled to currents at all other points.
\end{itemize}

The above discussion shows that mutual coupling is fundamentally governed by the radiation operator in \eqref{surface_current_Green}. In the continuous setting, the dyadic Green's function $\boldsymbol{\mathcal{G}}_{\omega}(\mathbf{s}, \mathbf{s}^{\prime})$ acts as a \emph{dyadic mutual-impedance kernel}, mapping the current at $\mathbf{s}^{\prime}$ to the field induced at $\mathbf{s}$. A scalar mutual-impedance coefficient is obtained only after projecting this dyadic kernel onto specific current and field directions. This viewpoint is consistent with the power expressions in \eqref{source_method_radiation} and \eqref{radiation_power_surface}. In particular, under the lossless background considered here, the real part $\Re\{\boldsymbol{\mathcal{G}}_{\omega}(\mathbf{s}, \mathbf{s}^{\prime})\}$ represents the radiation-resistance component of the coupling kernel, whereas the imaginary part is associated with reactive energy exchange. Owing to the dyadic nature of $\boldsymbol{\mathcal{G}}_{\omega}(\mathbf{s}, \mathbf{s}^{\prime})$, the coupling between two points generally depends not only on their separation, but also on their relative orientation and polarization. This is in sharp contrast to simplified scalar or isotropic models, in which the coupling is often assumed to depend only on the distance between points \cite{5446312, 11006094}. The full dyadic model, therefore, captures a much richer set of coupling behaviors that are essential for accurately analyzing and designing continuous-space systems.

\begin{figure}[!t]
    \centering
    \subfigure[Along the $\mathsf{x}$-axis]{
        \includegraphics[width=0.4\textwidth]{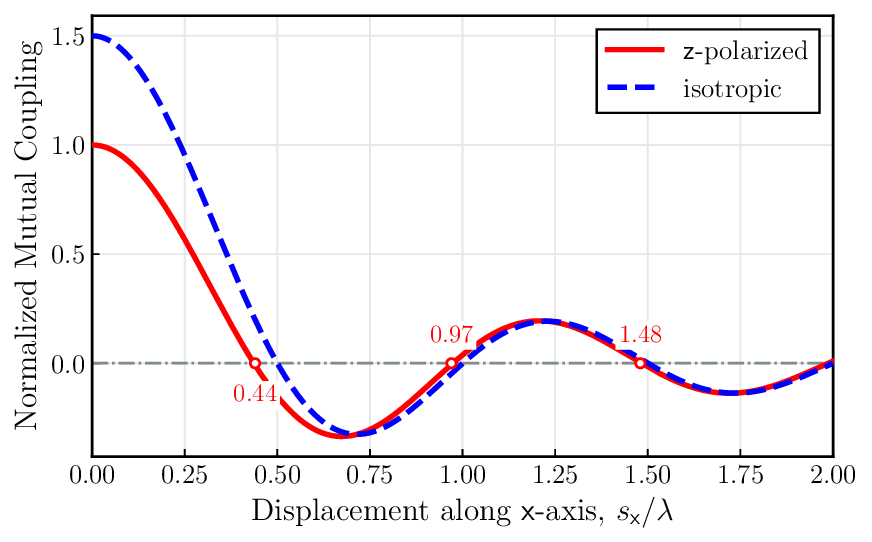}
        \label{fig:coupling_x}
    }
    \subfigure[Along the $\mathsf{z}$-axis]{
        \includegraphics[width=0.4\textwidth]{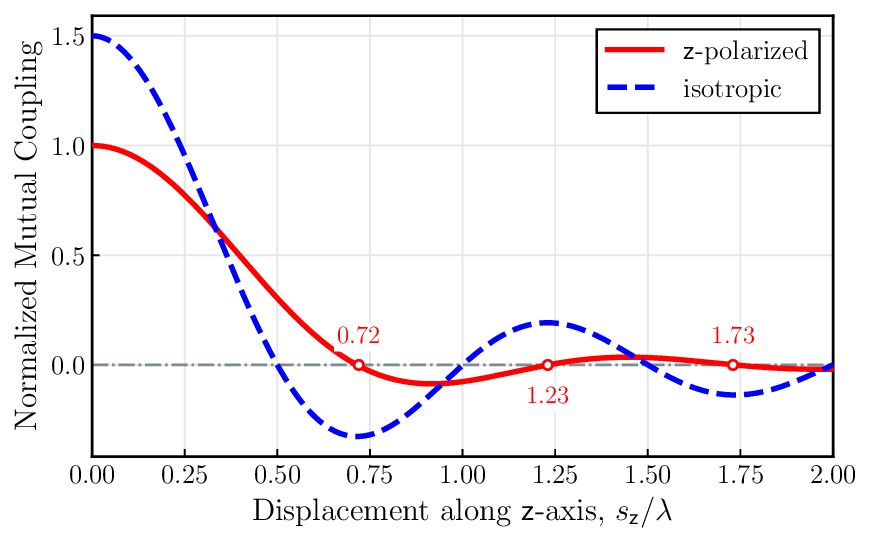}
        \label{fig:coupling_z}
    }
    \caption{Normalized radiation mutual coupling between two points as a function of their separation along the $\mathsf{x}$- and $\mathsf{z}$-axes. The ``$\mathsf{z}$-polarized'' curves are obtained by projecting the dyadic Green's function onto the $\mathsf{z}$-polarized current direction, while the ``isotropic'' curves correspond to an isotropic scalar coupling model. The different amplitudes and zero-crossing locations show that physically polarized coupling depends not only on distance but also on the relative orientation between the displacement vector and the current polarization.}
    \label{fig:radiation_coupling}
\end{figure}

To gain further insight, consider coupling along a single polarization direction, for example, the $\mathsf{z}$-direction with polarization vector $\hat{\bm{\mathsf{z}}} = [0,0,1]^{\mathsf{T}}$. Projecting the dyadic kernel onto this direction yields the scalar coupling coefficient \cite{wang2025mutual}
\begin{align}
    c_{\mathsf{z}}(\mathbf{s}, \mathbf{s}^{\prime}) \triangleq \, & -\hat{\bm{\mathsf{z}}}^{\mathsf{T}} \Re \big\{ \boldsymbol{\mathcal{G}}_{\omega}(\mathbf{s}, \mathbf{s}^{\prime}) \big\} \hat{\bm{\mathsf{z}}} \nonumber \\
    = \, & k_0 \eta_0 \left( \varphi(\mathbf{s} - \mathbf{s}^{\prime}) + \frac{1}{k_0^2} \partial^2_{\mathsf{z}} \varphi(\mathbf{s} - \mathbf{s}^{\prime}) \right),\label{eq:coupling_cz}
\end{align}
where
\begin{equation}
    \varphi(\mathbf{s}) = \frac{\sin(k_0 \|\mathbf{s}\|)}{4\pi \|\mathbf{s}\|}.
\end{equation}
This expression shows that the coupling coefficient consists of two terms: a distance-dependent $\mathrm{sinc}$-type term $\varphi(\cdot)$ and a polarization-dependent correction term $\partial^2_{\mathsf{z}} \varphi(\cdot)$. The first term resembles the form commonly encountered in scalar models, while the second term reveals that, for physically polarized currents, the coupling also depends on the relative observation direction with respect to the current orientation. Therefore, even for a fixed separation distance, the coupling strength can vary substantially with geometry and polarization.

\begin{remark}[\emph{Non-mutual-coupling condition}]
    \normalfont
    A common conclusion in the literature based on scalar or isotropic models is that \rev{the mutual coupling coefficient in \eqref{eq:coupling_cz}, i.e., the real part of the coupling that determines the radiated power,} vanishes whenever two points are separated by an integer multiple of half a wavelength, i.e., $\|\mathbf{s} - \mathbf{s}^{\prime}\| = m\lambda/2$ for \rev{$m = 1, 2, \dots$} \cite{11006094, 5446312, yordanov2009arrays, 6404701, friedlander2020extended}. This conclusion follows directly from the zeros of the $\mathrm{sinc}$-type term $\varphi(\cdot)$. However, such a condition is generally insufficient for physically polarized currents. Once the polarization-dependent derivative term $\partial^2_{\mathsf{z}} \varphi(\cdot)$ is retained, the coupling no longer depends solely on the separation distance, and the half-wavelength rule is no longer sufficient in general to eliminate coupling. The resulting behavior of the full coefficient $c_{\mathsf{z}}(\mathbf{s}, \mathbf{s}^{\prime})$ is illustrated in Fig.~\ref{fig:coupling_x} and Fig.~\ref{fig:coupling_z}.
    \rev{The figures further show that, in addition to the shifted nulls, mutual coupling remains non-negligible well beyond $\lambda/2$, in contrast to the common practice of neglecting it at larger separations.}
\end{remark}

The above discussion concerns the coupling between point-like sources or infinitesimal dipoles. In practical apertures, however, the current is distributed over a finite surface, and the resulting mutual coupling is governed by the full integral operator in \eqref{surface_current_Green}. Accordingly, the coupling between two finite apertures is obtained by integrating the dyadic kernel over both surfaces. In general, this interaction does not admit a simple closed-form expression and often requires full-wave numerical evaluation.

To make this explicit, consider two finite current distributions, $\boldsymbol{\mathcal{J}}_1(\mathbf{s})$ and $\boldsymbol{\mathcal{J}}_2(\mathbf{s})$, supported on two separate aperture surfaces $\mathcal{S}_1$ and $\mathcal{S}_2$, respectively. In this case, the coupling is no longer characterized by a single pointwise coefficient, but rather by the following \rev{integral}:
\begin{equation}\label{eq:C12}
C_{12}
=
- \int_{\mathcal{S}_1} \int_{\mathcal{S}_2}
\boldsymbol{\mathcal{J}}_1^{*}(\mathbf{s})
\cdot
\boldsymbol{\mathcal{G}}_{\omega}(\mathbf{s},\mathbf{s}^{\prime})
\boldsymbol{\mathcal{J}}_2(\mathbf{s}^{\prime})
\,\mathrm{d}\mathbf{s}^{\prime}\mathrm{d}\mathbf{s}.
\end{equation}
It measures the aggregate EM interaction between the two current distributions, rather than the interaction between two isolated points. If the excitation currents associated with $\boldsymbol{\mathcal{J}}_1(\mathbf{s})$ and $\boldsymbol{\mathcal{J}}_2(\mathbf{s})$ are denoted by $I_1$ and $I_2$, respectively, then the commonly used \emph{mutual impedance} in antenna theory can be defined as $Z_{12} = C_{12}/(\rev{I_1^*} I_2)$ \cite{orfanidis2016electromagnetic}. \rev{Because $C_{12}$ is normalized by the excitation currents, $Z_{12}$ depends only on the normalized current distributions $\boldsymbol{\mathcal{J}}_1(\mathbf{s})/I_1$ and $\boldsymbol{\mathcal{J}}_2(\mathbf{s})/I_2$ over the two surfaces.} These expressions show that finite-aperture coupling depends jointly on the aperture geometry, the current distributions or basis functions, the polarization of the currents, and the near-field interactions across the entire pair of surfaces. \rev{More generally, when an aperture is excited through $N$ ports or basis modes, evaluating $Z_{12}$ between every pair of normalized port current distributions yields the off-diagonal entries of the multiport impedance matrix in circuit theory, while the diagonal entries also include the surface-impedance loss in \eqref{boundary_condition}. This field-to-circuit link will be used in Sections~\ref{sec:circuit_to_field} and \ref{sec:power_constraints} to establish the multiport circuit-to-field and physical power models.}


\section{Fundamentals of CAPA Systems} \label{sec:fundamentals}

CAPAs provide a natural bridge between EM theory and communication signal models. Unlike conventional spatially discrete arrays, a CAPA models the aperture as a continuous radiating surface, thereby capturing the spatial structure of EM field generation more directly. This perspective is particularly valuable when the aperture is electrically large, densely integrated, or strongly coupled, in which case an element-wise description is no longer the most natural representation. In the following, we first introduce the continuous-space signal model for CAPAs, then discuss practical hardware realizations, and finally present a unified framework that connects the discrete circuit domain with the continuous EM domain.

\subsection{Continuous-Space Signal Model}

The preceding discussion shows that EM radiation and reception are inherently continuous in space. This naturally motivates communication architectures in which both the Tx and the Rx are implemented as continuous physical apertures. On the transmit side, the aperture supports a continuous source-current distribution $\bm{\mathcal{J}}_{\omega}(\mathbf{s})$, while on the receive side it observes a continuous electric field $\bm{\mathcal{E}}_{\omega}(\mathbf{r})$ over the aperture surface. We refer to a Tx or Rx with such a continuous aperture as a CAPA.

For a uni-polarized CAPA-based link operating over a narrowband, frequency-flat channel, the transmit and receive signals at a given time can be described by \cite{franceschetti2017wave}
\begin{align}
\label{basic_CAPA_model}
    y(\mathbf{r})=\int_{\mathcal{S}_{\mathrm{t}}} h(\mathbf{r},\mathbf{s}) x(\mathbf{s}) \,\mathrm{d}\mathbf{s}+ n(\mathbf{r}),
    \qquad \forall \mathbf{r}\in\mathcal{S}_{\mathrm{r}},
\end{align}
where $\mathcal{S}_{\mathrm{t}}$ and $\mathcal{S}_{\mathrm{r}}$ denote the transmit and receive apertures, respectively, $x(\mathbf{s})$ is the current-level transmit signal over the Tx-CAPA, $y(\mathbf{r})$ is the received signal over the Rx-CAPA, $h(\mathbf{r},\mathbf{s})$ is the spatial channel kernel from transmit point $\mathbf{s}$ to receive point $\mathbf{r}$, and $n(\mathbf{r})$ is the noise field over the receive aperture. This input-output relation is consistent with the EM formulation in \eqref{EM_model_green}.

For comparison, a conventional point-to-point SPDA-based MIMO system is described by
\begin{align}
    \mathbf{y}=\mathbf{H}\mathbf{x}+\mathbf{n},
\end{align}
where $\mathbf{H}\in\mathbb{C}^{N_{\mathrm{r}}\times N_{\mathrm{t}}}$ is the channel matrix, and $N_{\mathrm{t}}$ and $N_{\mathrm{r}}$ are the numbers of transmit and receive antennas, respectively.

The difference between the two models is fundamental. In an SPDA system, the channel is represented by a finite-dimensional matrix that maps a discrete transmit vector to a discrete receive vector. In a CAPA system, by contrast, the channel is a continuous linear operator that maps a source-current function over the transmit aperture to an induced field function over the receive aperture. As a result, the transmit signal, the receive signal, and the noise are all spatial functions, rather than finite-dimensional vectors. This leads to a different mathematical structure for the communication problem and indicates that conventional matrix-based tools developed for SPDAs must be reinterpreted or generalized in the CAPA systems.

\subsection{From Ideal CAPAs to Practical Implementations} \label{sec:implementations}

The idea of realizing a continuous, or nearly continuous, radiating aperture has a long history in microwaves and photonics. Early attempts can be traced back to 1965, when tightly coupled current sheets and monolayer metallic structures were proposed to emulate continuous radiators \cite{wheeler1965simple}. More recently, advances in reconfigurable antennas and metasurfaces have brought several hardware architectures closer to this ideal. For example, \cite{hwang2020binary} developed a binary meta-hologram based on a leaky waveguide for radiation toward prescribed directions, \cite{badawe2016true} considered dense arrangements of electrically small metamaterial elements based on resonant inclusions, \cite{hu2022arbitrary} demonstrated an electrically driven metasurface capable of generating waves with arbitrary polarization, and \cite{10068425} employed loaded $N$-port structures to approximate continuous current distributions.

Even so, truly continuous control of the source current remains difficult in practice. Most implementations, therefore, approximate the continuous current distribution by a finite set of basis modes:
\begin{equation}
    x(\mathbf{s}) = \sum_{n=1}^{N} \psi_n(\mathbf{s}) i_n,
    \quad \mathbf{s}\in \mathcal{S}_{\mathrm{t}},
\end{equation}
where $N$ is the number of controllable modes, $\psi_n(\mathbf{s})$ is the $n$-th basis current mode, and $i_n$ is its complex excitation coefficient. In this representation, the infinite-dimensional current distribution over the aperture is projected onto an $N$-dimensional controllable subspace. Hence, the synthesis of a continuous current profile is transformed into the design of a finite set of coefficients $\{i_n\}_{n=1}^N$. Depending on how the basis functions are chosen and physically realized, practical CAPA implementations can be broadly divided into two classes.

\begin{itemize}
    \item \emph{Pixel-based implementation:}  
    In this approach, the aperture is synthesized by a large number of localized antenna ``pixels.'' The basis function $\psi_n(\mathbf{s})$ corresponds to the current mode primarily generated by the $n$-th local element and is mainly supported on a subregion $\mathcal{S}_{\mathrm{t},n}\subset\mathcal{S}_{\mathrm{t}}$. This implementation is conceptually close to a densely packed SPDA and is therefore relatively easy to connect with conventional array models. Its main challenge is that accurately approximating a continuous current distribution often requires either a very large number of subwavelength elements or highly reconfigurable local responses. This in turn leads to strong mutual coupling, high hardware complexity, increased power consumption, and substantial control overhead.

    \item \emph{Pattern-based implementation:}  
    This approach employs a set of global basis functions that extend over the entire aperture. Each $\psi_n(\mathbf{s})$ represents a resonant mode or global current pattern supported by the structure, and the resulting mode set spans an $N$-dimensional subspace of $L^2(\mathcal{S}_{\mathrm{t}})$. The main advantage is that the system controls global modes rather than individual local pixels, which can reduce the number of required RF chains and alleviate severe local coupling. The tradeoff is that generating, selecting, and exciting these global modes typically requires more sophisticated circuit structures, such as loaded $N$-port networks \cite{10068425}, together with more involved calibration and signal processing.
\end{itemize}
\rev{In both implementations, the coefficients $\{i_n\}_{n=1}^N$ can be chosen arbitrarily only when each mode is driven by a dedicated RF chain. To reduce hardware cost, practical architectures often rely on reconfigurable elements that constrain the feasible values of $\{i_n\}_{n=1}^N$. For example, analog phase shifters adjust only the phase, the tunable elements of DMAs obey a Lorentzian-constrained response \cite{shlezinger2021dynamic}, and RHSs control only the radiation amplitude through a holographic pattern \cite{deng2021reconfigurable}. Such constraints further restrict the set of realizable current distributions. In the remainder of this tutorial, unless otherwise specified, the term CAPA refers to the ideal aperture whose current distribution can be controlled continuously, whereas pixel-based and pattern-based implementations, as well as densely spaced SPDAs, approximate it with a finite number of controllable modes.}

\begin{figure*}[!t]
 \centering
\includegraphics[width=0.7\textwidth]{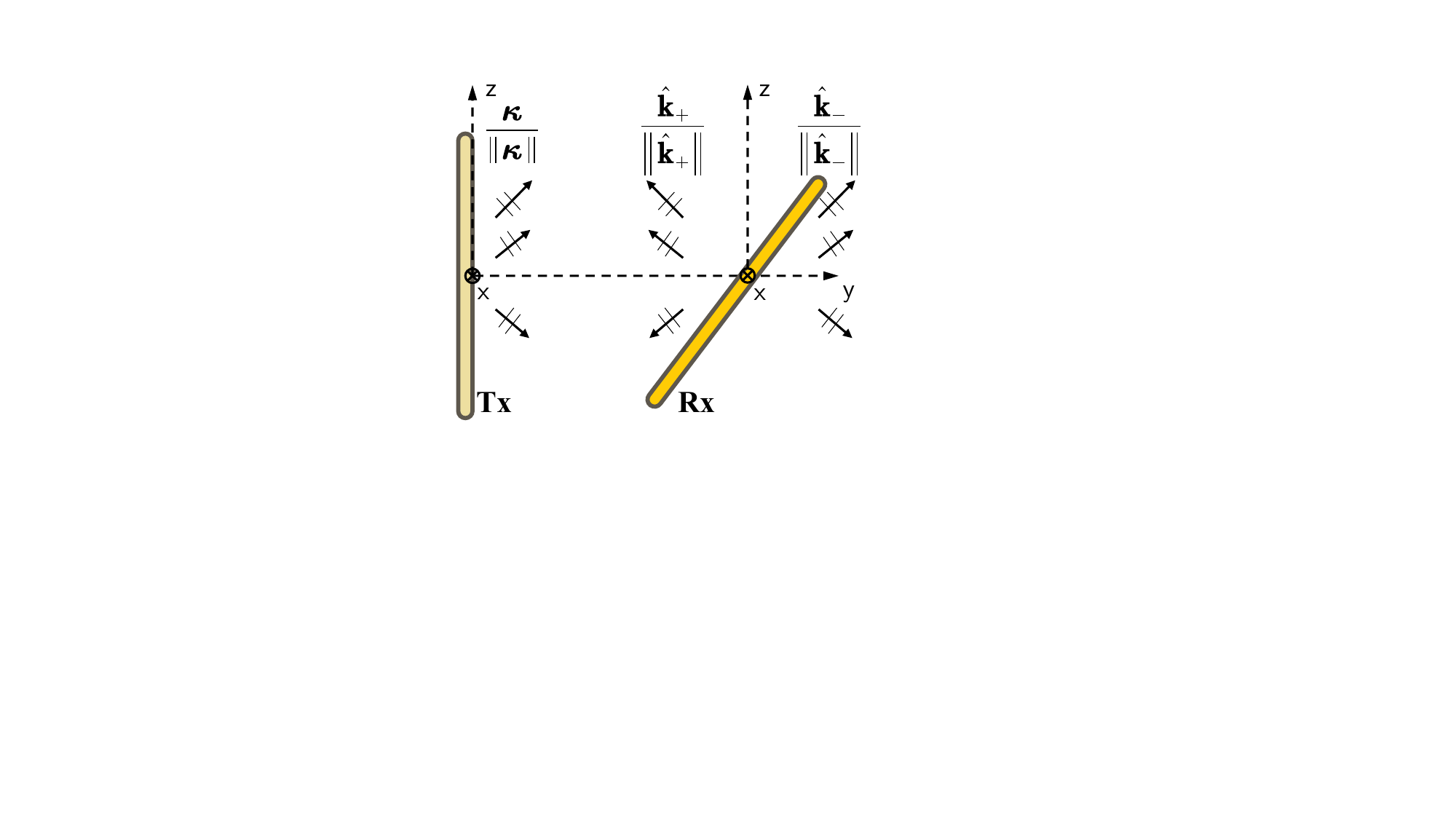}
\caption{Illustration of the circuit-to-field model for a Tx-CAPA. The finite-dimensional circuit domain consists of controllable source voltages and port currents, which are related through the \rev{multiport} circuit network. These port currents excite basis current modes on the physical surface, producing a continuous source-current distribution over the aperture. The resulting current then radiates an EM field into space through the Green's-function radiation operator, thereby connecting practical hardware excitations with the continuous-space transmit signal model.}
\label{Figure: circuit-to-field}
\end{figure*}

An analogous issue arises at the Rx. A practical CAPA cannot observe the continuous field $y(\mathbf{r})$ pointwise for all $\mathbf{r}\in\mathcal{S}_{\mathrm{r}}$. Instead, it extracts a finite number of circuit-domain observables by projecting the receive signal onto a set of observation modes:
\begin{equation}
    v_m = \int_{\mathcal{S}_{\mathrm{r}}} \phi_m^*(\mathbf{r}) y(\mathbf{r}) \,\mathrm{d}\mathbf{r},
    \quad m=1,\dots,M,
\end{equation}
where $M$ is the number of accessible observation modes, $\phi_m(\mathbf{r})$ is the $m$-th observation mode or combining pattern, and $v_m$ is the corresponding analog observation. In this way, continuous-aperture reception is transformed into the acquisition of a finite number of coefficients. Equivalently, the continuous field over the Rx aperture is projected onto an $M$-dimensional observation subspace that can be accessed by the hardware.

The same implementation taxonomy applies at the Rx. In a pixel-based Rx, $\phi_m(\mathbf{r})$ is mainly localized on a subregion $\mathcal{S}_{\mathrm{r},m}\subset\mathcal{S}_{\mathrm{r}}$, so each coefficient captures the field over a local neighborhood. In a pattern-based Rx, $\phi_m(\mathbf{r})$ is a global mode supported by the entire aperture, so the Rx directly extracts a low-dimensional set of dominant spatial patterns. Therefore, on both the Tx and Rx sides, practical CAPAs operate by interfacing the underlying continuous EM aperture with a finite-dimensional circuit domain. Once the transmit synthesis and receive observation are both reduced to finite-dimensional modal operations, practical CAPAs admit a unified circuit-to-field description.

\subsection{A Unified Circuit-to-Field Model} \label{sec:circuit_to_field}

Although the above implementations differ in hardware form, they can all be described within a unified mathematical framework that links the discrete circuit domain to the continuous EM domain, as illustrated in Fig.~\ref{Figure: circuit-to-field}. The key idea is that practical CAPAs do not manipulate a continuous aperture directly. Instead, they operate through a finite number of circuit ports or modes, which in turn synthesize or observe continuous EM fields over the aperture. In the following, we present this framework separately for the Tx and Rx sides.

\subsubsection{Tx-Side Circuit-to-Field Model}

Consider a practical CAPA Tx with $N$ programmable RF ports and $K$ digital data streams collected in $\mathbf{c}\in\mathbb{C}^{K\times 1}$. Under the usual narrowband complex-baseband model, the digital precoder, together with the associated RF chains, maps these communication streams into the equivalent source voltages applied to the aperture ports. Denoting the resulting source-voltage vector by $\mathbf{v}\in\mathbb{C}^{N\times 1}$ and the corresponding precoding matrix by $\mathbf{W}\in\mathbb{C}^{N\times K}$, we have
\begin{equation}
    \mathbf{v}=\mathbf{W}\mathbf{c}.
\end{equation}
Here, the $n$-th entry of $\mathbf{v}$ is the complex source voltage applied to the $n$-th port, while the $k$-th column of $\mathbf{W}$ specifies how the $k$-th data stream contributes to the excitation of all ports. For notational simplicity, any effective source impedance can be absorbed into the circuit response, so that the voltages in $\mathbf{v}$ can be understood as the effective port-driving voltages.

Once these source voltages are applied to the CAPA structure, the actual port currents are determined by the coupled multiport circuit rather than chosen independently. At a fixed operating frequency, this circuit response can be represented by an effective loaded admittance matrix $\mathbf{Y}\in\mathbb{C}^{N\times N}$, which captures the self-response of each port together with the effects of mutual coupling, loading, and matching within the hardware. The resulting port-current vector $\mathbf{i} = [i_1,\dots,i_N]^{\mathsf{T}} \in\mathbb{C}^{N\times 1}$ therefore satisfies
\begin{equation}
    \mathbf{i}=\mathbf{Y}\mathbf{v}=\mathbf{Y}\mathbf{W}\mathbf{c}.
\end{equation}

These discrete port currents then excite the physical aperture and generate a continuous source-current distribution over $\mathcal{S}_{\mathrm{t}}$. The resulting aperture current can thus be expressed as the superposition of the current patterns associated with the individual ports. Let $\boldsymbol{\psi}(\mathbf{s})=[\psi_1(\mathbf{s}),\dots,\psi_N(\mathbf{s})]^{\mathsf{T}}$ collect the basis current modes associated with the $N$ ports, where $\psi_n(\mathbf{s})$ denotes the continuous aperture-current pattern induced by the $n$-th port under unit current excitation. Then, the continuous transmit signal can be written as
\begin{equation}\label{eq:circuit_to_EM_mapping_rewrite}
    x(\mathbf{s})=\boldsymbol{\psi}^{\mathsf{T}}(\mathbf{s})\mathbf{i}
    =\boldsymbol{\psi}^{\mathsf{T}}(\mathbf{s})\mathbf{Y}\mathbf{W}\mathbf{c},
    \qquad \mathbf{s}\in\mathcal{S}_{\mathrm{t}}.
\end{equation}
Hence, the overall Tx chain proceeds in three stages: $\mathbf{c}\;\rightarrow\;\mathbf{v}\;\rightarrow\;\mathbf{i}\;\rightarrow\;x(\mathbf{s})$. Substituting \eqref{eq:circuit_to_EM_mapping_rewrite} into the continuous channel model yields
\begin{equation}
    y(\mathbf{r})
    =\int_{\mathcal{S}_{\mathrm{t}}}
    h(\mathbf{r},\mathbf{s})
    \boldsymbol{\psi}^{\mathsf{T}}(\mathbf{s})\mathbf{Y}\mathbf{W}\mathbf{c}
    \,\mathrm{d}\mathbf{s}
    +n(\mathbf{r}).
\end{equation}

This expression also reveals an equivalent continuous beamforming interpretation. Since the mapping from $\mathbf{c}$ to $x(\mathbf{s})$ is linear, the transmit signal can be expressed as
\begin{equation}
    x(\mathbf{s})=\sum_{k=1}^{K} w_k(\mathbf{s}) c_k,
\end{equation}
where $c_k$ is the $k$-th data symbol and the corresponding continuous beamformer is
\begin{equation}\label{eq:effective_beamformer_rewrite}
    w_k(\mathbf{s})
    =\boldsymbol{\psi}^{\mathsf{T}}(\mathbf{s})\mathbf{Y}\mathbf{w}_k,
\end{equation}
with $\mathbf{w}_k$ denoting the $k$-th column of $\mathbf{W}$. Therefore, $w_k(\mathbf{s})$ is the effective continuous transmit beamforming pattern generated by the $k$-th stream after passing through both the circuit response and the aperture field synthesis. This model provides a direct bridge between the circuit variables $\{\mathbf{v},\mathbf{i},\mathbf{Y},\mathbf{W}\}$ and the continuous EM quantities $\{x(\mathbf{s}),w_k(\mathbf{s})\}$. It also shows that the continuous beamformers studied in the following sections are not abstract mathematical objects; rather, they are the effective spatial patterns produced jointly by the digital precoder and the underlying CAPA hardware.

\rev{In practice, for a given hardware structure, $\mathbf{Y}$ can be obtained from measured scattering parameters of a prototype or from full-wave EM simulations that evaluate the mutual-impedance integrals in \eqref{eq:C12}, whereas $\{\psi_n(\mathbf{s})\}$ are obtained from the simulated port current distributions or from near-field measurements. Once these parameters are obtained, a target continuous current $x^{\star}(\mathbf{s})$ can be synthesized in the least-squares sense via the port currents
\begin{equation}\label{eq:port_current_projection}
\mathbf{i}^{\star}=\mathbf{G}_{\psi}^{-1}\int_{\mathcal{S}_{\mathrm{t}}}\boldsymbol{\psi}^*(\mathbf{s})x^{\star}(\mathbf{s})\,\mathrm{d}\mathbf{s},
\end{equation}
where $\mathbf{G}_{\psi}=\int_{\mathcal{S}_{\mathrm{t}}}\boldsymbol{\psi}^*(\mathbf{s})\boldsymbol{\psi}^{\mathsf{T}}(\mathbf{s})\,\mathrm{d}\mathbf{s}$ is the Gram matrix of the modes, and the corresponding source excitation voltages are $\mathbf{v}^{\star}=\mathbf{Y}^{-1}\mathbf{i}^{\star}$.}

\subsubsection{Rx-Side Field-to-Circuit Model}

The Rx model can be understood by following the reverse path of the Tx model. Under the uni-polarized model, the scalar field $y(\mathbf{r})$ represents the effective receive electric field over the aperture. A practical CAPA Rx does not observe this field pointwise. Instead, it first projects the continuous field onto a finite set of observation modes associated with the receive ports. Let $\boldsymbol{\phi}(\mathbf{r})=[\phi_1(\mathbf{r}),\dots,\phi_M(\mathbf{r})]^{\mathsf{T}}$ collect the observation modes associated with the $M$ observation ports. This incident field first induces the open-circuit port-voltage vector $\mathbf{v}_{\mathrm{oc}}\in\mathbb{C}^{M\times 1}$ given by
\begin{equation}
    \mathbf{v}_{\mathrm{oc}}
    =\int_{\mathcal{S}_{\mathrm{r}}}\boldsymbol{\phi}^{*}(\mathbf{r})y(\mathbf{r})\,\mathrm{d}\mathbf{r},
\end{equation}
where the $m$-th entry is the voltage that would appear at the $m$-th port if it were left unloaded. These open-circuit voltages are then transformed by the Rx hardware, including the matching/loading network and any analog RF processing, into the RF voltage vector $\mathbf{v}_{\mathrm{RF}}\in\mathbb{C}^{M\times 1}$. Representing this linear circuit response by an effective transfer matrix $\mathbf{T}\in\mathbb{C}^{M\times M}$, we have
\begin{equation}
    \mathbf{v}_{\mathrm{RF}}
    =\mathbf{T}\mathbf{v}_{\mathrm{oc}}
    =\mathbf{T}\int_{\mathcal{S}_{\mathrm{r}}}\boldsymbol{\phi}^{*}(\mathbf{r})y(\mathbf{r})\,\mathrm{d}\mathbf{r}.
\end{equation}
Finally, after downconversion, the Rx extracts $K$ data streams through the digital combiner $\mathbf{U}=[\mathbf{u}_1,\dots,\mathbf{u}_K]\in\mathbb{C}^{M\times K}$, leading to
\begin{equation}
    \hat{\mathbf{c}}=\mathbf{U}^{\mathsf{H}}\mathbf{v}_{\mathrm{RF}}.
\end{equation}
Hence, the Rx chain proceeds as $y(\mathbf{r})\;\rightarrow\;\mathbf{v}_{\mathrm{oc}}\;\rightarrow\;\mathbf{v}_{\mathrm{RF}}\;\rightarrow\;\hat{\mathbf{c}}$.

This expression reveals the Rx-side counterpart of the continuous beamforming interpretation. In particular, the $k$-th detected stream can be written as
\begin{equation}
    \hat{c}_k=\int_{\mathcal{S}_{\mathrm{r}}} b_k^*(\mathbf{r}) y(\mathbf{r})\,\mathrm{d}\mathbf{r},
\end{equation}
where the corresponding equivalent continuous receive beamformer is
\begin{equation}\label{eq:effective_combiner}
    b_k(\mathbf{r})
    =\boldsymbol{\phi}^{\mathsf{T}}(\mathbf{r})\mathbf{T}^{\mathsf{H}}\mathbf{u}_k.
\end{equation}
Hence, $b_k(\mathbf{r})$ is the effective continuous receive pattern jointly synthesized by the physical observation modes, the Rx circuit response, and the digital combiner. Substituting the continuous channel model gives
\begin{equation}
    \hat{c}_k
    =\int_{\mathcal{S}_{\mathrm{r}}}\int_{\mathcal{S}_{\mathrm{t}}}
    b_k^*(\mathbf{r})h(\mathbf{r},\mathbf{s})x(\mathbf{s})
    \,\mathrm{d}\mathbf{s}\,\mathrm{d}\mathbf{r}
    +z_k,
\end{equation}
where $z_k \triangleq \int_{\mathcal{S}_{\mathrm{r}}} b_k^*(\mathbf{r})n(\mathbf{r})\,\mathrm{d}\mathbf{r}$ is the effective noise after Rx processing. This is the Rx-side counterpart of the transmit beamformer $w_k(\mathbf{s})$.

\subsubsection{End-to-End Model}

Combining the Tx and Rx models yields the end-to-end transmission relation
\begin{equation}
    \hat{\mathbf{c}}
    =\int_{\mathcal{S}_{\mathrm{r}}}\int_{\mathcal{S}_{\mathrm{t}}}
    \mathbf{b}^{\mathsf{H}}(\mathbf{r})h(\mathbf{r},\mathbf{s})\mathbf{w}(\mathbf{s})\mathbf{c}
    \,\mathrm{d}\mathbf{s}\,\mathrm{d}\mathbf{r}
    + \mathbf{z},
\end{equation}
where $\mathbf{b}(\mathbf{r}) = [b_1(\mathbf{r}), \dots, b_K(\mathbf{r})]$, $\mathbf{w}(\mathbf{s}) = [w_1(\mathbf{s}), \dots, $ $w_K(\mathbf{s})]$, and $\boldsymbol{z} = [z_1, \dots, z_K]^{\mathsf{T}}$ is the effective noise vector after Rx processing. This expression provides a unified end-to-end model that connects the digital data streams, the circuit-domain hardware response, and the continuous EM fields. It also serves as the basic system model for the beamforming design, channel modeling, and performance analysis developed in the following sections.

\subsection{Power Constraints in CAPA Systems} \label{sec:power_constraints}

Following the surface-current power model in Section~\ref{sec:power}, \rev{consider a uni-polarized Tx-CAPA with surface current $\boldsymbol{\mathcal{J}}_{\omega}(\mathbf{s}) = x(\mathbf{s})\hat{\mathbf{p}}_{\mathrm{t}}$, where the unit vector $\hat{\mathbf{p}}_{\mathrm{t}}$ is tangential to $\mathcal{S}_{\mathrm{t}}$, and let $e(\mathbf{s})\triangleq\hat{\mathbf{p}}_{\mathrm{t}}^{\mathsf{T}}\boldsymbol{\mathcal{E}}_{\omega}(\mathbf{s})$ denote the reradiated field along $\hat{\mathbf{p}}_{\mathrm{t}}$.} Then, the radiated power and dissipated power can be expressed as
\begin{align}
    P_{\mathrm{rad}}
    &=
    -\int_{\mathcal{S}_{\mathrm{t}}}
    \frac{1}{2}\Re\!\left\{x^{*}(\mathbf{s})e(\mathbf{s})\right\}\,\mathrm{d}\mathbf{s}, \\
    P_{\mathrm{loss}}
    &=
    \int_{\mathcal{S}_{\mathrm{t}}}
    \frac{1}{2}\Re\!\left\{Z_s(\mathbf{s})\right\}\lvert x(\mathbf{s})\rvert^2\,\mathrm{d}\mathbf{s}.
\end{align}
Here, $P_{\mathrm{rad}}$ characterizes the net power carried away from the aperture by radiation, whereas $P_{\mathrm{loss}}$ accounts for the ohmic power dissipated over the aperture through the resistive part of the surface impedance.

Moreover, the reradiated field at each point is induced by the entire aperture current distribution through
\begin{equation}
    e(\mathbf{s})
    =
    \int_{\mathcal{S}_{\mathrm{t}}}
    \rev{\mathcal{G}_{\mathrm{p}}(\mathbf{s},\mathbf{s}')}x(\mathbf{s}')\,\mathrm{d}\mathbf{s}',
\end{equation}
where \rev{$\mathcal{G}_{\mathrm{p}}(\mathbf{s},\mathbf{s}')\triangleq\hat{\mathbf{p}}_{\mathrm{t}}^{\mathsf{T}}\boldsymbol{\mathcal{G}}_{\omega}(\mathbf{s},\mathbf{s}')\hat{\mathbf{p}}_{\mathrm{t}}$} is the effective scalar radiation kernel obtained from the projected dyadic Green's function. This relation makes explicit that the radiated field at one point depends on the current distribution over the entire aperture, thereby capturing mutual coupling in the continuous setting. Substituting this expression into $P_{\mathrm{rad}}$ yields the equivalent form
\begin{equation} \label{eq:power_radiation_integral}
    P_{\mathrm{rad}}
    =
    -\int_{\mathcal{S}_{\mathrm{t}}}\int_{\mathcal{S}_{\mathrm{t}}}
    \frac{1}{2}x^{*}(\mathbf{s})\Re\!\left\{\rev{\mathcal{G}_{\mathrm{p}}(\mathbf{s},\mathbf{s}')}\right\}x(\mathbf{s}')
    \,\mathrm{d}\mathbf{s}'\mathrm{d}\mathbf{s}.
\end{equation}
This expression shows that the radiated power is generally not determined pointwise by $|x(\mathbf{s})|^2$, but instead by a quadratic functional of the entire current distribution.

\subsubsection{\rev{Simplified Power Metric}}
\rev{The physical radiated-power model in \eqref{eq:power_radiation_integral} requires knowledge of the coupling kernel $\mathcal{G}_{\mathrm{p}}(\mathbf{s},\mathbf{s}')$, which complicates analysis and optimization. For this reason, existing studies on CAPA design commonly adopt the simplified power metric \cite{9906802,10158997,10910020,10938678}
\begin{equation}\label{eq:simplified_power}
    \int_{\mathcal{S}_{\mathrm{t}}}|x(\mathbf{s})|^2\,\mathrm{d}\mathbf{s}\le P_{\mathrm{t}},
\end{equation}
which is the continuous counterpart of $\|\mathbf{w}\|^2\le P$ in SPDA systems. To clarify the physical meaning of \eqref{eq:simplified_power}, we next examine its connection to the radiated and dissipated powers from two complementary perspectives: an angular-domain upper bound and a wavenumber-domain approximation. Both perspectives build on the flux method in \eqref{power_farfield_sphere}. Specifically, according to \eqref{far_field_E_field}, the far-field electric field generated by the aperture is given by}
\begin{align}\label{eq:capa_far_field}
    \boldsymbol{\mathcal{E}}_{\omega}(\mathbf{r})
    =&
    -\mathrm{j}k_0\eta_0
    \frac{\mathrm{e}^{-\mathrm{j}k_0 r}}{4\pi r}
    \left(\mathbf{I}-\hat{\boldsymbol{\kappa}}\hat{\boldsymbol{\kappa}}^{\mathsf{T}}\right)\hat{\mathbf{p}}_{\mathrm{t}}
    \nonumber\\
    &\times
    \int_{\mathcal{S}_{\mathrm{t}}}
    \mathrm{e}^{\mathrm{j}k_0\hat{\boldsymbol{\kappa}}^{\mathsf{T}}(\mathbf{s}-\mathbf{s}_{\mathrm{ref}})}
    x(\mathbf{s})\,\mathrm{d}\mathbf{s},
\end{align}
where $\hat{\boldsymbol{\kappa}}$ denotes the propagation direction in the far field and the equality holds as $r \rightarrow \infty$. Substituting this expression into \eqref{power_farfield_sphere} and using $\mathrm{d}A=r^2\mathrm{d}\Omega$ gives
\begin{align}\label{eq:prad_angular}
    P_{\mathrm{rad}}
    =&\frac{k_0^2\eta_0}{32\pi^2}
    \int_{4\pi}
    \left\|
    \left(\mathbf{I}-\hat{\boldsymbol{\kappa}}\hat{\boldsymbol{\kappa}}^{\mathsf{T}}\right)
    \hat{\mathbf{p}}_{\mathrm{t}}
    \right\|^2
    \nonumber\\
    &\times
    \left|
    \int_{\mathcal{S}_{\mathrm{t}}}
    \mathrm{e}^{\mathrm{j}k_0\hat{\boldsymbol{\kappa}}^{\mathsf{T}}(\mathbf{s}-\mathbf{s}_{\mathrm{ref}})}
    x(\mathbf{s})\,\mathrm{d}\mathbf{s}
    \right|^2
    \mathrm{d}\Omega.
\end{align}

\paragraph{\rev{Upper-Bound Perspective}}
By the Cauchy-Schwarz inequality and the fact that $\left|\mathrm{e}^{\mathrm{j}\theta}\right|=1$, we have for every direction $\hat{\boldsymbol{\kappa}}$ that
\begin{equation}\label{eq:cauchy_schwarz}
    \left|
    \int_{\mathcal{S}_{\mathrm{t}}}
    \mathrm{e}^{\mathrm{j}k_0\hat{\boldsymbol{\kappa}}^{\mathsf{T}}(\mathbf{s}-\mathbf{s}_{\mathrm{ref}})}
    x(\mathbf{s})\,\mathrm{d}\mathbf{s}
    \right|^2
    \le
    |\mathcal{S}_{\mathrm{t}}|
    \int_{\mathcal{S}_{\mathrm{t}}}|x(\mathbf{s})|^2\,\mathrm{d}\mathbf{s}.
\end{equation}
Then, using the identity $\left\|
\left(\mathbf{I}-\hat{\boldsymbol{\kappa}}\hat{\boldsymbol{\kappa}}^{\mathsf{T}}\right)\hat{\mathbf{p}}_{\mathrm{t}}
\right\|^2
=
1-|\hat{\boldsymbol{\kappa}}^{\mathsf{T}}\hat{\mathbf{p}}_{\mathrm{t}}|^2$, we obtain
\begin{align}
    P_{\mathrm{rad}}
    \le&
    \frac{k_0^2\eta_0}{32\pi^2}
    |\mathcal{S}_{\mathrm{t}}|
    \int_{\mathcal{S}_{\mathrm{t}}}|x(\mathbf{s})|^2\,\mathrm{d}\mathbf{s}
    \int_{4\pi}
    \left(1-|\hat{\boldsymbol{\kappa}}^{\mathsf{T}}\hat{\mathbf{p}}_{\mathrm{t}}|^2\right)
    \mathrm{d}\Omega
    \nonumber\\
    =&
    \frac{k_0^2\eta_0}{12\pi}
    |\mathcal{S}_{\mathrm{t}}|
    \int_{\mathcal{S}_{\mathrm{t}}}|x(\mathbf{s})|^2\,\mathrm{d}\mathbf{s},
    \label{eq:prad_upper_bound}
\end{align}
where the second equality follows from the rotational symmetry of the sphere, which implies $\int_{4\pi}\left(1-|\hat{\boldsymbol{\kappa}}^{\mathsf{T}}\hat{\mathbf{p}}_{\mathrm{t}}|^2\right)\mathrm{d}\Omega
= 8\pi/3$ for any unit vector $\hat{\mathbf{p}}_{\mathrm{t}}$. \rev{The bound \eqref{eq:prad_upper_bound} guarantees that any current satisfying \eqref{eq:simplified_power} radiates no more than $\frac{k_0^2\eta_0}{12\pi}|\mathcal{S}_{\mathrm{t}}|P_{\mathrm{t}}$, and thus provides a conservative transmit-power constraint that depends only on the aperture area and the $L^2$-norm of the current distribution. However, it is tight only for electrically small apertures with $|\mathcal{S}_{\mathrm{t}}|\ll\lambda^2$. Particularly, for a uniform current over a square aperture, the ratio between \eqref{eq:prad_upper_bound} and the actual radiated power is about $7$~dB for $|\mathcal{S}_{\mathrm{t}}|=\lambda^2$ and $26$~dB for $|\mathcal{S}_{\mathrm{t}}|=100\lambda^2$, and approaches $4\pi|\mathcal{S}_{\mathrm{t}}|/(3\lambda^2)$ for electrically large apertures. This is because the Cauchy-Schwarz inequality \eqref{eq:cauchy_schwarz} becomes an equality for a given propagation direction $\hat{\boldsymbol{\kappa}}$ only when the current $x(\mathbf{s})$ has a uniform magnitude and a phase that compensates the propagation phase $\mathrm{e}^{\mathrm{j}k_0\hat{\boldsymbol{\kappa}}^{\mathsf{T}}(\mathbf{s} - \mathbf{s}_{\mathrm{ref}})}$. The upper bound in \eqref{eq:prad_upper_bound} assumes such coherent addition in all directions simultaneously. This condition can be approximately satisfied for electrically small apertures, where the propagation phase varies negligibly across the aperture. In contrast, for electrically large apertures, coherent addition is confined to a narrow angular region rather than occurring in all directions.}

\paragraph{\rev{Wavenumber-Domain Perspective}}
\rev{
For planar apertures, the relationship between \eqref{eq:simplified_power} and the actual radiated power can be characterized more precisely in the wavenumber domain. Consider a Tx-CAPA lying on the $\mathsf{x}$-$\mathsf{z}$ plane, with $\mathbf{s}=[s_{\mathsf{x}},0,s_{\mathsf{z}}]^{\mathsf{T}}$, $\mathbf{s}_{\mathrm{ref}}=\mathbf{0}$, and a tangential polarization vector $\hat{\mathbf{p}}_{\mathrm{t}}$. A propagation direction in either half-space can be parameterized as $\hat{\boldsymbol{\kappa}}=[\hat{\kappa}_{\mathsf{x}},\hat{\kappa}_{\mathsf{y}},\hat{\kappa}_{\mathsf{z}}]^{\mathsf{T}}=[\sin\theta\cos\phi,\pm\cos\theta,\sin\theta\sin\phi]^{\mathsf{T}}$, where $\theta\in[0,\pi/2]$ denotes the angle from the aperture normal in the corresponding half-space, and $\phi\in[0,2\pi)$ is the azimuth angle. Define the transverse wavenumber vector as $\boldsymbol{\kappa}=[\kappa_{\mathsf{x}},\kappa_{\mathsf{z}}]^{\mathsf{T}}\triangleq-k_0[\hat{\kappa}_{\mathsf{x}},\hat{\kappa}_{\mathsf{z}}]^{\mathsf{T}}$. The spatial integral in \eqref{eq:prad_angular} can then be expressed as the two-dimensional Fourier transform of the current distribution, given by $X(\boldsymbol{\kappa})\triangleq\int_{\mathcal{S}_{\mathrm{t}}}x(\mathbf{s})\mathrm{e}^{-\mathrm{j}(\kappa_{\mathsf{x}}s_{\mathsf{x}}+\kappa_{\mathsf{z}}s_{\mathsf{z}})}\,\mathrm{d}\mathbf{s}$. Transforming the angular integration in \eqref{eq:prad_angular} into the transverse wavenumber domain yields
\begin{align}
    P_{\mathrm{rad}}
    \overset{(a)}{=}&
    2\times\frac{k_0^2\eta_0}{32\pi^2}
    \iint_{\kappa_{\mathsf{x}}^2+\kappa_{\mathsf{z}}^2\le k_0^2}
    \left(1-\frac{(\hat{\mathbf{p}}_{\mathrm{t}}^{\mathsf{T}}\tilde{\boldsymbol{\kappa}})^2}{k_0^2}\right)
    |X(\boldsymbol{\kappa})|^2\,\frac{\mathrm{d}\kappa_{\mathsf{x}}\mathrm{d}\kappa_{\mathsf{z}}}{k_0\gamma(\boldsymbol{\kappa})}
    \nonumber\\
    =&
    \frac{\eta_0}{4}
    \iint_{\kappa_{\mathsf{x}}^2+\kappa_{\mathsf{z}}^2\le k_0^2}
    \vartheta(\boldsymbol{\kappa})\,|X(\boldsymbol{\kappa})|^2\,\frac{\mathrm{d}\kappa_{\mathsf{x}}\mathrm{d}\kappa_{\mathsf{z}}}{(2\pi)^2},
    \label{eq:prad_wavenumber}
\end{align}
where $\tilde{\boldsymbol{\kappa}}\triangleq[\kappa_{\mathsf{x}},0,\kappa_{\mathsf{z}}]^{\mathsf{T}}$, $\gamma(\boldsymbol{\kappa})\triangleq\sqrt{k_0^2-\kappa_{\mathsf{x}}^2-\kappa_{\mathsf{z}}^2}=k_0\cos\theta$ is the normal wavenumber, and $\vartheta(\boldsymbol{\kappa})\triangleq\big(k_0^2-(\hat{\mathbf{p}}_{\mathrm{t}}^{\mathsf{T}}\tilde{\boldsymbol{\kappa}})^2\big)/\big(k_0\gamma(\boldsymbol{\kappa})\big)$ captures the direction-dependent weighting of the radiated power. Step $(a)$ follows from the Jacobian relation $\mathrm{d}\kappa_{\mathsf{x}}\mathrm{d}\kappa_{\mathsf{z}}=k_0\gamma(\boldsymbol{\kappa})\,\mathrm{d}\Omega$. The factor of $2$ accounts for the two half-spaces, which map onto the same wavenumber disk $\kappa_{\mathsf{x}}^2+\kappa_{\mathsf{z}}^2\le k_0^2$ and contribute equally to the radiated power.

To characterize the weighting factor, let $\hat{\mathbf{p}}_{\mathrm{t}}=[\cos\phi_{\mathrm{p}},0,\sin\phi_{\mathrm{p}}]^{\mathsf{T}}$. Since $\hat{\mathbf{p}}_{\mathrm{t}}^{\mathsf{T}}\tilde{\boldsymbol{\kappa}}=-k_0\sin\theta\cos(\phi-\phi_{\mathrm{p}})$, we obtain
\begin{equation}\label{eq:vartheta_closed}
    \vartheta(\boldsymbol{\kappa})=\frac{1-\sin^2\theta\cos^2(\phi-\phi_{\mathrm{p}})}{\cos\theta},
\end{equation}
which is shown in Fig.~\ref{fig:vartheta}. It can be observed that when $\theta$ is not too large, i.e., when the radiation direction is not far from the aperture normal, the weighting factor varies slowly and remains close to one, e.g., within about $1.5$~dB for $\theta\le45^\circ$.

\begin{figure}[!t]
\centering
\includegraphics[width=0.4\textwidth]{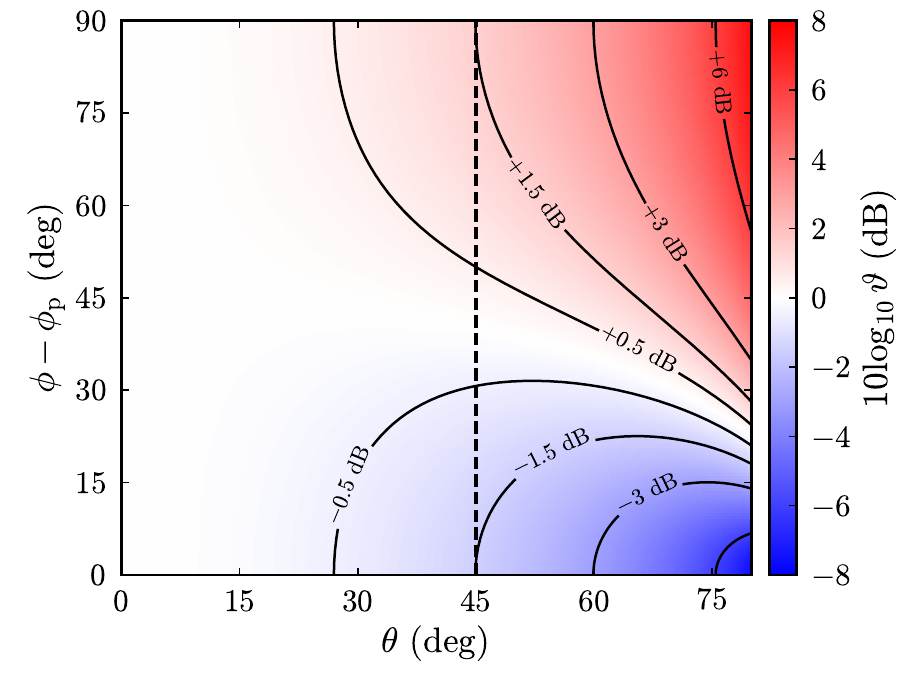}
\caption{Direction-dependent weight $\vartheta(\boldsymbol{\kappa})$ in \eqref{eq:vartheta_closed} versus the angle $\theta$ from the aperture normal and the azimuth $\phi-\phi_{\mathrm{p}}$ relative to the polarization direction. }
\label{fig:vartheta}
\end{figure} 

Using \eqref{eq:prad_wavenumber}, the radiated power can be approximated as
\begin{align}
    P_{\mathrm{rad}}
    \overset{(a)}{\approx}&
    \frac{\eta_0}{4}\bar{\vartheta}
    \iint_{\kappa_{\mathsf{x}}^2+\kappa_{\mathsf{z}}^2\le k_0^2}
    |X(\boldsymbol{\kappa})|^2\,\frac{\mathrm{d}\kappa_{\mathsf{x}}\mathrm{d}\kappa_{\mathsf{z}}}{(2\pi)^2}
    \nonumber\\
    \overset{(b)}{\approx}&
    \frac{\eta_0}{4}\bar{\vartheta}
    \iint_{\mathbb{R}^2}
    |X(\boldsymbol{\kappa})|^2\,\frac{\mathrm{d}\kappa_{\mathsf{x}}\mathrm{d}\kappa_{\mathsf{z}}}{(2\pi)^2}
    =
    \frac{\eta_0}{4}\bar{\vartheta}
    \int_{\mathcal{S}_{\mathrm{t}}}|x(\mathbf{s})|^2\,\mathrm{d}\mathbf{s},
    \label{eq:prad_wavenumber_approx}
\end{align}
where the last equality follows from Parseval's identity. To obtain step $(a)$, consider a beam steered toward $\boldsymbol{\kappa}_0$. For an electrically large aperture, $|X(\boldsymbol{\kappa})|^2$ is concentrated in a narrow region around $\boldsymbol{\kappa}_0$, with a spectral width inversely proportional to the aperture dimensions. Provided that the beam direction is not far from the aperture normal, $\vartheta(\boldsymbol{\kappa})$ varies slowly across this region, cf. Fig.~\ref{fig:vartheta}, and can be approximated by $\bar{\vartheta}=\vartheta(\boldsymbol{\kappa}_0)$. For multiple well-separated beams, $\bar{\vartheta}$ can be approximated by a spectral-power-weighted average of the corresponding directional weights. Step $(b)$ follows from the distinction between radiating and evanescent waves. As will be discussed in Remark~\ref{remark_evanescent_wave}, components within the disk $\kappa_{\mathsf{x}}^2+\kappa_{\mathsf{z}}^2\le k_0^2$ correspond to radiating waves, whereas those outside the disk correspond to evanescent waves that do not contribute to the far-field radiated power. Therefore, extending the integration domain to $\mathbb{R}^2$ introduces little error when the current spectrum is predominantly confined to the radiating region.
}

\subsubsection{Circuit-Field Equivalence}
The above EM-domain power model connects directly to the circuit-level description in the previous subsection. Using the relation $x(\mathbf{s})=\boldsymbol{\psi}^{\mathsf{T}}(\mathbf{s})\mathbf{i}$, the radiated and dissipated powers can be rewritten as quadratic forms in the port-current vector as follows:
\begin{align}
    P_{\mathrm{rad}} =\frac{1}{2}\mathbf{i}^{\mathsf{H}}\mathbf{R}_{\mathrm{rad}}\mathbf{i}, \quad
    P_{\mathrm{loss}} =\frac{1}{2}\mathbf{i}^{\mathsf{H}}\mathbf{R}_{\mathrm{loss}}\mathbf{i},
\end{align}
where
\begin{align}
    [\mathbf{R}_{\mathrm{rad}}]_{m,n}
    &\triangleq
    -\int_{\mathcal{S}_{\mathrm{t}}}\int_{\mathcal{S}_{\mathrm{t}}}
    \psi_m^{*}(\mathbf{s})\Re\!\left\{\rev{\mathcal{G}_{\mathrm{p}}(\mathbf{s},\mathbf{s}')}\right\}\psi_n(\mathbf{s}')
    \,\mathrm{d}\mathbf{s}'\mathrm{d}\mathbf{s},\\
    [\mathbf{R}_{\mathrm{loss}}]_{m,n}
    &\triangleq
    \int_{\mathcal{S}_{\mathrm{t}}}
    \Re\!\left\{Z_s(\mathbf{s})\right\}\psi_m^{*}(\mathbf{s})\psi_n(\mathbf{s})
    \,\mathrm{d}\mathbf{s}.
\end{align}
Here, $\mathbf{R}_{\mathrm{rad}}$ is the radiation-resistance matrix induced by the EM coupling kernel, and $\mathbf{R}_{\mathrm{loss}}$ is the aperture-loss matrix induced by the resistive part of the surface impedance. Together, they provide the circuit-domain counterparts of radiation and dissipation in the continuous EM model. \rev{In particular, $[\mathbf{R}_{\mathrm{rad}}]_{m,n}$ is the mutual radiation resistance between ports $m$ and $n$, i.e., the real part of the mutual impedance $Z_{12}$ in Section~\ref{sec:mutual_coupling} evaluated with the port modes $\psi_m(\mathbf{s})$ and $\psi_n(\mathbf{s})$.}

If the effective admittance matrix is nonsingular and the corresponding impedance matrix is defined as $\mathbf{Z}\triangleq\mathbf{Y}^{-1}$, then the circuit-level source power is given by
\begin{equation}
    P_{\mathrm{src}}
    =
    \frac{1}{2}\Re\!\left\{\mathbf{i}^{\mathsf{H}}\mathbf{Z}\mathbf{i}\right\}
    =
    \frac{1}{2}\Re\!\left\{\mathbf{v}^{\mathsf{H}}\mathbf{i}\right\}.
\end{equation}
Beyond the radiation and aperture-loss effects represented explicitly by $\mathbf{R}_{\mathrm{rad}}$ and $\mathbf{R}_{\mathrm{loss}}$, the impedance matrix $\mathbf{Z}$, or equivalently the admittance matrix $\mathbf{Y}$, also captures the self-impedance of each port and the mutual impedance induced by loading and matching within the circuit network. Consequently, the source power $P_{\mathrm{src}}$ generally includes not only the power radiated into free space and dissipated over the aperture, but also the power dissipated in the internal circuit. Therefore, in general, we have $P_{\mathrm{src}} \ge P_{\mathrm{rad}} + P_{\mathrm{loss}}$\rev{{, with equality when the matching and loading networks are lossless, so that $\Re\{\mathbf{Z}\}=\mathbf{R}_{\mathrm{rad}}+\mathbf{R}_{\mathrm{loss}}$}}. This inequality highlights an important modeling distinction, i.e., the EM-domain powers $P_{\mathrm{rad}}$ and $P_{\mathrm{loss}}$ quantify only the power associated with the aperture itself, whereas $P_{\mathrm{src}}$ measures the total power drawn from the driving sources. \rev{In summary, the three power models differ in the level of hardware detail that they require. The source power $P_{\mathrm{src}}$ depends on the impedance matrix $\mathbf{Z}$ of the entire circuit, including the matching and loading networks. It thus measures the total power drawn from the sources and is the relevant metric for the energy efficiency of a specific hardware implementation. The aperture power $P_{\mathrm{rad}}+P_{\mathrm{loss}}$ depends only on the current distribution through the coupling kernel and the surface impedance, regardless of how the current is fed. It thus captures radiation and mutual coupling exactly at the aperture level and is suited to designing the current distribution under physical power constraints. The simplified metric \eqref{eq:simplified_power} depends only on the $L^2$-norm of the current, which offers analytical tractability.}

\section{Channel Models of CAPA Systems} \label{sec:channel_model}

Building on the EM foundations introduced above, this section translates physical field propagation into channel models suitable for communication analysis. The objective is not to replace Maxwell's equations, but to express their implications in forms that are amenable to capacity analysis, beamforming design, and channel estimation. Table~\ref{tab:channel_models_summary} summarizes the channel models discussed in this section.

\begin{table*}[!t]
\centering
\setlength{\abovecaptionskip}{0pt}
\caption{Summary of the CAPA Channel Models in Section~\ref{sec:channel_model}.}
\label{tab:channel_models_summary}
\footnotesize
\setlength{\tabcolsep}{3pt}
\renewcommand{\arraystretch}{1.15}
\begin{tabular}{|>{\centering\arraybackslash}p{2cm}|>{\centering\arraybackslash}p{3cm}|>{\centering\arraybackslash}p{2.2cm}|>{\raggedright\arraybackslash}p{8cm}|}
\hline
\textbf{Category} & \textbf{Channel Model} & \textbf{Key expressions} & \textbf{Characteristics} \\ \hline
\multirow{5}{*}{\parbox{2.0cm}{\centering Frequency-flat}} & Tri-polarized LoS & \eqref{Tri_Polarized_Signal_Model}, \eqref{dy_GF_LoS_Model} & Dyadic vector-field LoS model with tri-polarized coupling. \\ \cline{2-4}
& Uni-polarized LoS & \eqref{Uni_Polarized_Signal_Model_Expression}, \eqref{dyadic Green's function_Standard_Scalar_Often_Used} & Scalarized LoS projection onto one polarization direction. \\ \cline{2-4}
& Physics-based multipath & \eqref{Target_Response_General}, \eqref{eq:physics_based_multipath_model} & Multipath model with explicit scatterers and path geometry. \\ \cline{2-4}
& Correlation-based multipath & \multirow{2}{*}{\eqref{4FPWD_Model_Final}, \eqref{Angular_Response_Version_2}} & \multirow{2}{*}{Multipath model with angular statistics.} \\ \hline
\multirow{5}{*}{\parbox{2.0cm}{\centering Frequency-selective}} & Time-domain channel & \eqref{eq:time_domain_baseband_received} & Space-time convolution model. \\ \cline{2-4}
& Physics-based doubly dispersive multipath & \multirow{2}{*}{\eqref{eq:physics_based_doubly_dispersive_model}} & \multirow{2}{*}{Multipath model with explicit scatterers and delay-Doppler factors.} \\ \cline{2-4}
& Correlation-based doubly dispersive multipath & \multirow{2}{*}{\eqref{eq:delay_doppler_angular_response}, \eqref{eq:time_varying_angular_response}, \eqref{Doubly_Dispersive_Correlation_Model_New}} & \multirow{2}{*}{Multipath model with angular, delay, and Doppler statistics.} \\ \hline
\end{tabular}
\vspace{-10pt}
\end{table*}

\subsection{Line-of-Sight Channel Model}\label{Section: Channel Modeling: Line-of-Sight Channel Model}

LoS propagation is the simplest setting in which the continuous-aperture nature of CAPA channels appears explicitly. It also serves as the basic reference model for the multipath, correlation, and capacity analyses developed later. We therefore begin with the LoS case.

\subsubsection{Geometrical Setup}

\begin{figure}[!t]
 \centering
\includegraphics[width=0.45\textwidth]{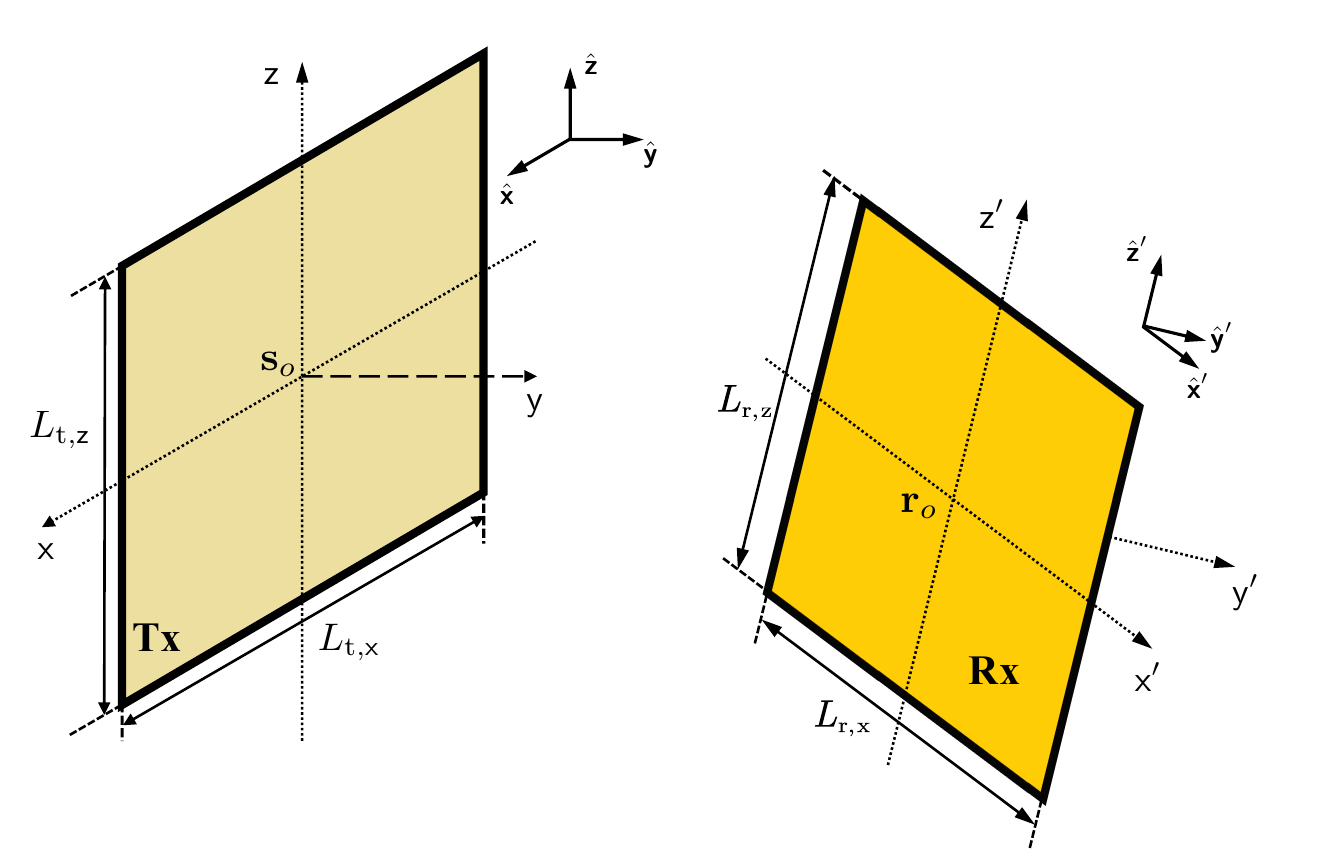}
\caption{Geometry setup of the considered CAPA system.}
\label{Figure: Line-of-Sight Channel Model: System_Model}
\end{figure}

Consider a point-to-point wireless link in which both the Tx and the Rx are equipped with planar CAPAs, as illustrated in Fig.~\ref{Figure: Line-of-Sight Channel Model: System_Model}. In particular, the Tx-CAPA lies on the $\mathsf{x}$-$\mathsf{z}$ plane and is centered at $\mathbf{s}_o=[0,0,0]^{\mathsf{T}}$, with dimensions $L_{{\rm{t}},\mathsf{x}}$ and $L_{{\rm{t}},\mathsf{z}}$ along the $\mathsf{x}$- and $\mathsf{z}$-axes, respectively. The Rx-CAPA is centered at $\mathbf{r}_o=[r_{o,\mathsf{x}},r_{o,\mathsf{y}},r_{o,\mathsf{z}}]^{\mathsf{T}}$ and has aperture size $L_{{\rm{r}},\mathsf{x}}\times L_{{\rm{r}},\mathsf{z}}$. Its local principal axes are denoted by $\hat{\bm{\mathsf{x}}}'$, $\hat{\bm{\mathsf{y}}}'$, and $\hat{\bm{\mathsf{z}}}'$, where the aperture edges are parallel to $\hat{\bm{\mathsf{x}}}'$ and $\hat{\bm{\mathsf{z}}}'$, while the aperture normal is $\hat{\bm{\mathsf{y}}}'$. These vectors form an orthonormal basis satisfying $\mathbf{C}^{\mathsf{T}}\mathbf{C}=\mathbf{C}\mathbf{C}^{\mathsf{T}}={\mathbf{I}}_3, \quad \mathbf{C}\triangleq [\hat{\bm{\mathsf{x}}}',\hat{\bm{\mathsf{y}}}',\hat{\bm{\mathsf{z}}}']\in{\mathbb{R}}^{3\times3}$. Using this basis, we define a local $\mathsf{x}'\mathsf{y}'\mathsf{z}'$ Cartesian coordinate system with origin at $\mathbf{r}_o$. The matrix $\mathbf{C}$ maps coordinates from the local Rx frame to the global $\mathsf{x}\mathsf{y}\mathsf{z}$ frame. Hence, if ${\mathbf{r}}'=[r_{\mathsf{x}}',r_{\mathsf{y}}',r_{\mathsf{z}}']^{\mathsf{T}}$ denotes the local coordinates of a point on or near the Rx-CAPA, then its global coordinates are given by
\begin{align}\label{Transformation_Coordinate}
{\mathbf{r}}=\mathbf{r}_o+\mathbf{C}{\mathbf{r}}'.
\end{align}
The Tx and Rx aperture regions can therefore be written as
\begin{subequations}
\begin{align}
{\mathcal{S}}_{\mathrm{t}}&=\left\{\mathbf{s} \in \mathbb{R}^3 \, : \, |s_{\mathsf{x}}|\le \frac{L_{{\rm{t}},\mathsf{x}}}{2}, |s_{\mathsf{z}}|\le \frac{L_{{\rm{t}},\mathsf{z}}}{2}, s_{\mathsf{y}} = 0 \right\},\\
{\mathcal{S}}_{\mathrm{r}}&=\left\{(\mathbf{r}_o+\mathbf{C} \mathbf{r}') \in \mathbb{R}^3 \, : \, |r'_{\mathsf{x}}|\le \frac{L_{{\rm{r}},\mathsf{x}}}{2}, |r'_{\mathsf{z}}|\le \frac{L_{{\rm{r}},\mathsf{z}}}{2}, r'_{\mathsf{y}} = 0  \right\},
\end{align}
\end{subequations} 
That is, the Tx aperture is described directly in the global frame, whereas the Rx aperture is described in its own local frame and then mapped into the global coordinates through the matrix $\mathbf{C}$.

\subsubsection{Tri-Polarized Model}

We first consider the most general idealized case. If each point on the Tx and Rx CAPAs is equipped with three mutually orthogonal dipoles, then the aperture can excite and sense arbitrary electric-field polarizations. This idealized antenna configuration is commonly referred to as a \emph{tripole}. For a frequency-nonselective channel between two tri-polarized CAPAs, the transmit and receive vector fields at a fixed time are related by
\begin{align}\label{Tri_Polarized_Signal_Model}
{\mathbf{y}}({\mathbf{r}})=\int_{{\mathcal{S}}_{\mathrm{t}}}{\mathbf{H}}({\mathbf{r}},{\mathbf{s}}){\mathbf{x}}({\mathbf{s}})\,{\rm{d}}\mathbf{s}+{\mathbf{z}}({\mathbf{r}}),
\end{align}
where ${\mathbf{x}}({\mathbf{s}})\in{\mathbb{C}}^{3\times1}$ is the transmit vector signal at point ${\mathbf{s}}\in{{\mathcal{S}}_{\mathrm{t}}}$, ${\mathbf{y}}({\mathbf{r}})\in{\mathbb{C}}^{3\times1}$ is the received vector signal at point ${\mathbf{r}}\in{{\mathcal{S}}_{\mathrm{r}}}$, and ${\mathbf{z}}({\mathbf{r}})$ is the additive noise field. The channel response ${\mathbf{H}}({\mathbf{r}},{\mathbf{s}})\in{\mathbb{C}}^{3\times3}$ is a dyadic integral kernel whose entries describe not only spatial propagation from ${\mathbf{s}}$ to ${\mathbf{r}}$, but also polarization coupling among the three orthogonal components.

In a free-space LoS environment, ${\mathbf{H}}({\mathbf{r}},{\mathbf{s}})$ is governed by the dyadic Green's function $\boldsymbol{\mathcal{G}}_{\omega}(\mathbf{r}, \mathbf{s})$ introduced in Section~\ref{sec:theory}. When $\lVert{\mathbf{r}}-{\mathbf{s}}\rVert \gg \lambda$, the reactive terms are negligible for most communication metrics, and the radiative component provides the LoS channel response as follows:
\begin{align}\label{dy_GF_LoS_Model}
{\mathbf{H}}_{\rm{LoS}}({\mathbf{r}},{\mathbf{s}})
&= \boldsymbol{\mathcal{G}}_{\omega}(\mathbf{r}, \mathbf{s}) \approx \frac{-{\rm{j}}k_0\eta_0{\rm{e}}^{-{\rm{j}}k_0\lVert{\mathbf{r}}-{\mathbf{s}}\rVert}}{4\pi\lVert{\mathbf{r}}-{\mathbf{s}}\rVert}
\left({\mathbf{I}}_3-\hat{\mathbf{u}}\hat{\mathbf{u}}^{\mathsf{T}}\right),
\end{align}
where $\hat{\mathbf{u}} = ({\mathbf{r}}-{\mathbf{s}})/\lVert{\mathbf{r}}-{\mathbf{s}}\rVert$ is the unit propagation-direction vector from the source point to the observation point. The projection matrix $\left({\mathbf{I}}_3-\hat{\mathbf{u}}\hat{\mathbf{u}}^{\mathsf{T}}\right)$ reflects the transverse nature of radiated EM waves in the radiative region.

\subsubsection{Uni-Polarized Model}

We next specialize the model to the more common case in which each point on the Tx and Rx CAPAs excites or senses only one prescribed electric-field component. This is the uni-polarized configuration introduced in Section~\ref{sec:polarization}. It can be viewed as a scalar projection of the tri-polarized model onto the chosen Tx and Rx polarization directions.

By defining the noise field as $z({\mathbf{r}})\triangleq{\hat{\mathbf{p}}_{\mathrm{r}}^{\mathsf{T}}}{\mathbf{z}}({\mathbf{r}})\in{\mathbb{C}}$ and denoting $x({\mathbf{s}})$ as the transmit signal carried along the polarization direction $\hat{\mathbf{p}}_{\mathrm{t}}$, the uni-polarized received signal $y({\mathbf{r}}) \triangleq \hat{\mathbf{p}}_{\mathrm{r}}^{\mathsf{T}} \mathbf{y}(\mathbf{r})$ simplifies from \eqref{Tri_Polarized_Signal_Model} to
\begin{align}\label{Uni_Polarized_Signal_Model_Expression}
y({\mathbf{r}})=\int_{{\mathcal{S}}_{\mathrm{t}}}h({\mathbf{r}},{\mathbf{s}})x({\mathbf{s}})\,{\rm{d}}\mathbf{s}+z({\mathbf{r}}),
\end{align}
where $h({\mathbf{r}},{\mathbf{s}})
    \triangleq
    {\hat{\mathbf{p}}_{\mathrm{r}}^{\mathsf{T}}}{\mathbf{H}}({\mathbf{r}},{\mathbf{s}}){\hat{\mathbf{p}}_{\mathrm{t}}}$ 
characterizes the uni-polarized channel response between ${\mathbf{r}}\in{\mathcal{S}}_{\mathrm{r}}$ and ${\mathbf{s}}\in{\mathcal{S}}_{\mathrm{t}}$.

In a free-space LoS environment, this channel response becomes
\begin{align}\label{eq:uni_pol_LoS_rho}
    & h_{\rm{LoS}}({\mathbf{r}},{\mathbf{s}})
    =
    \hat{\mathbf{p}}_{\mathrm{r}}^{\mathsf{T}} \mathbf{H}_{\rm{LoS}}({\mathbf{r}},{\mathbf{s}}) \hat{\mathbf{p}}_{\mathrm{t}} \nonumber \\
    &\approx
    \frac{-{\rm{j}}\eta_0k_0{\rm{e}}^{-{\rm{j}}k_0\lVert{\mathbf{r}}-{\mathbf{s}}\rVert}}{4\pi\lVert{\mathbf{r}}-{\mathbf{s}}\rVert} \!\!\!
    \underbrace{\hat{\mathbf{p}}_{\mathrm{r}}^{\mathsf{T}} \left( \mathbf{I}_3 - \hat{\mathbf{u}}\hat{\mathbf{u}}^{\mathsf{T}} \right) \hat{\mathbf{p}}_{\mathrm{t}}}_{\text{polarization matching factor } \rho_{\mathrm{pol}} }.
\end{align}
The factor $\rho_{\mathrm{pol}}$ quantifies the polarization compatibility between the Tx excitation and the Rx observation after accounting for the transverse projection imposed by propagation. In particular, it captures the loss caused by polarization mismatch and can even become zero for certain Tx/Rx polarization choices and propagation directions.

Many simplified CAPA channel models omit $\rho_{\mathrm{pol}}$ to focus on spatial propagation, implicitly assuming that the Tx and Rx polarizations are aligned and both are transverse to the propagation direction \cite{bjornson2021primer}. Under this assumption, the LoS channel reduces to
\begin{equation}\label{dyadic Green's function_Standard_Scalar_Often_Used}
h_{\rm{LoS}}({\mathbf{r}},{\mathbf{s}})\approx 
\frac{-{\rm{j}}\eta_0k_0{\rm{e}}^{-{\rm{j}}k_0\lVert{\mathbf{r}}-{\mathbf{s}}\rVert}}{4\pi\lVert{\mathbf{r}}-{\mathbf{s}}\rVert}.
\end{equation}
\rev{This scalar model describes the power of the electric field averaged over all polarization directions. Indeed, since $\mathrm{tr}\{(\mathbf{I}_3-\hat{\mathbf{u}}\hat{\mathbf{u}}^{\mathsf{T}})(\mathbf{I}_3-\hat{\mathbf{u}}\hat{\mathbf{u}}^{\mathsf{T}})^{\mathsf{T}}\}=2$ is a constant, the average power does not depend on the propagation direction $\hat{\mathbf{u}}$ when the source current is equally distributed over all polarization directions \cite{wan2024near}. In practice, however, each antenna transmits and receives along a specific polarization, so that only the corresponding field component contributes to the channel, as captured by $\rho_{\mathrm{pol}}$ in \eqref{eq:uni_pol_LoS_rho}. For example, $\rho_{\mathrm{pol}}$ is close to zero for cross-polarized Tx and Rx, for which the scalar model overestimates the channel strength. Moreover, when the Tx-Rx distance is comparable with the aperture sizes, the waves from different parts of the Tx aperture arrive at the Rx from markedly different directions, so that $\rho_{\mathrm{pol}}$ varies across the apertures~\cite{bjornson2021primer}.}

\rev{Unless otherwise specified, the scalar model in \eqref{dyadic Green's function_Standard_Scalar_Often_Used} will be adopted in the remainder of this article, since it captures the spatial propagation effects that are central to CAPA systems. Whenever the above polarization effects need to be considered, $\rho_{\mathrm{pol}}$ can be incorporated into the channel kernel.}

\subsection{Physics-Based Multipath Channel Models}\label{Section: Multipath Fading Channel Models: Physics-Based}

LoS models are useful for isolating geometric propagation effects, but practical wireless channels generally also contain reflections, diffraction, and scattering caused by surrounding objects. As a result, the received field is typically the superposition of a deterministic LoS component and many non-line-of-sight (NLoS) components. In CAPA systems, multipath channel models can be broadly divided into \emph{physics-based} models, which explicitly describe scatterers and propagation paths in a particular physical setup, and \emph{correlation-based} models, which describe the channel statistically through second-order structures fitted to the typical behaviors in a scenario family. We begin with the physics-based view, since it preserves the propagation geometry most explicitly.

\begin{figure}[!t]
 \centering
\includegraphics[width=0.45\textwidth]{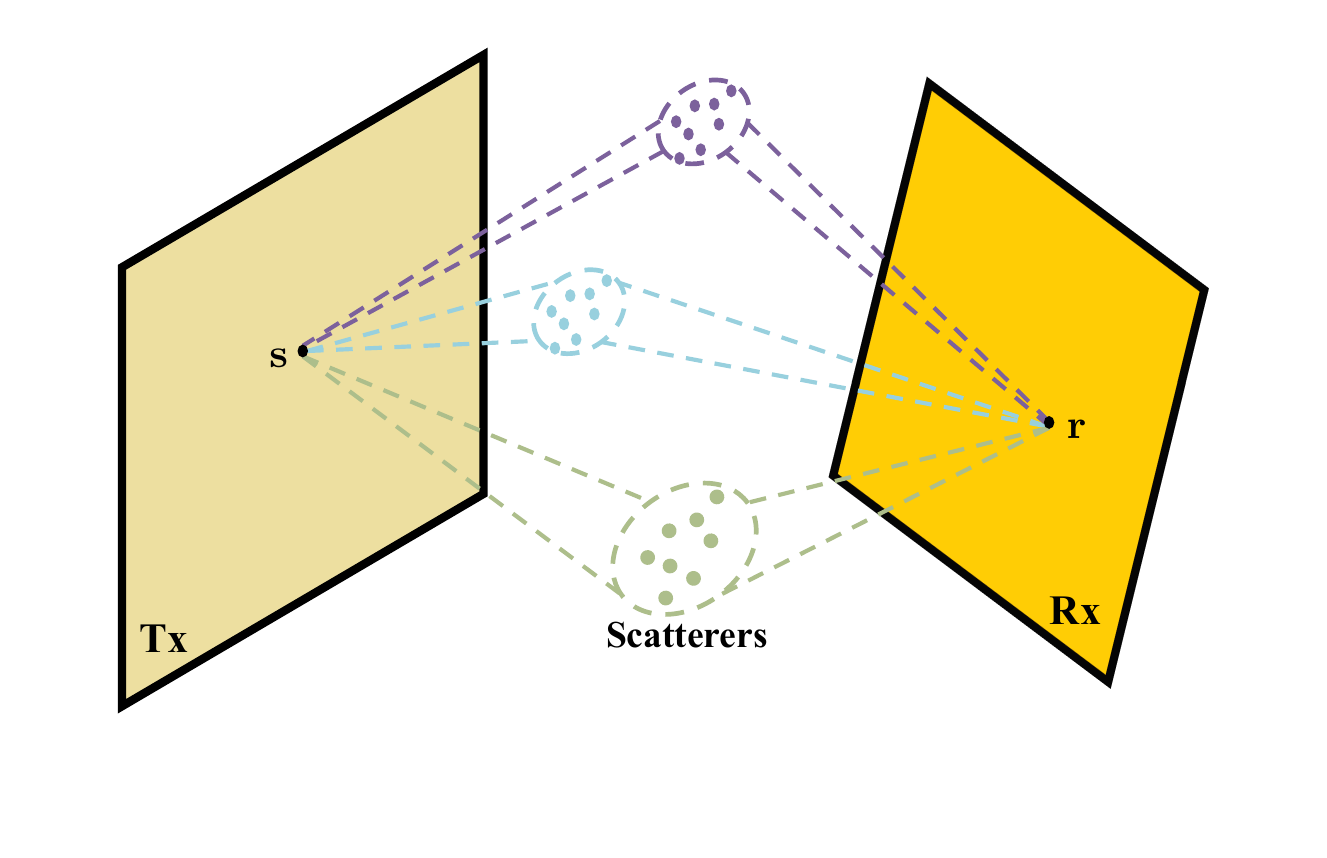}
\caption{Illustration of the physics-based multipath model.}
\label{Figure: Multipath Channel Models: Physics-Based}
\end{figure}

In the uni-polarized setting, a physics-based multipath channel can be written as
\begin{align}
h(\mathbf{r},{\mathbf{s}})
=
\underbrace{h_{\rm{LoS}}({\mathbf{r}},{\mathbf{s}})}_{\text{LoS component}}
+
\underbrace{h_{\rm{NLoS}}({\mathbf{r}},{\mathbf{s}})}_{\text{NLoS component}},
\end{align}
where the NLoS term collects the contributions from reflections and scattering through the environment, as illustrated in Fig.~\ref{Figure: Multipath Channel Models: Physics-Based}. In a general continuous-scattering description, the NLoS component can be decomposed into three responses~\cite{poon2005degrees,pizzo2022spatial} as follows:
\begin{align}
h_{\rm{NLoS}}({\mathbf{r}},{\mathbf{s}})
=
\int_{\varOmega}
a_{\rm{r}}(\mathbf{r},\mathbf{q})\Gamma_{\rm{s}}(\mathbf{q})a_{\rm{t}}(\mathbf{q},\mathbf{s})
\, {\rm{d}}\mathbf{q},
\end{align}
where $\varOmega$ denotes the set of scatterer locations or, more generally, the scattering region in space. The kernels $a_{\rm{t}}(\mathbf{q},\mathbf{s})$, $a_{\rm{r}}(\mathbf{r},\mathbf{q})$, and $\Gamma_{\rm{s}}(\mathbf{q})$ have the following physical meanings:
\begin{itemize}
    \item $a_{\rm{t}}(\mathbf{q},\mathbf{s})$ is the transmit response from the Tx point ${\mathbf{s}}\in{\mathcal{S}}_{\mathrm{t}}$ to the scatterer at ${\mathbf{q}}\in\varOmega$;
    \item $a_{\rm{r}}(\mathbf{r},\mathbf{q})$ is the receive response from the scatterer at ${\mathbf{q}}$ to the Rx point ${\mathbf{r}}\in{\mathcal{S}}_{\mathrm{r}}$;
    \item $\Gamma_{\rm{s}}(\mathbf{q})$ is the scattering response at ${\mathbf{q}}$, which characterizes how the incident wave is attenuated, phase shifted, and reradiated by the local object.
\end{itemize}
Hence, the NLoS field is obtained by integrating over all scatterer locations, with each point contributing through a cascaded Tx-to-scatterer response, local scattering response, and scatterer-to-Rx response.

A common modeling choice is to characterize the Tx and Rx responses using the corresponding LoS propagation kernels~\cite{poon2005degrees,pizzo2022spatial}, namely
\begin{align}
a_{\rm{r}}(\mathbf{r},\mathbf{q})=h_{\rm{LoS}}(\mathbf{r},\mathbf{q}),\quad
a_{\rm{t}}(\mathbf{q},\mathbf{s})=h_{\rm{LoS}}(\mathbf{q},\mathbf{s}).
\end{align} 
This means that each NLoS contribution can be viewed as a cascade of two free-space propagation segments connected by a local scattering interaction. To obtain a more explicit channel representation, one may further adopt a ray-tracing interpretation \cite{zwick2002stochastic} and group the scatterers into clusters as in \cite{spencer2000modeling,gustafson2013mm}. In this case, the scattering response is given by
\begin{align}\label{Target_Response_General}
\Gamma_{\rm{s}}(\mathbf{q})
=
\sum_{l=1}^{N_{l}}
\sum_{j\in\varOmega_{l}}
\Gamma_{l,j}\delta({\mathbf{q}}-{\mathbf{q}}_{l,j}),
\end{align}
where $\Gamma_{l,j}$ describes the complex reflectivity of the $(l,j)$-th scatterer, $\varOmega_{l}$ denotes the index set of scatterers in the $l$-th cluster, and $N_{l}$ is the total number of clusters.

Substituting \eqref{Target_Response_General} into the continuous-scattering model gives the following physics-based channel model:
\begin{align}\label{eq:physics_based_multipath_model}
h(\mathbf{r},{\mathbf{s}})
=
&h_{\rm{LoS}}({\mathbf{r}},{\mathbf{s}})
\nonumber\\
&+
\sum_{l=1}^{N_{l}}
\sum_{j\in\varOmega_{l}}
\Gamma_{l,j}
h_{\rm{LoS}}(\mathbf{r},\mathbf{q}_{l,j})
h_{\rm{LoS}}(\mathbf{q}_{l,j},\mathbf{s}).
\end{align}

The above model can be extended directly to the tri-polarized case as \cite{poon2005degrees}
\begin{align}
{\mathbf{H}}({\mathbf{r}},{\mathbf{s}})
=
&{\mathbf{H}}_{\rm{LoS}}({\mathbf{r}},{\mathbf{s}})
\nonumber\\
&+
\sum_{l=1}^{N_{l}}
\sum_{j\in\varOmega_{l}}
{\mathbf{H}}_{\rm{LoS}}(\mathbf{r},{\mathbf{q}_{l,j}})
{{\mathbf{\Gamma}}_{l,j}}
{\mathbf{H}}_{\rm{LoS}}({\mathbf{q}_{l,j}},\mathbf{s}),
\end{align}
where ${\mathbf{H}}_{\rm{LoS}}(\cdot,\cdot)$ is defined in \eqref{dy_GF_LoS_Model}, and ${{\mathbf{\Gamma}}_{l,j}}\in{\mathbb{C}}^{3\times3}$ represents the attenuation, phase shift, and polarization transformation introduced by the $(l,j)$-th scatterer. In contrast to the scalar model, this matrix-valued scattering response can couple different polarization components and is therefore able to capture depolarization effects induced by the environment.

Physics-based multipath models describe the radio environment through explicit scatterers and propagation paths. Therefore, they retain a clear geometric interpretation, which makes them especially suitable for site-specific modeling and for linking communication performance to physical propagation mechanisms. \rev{Although the model in \eqref{eq:physics_based_multipath_model} considers only single-bounce point scatterers, it can be extended to multiple bounces by cascading several scattering responses, to diffuse scattering and finite-sized objects by adopting a spatially extended $\Gamma_{\rm{s}}(\mathbf{q})$ computed via ray tracing or full-wave solvers, and to blockage by removing the obstructed paths~\cite{zwick2002stochastic,degli2007measurement}.} The main drawback of physics-based models is analytical and computational complexity because the number of paths and clusters may be large \rev{and} the parameters are often site-specific. The model may be cumbersome for theoretical analysis and results obtained using the model only \rev{represent} the considered setup. Hence, these models are particularly useful for case study analysis, deployment planning, ray-tracing-based evaluation, and optimization of reference scenarios \cite{imoize2021standard}.

\subsection{Correlation-Based Multipath Channel Models}\label{Section: Multipath Fading Channel Models: Correlation-Based}
Correlation-based models take a complementary view. Instead of tracking individual scatterers, they characterize the channel through second-order statistics, such as angular power spectra and spatial correlation functions. This abstraction is less site-specific and is often preferred for link- and system-level simulations because it is simpler to generate and calibrate than a full physics-based model~\cite{imoize2021standard}. These models are meant for generating many channel realizations that collectively describe the typical performance in a particular deployment scenario (e.g., outdoor dense urban propagation), although the individual channel realizations lack physical interpretations.

We follow the methodology in \cite{pizzo2022spatial}, which combines Weyl's decomposition of spherical waves into plane waves \cite{chew1999waves,weyl1919ausbreitung} with scattering matrix theory \cite{saxon1955tensor,gerjuoy1954variational,kerns1976plane,nieto1986generalized}. For conciseness, we focus on the NLoS component $h_{\rm{NLoS}}({\mathbf{r}},{\mathbf{s}})$.

\begin{figure}[!t]
 \centering
\includegraphics[width=0.45\textwidth]{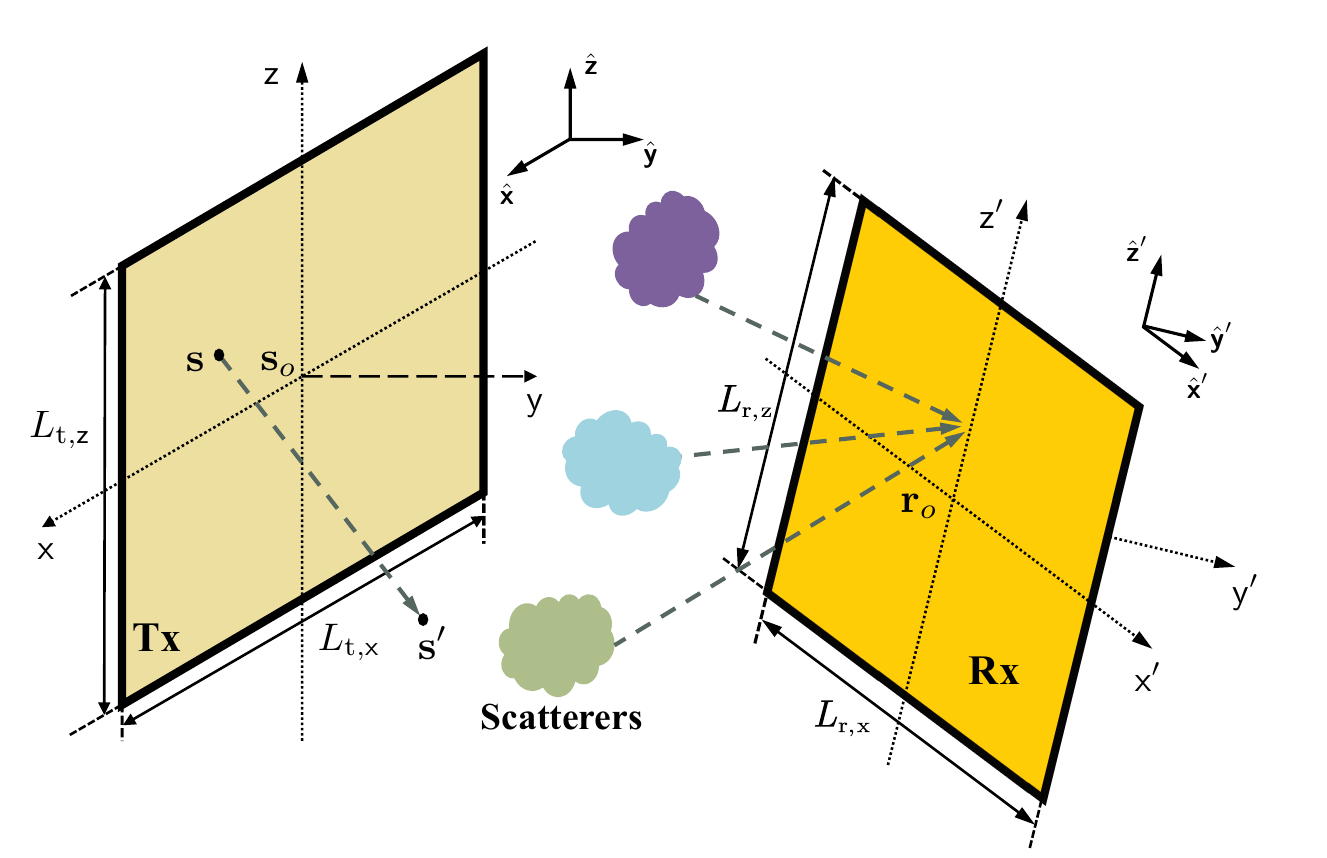}
\caption{Illustration of the correlation-based multipath model.}
\label{Figure: Multipath Channel Models: Correlation-Based}
\end{figure}

\subsubsection{Transmitted Field}
\begin{figure}[!t]
\centering
    \subfigure[$\mathsf{xyz}$.]
    {
        \includegraphics[width=0.225\textwidth]{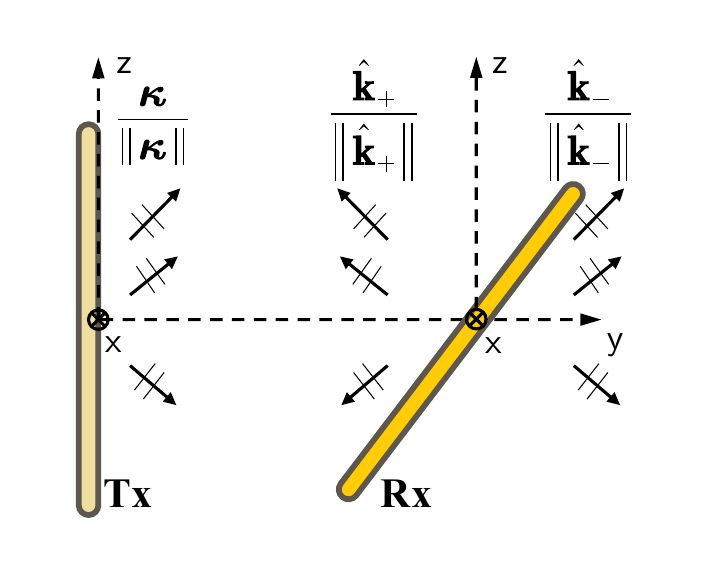}
	   \label{fig_direction1}	
    }
    \subfigure[$\mathsf{xyz}$-$\mathsf{x'y'z'}$.]
    {
        \includegraphics[width=0.225\textwidth]{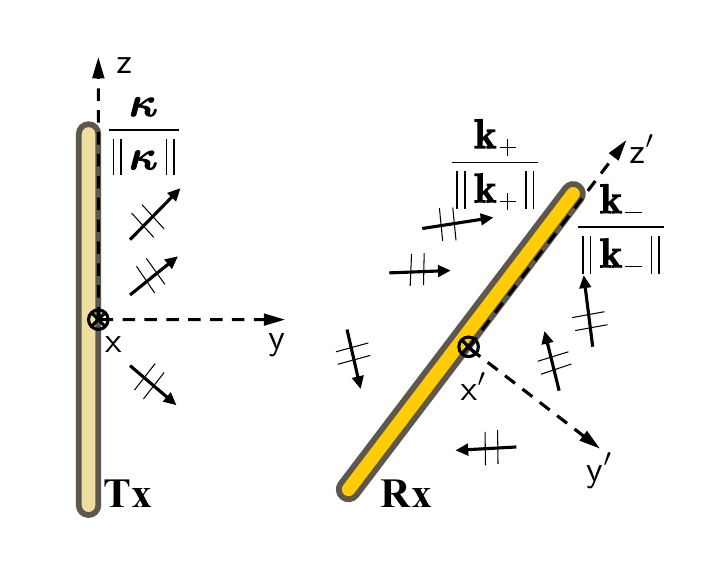}
	   \label{fig_direction2}	
    }
\caption{Projected view of the coordinate systems used in the correlation-based CAPA channel model. (a) shows the global $\mathsf{x}\mathsf{y}\mathsf{z}$ frame for describing the propagation direction of the plane-wave components, where the sign of the normal wavenumber component distinguishes waves associated with the two half-spaces. (b) illustrates the transformation from the global frame to the local Rx frame $\mathsf{x}'\mathsf{y}'\mathsf{z}'$, which is used to express the received plane-wave spectrum and its angular-domain representation.}
\label{Figure_direction}
\end{figure}

We first evaluate the \emph{transmitted field} generated by the Tx aperture at an intermediate point $\mathbf{s}'=[s'_{\mathsf{x}},s'_{\mathsf{y}},s'_{\mathsf{z}}]^{\mathsf{T}}\in{\mathbb{R}}^{3\times1}$ before interaction with the scatterers, as illustrated in Fig.~\ref{Figure: Multipath Channel Models: Correlation-Based}. This field is the free-space response to the source current:
\begin{align}\label{Transmit_Field_Basic}
e_{\rm{t}}({\mathbf{s}}')=\int_{{\mathcal{S}}_{\mathrm{t}}}h_{\rm{LoS}}({\mathbf{s}}',{\mathbf{s}})x({\mathbf{s}}) \, {\rm{d}}{\mathbf{s}},
\end{align}  
where $h_{\rm{LoS}}({\mathbf{s}}',{\mathbf{s}})$ represents the free-space EM propagation from a point $\mathbf{s}$ on the Tx array to $\mathbf{s}'$. Referring to \eqref{dyadic Green's function_Standard_Scalar_Often_Used}, this propagation channel response is mathematically characterized by a scalar Green's function:
\begin{align}\label{Green_Function}
h_{\rm{LoS}}({\mathbf{s}}',{\mathbf{s}})
=\frac{-{\rm{j}}\eta_0k_0{\rm{e}}^{-{\rm{j}}k_0\lVert{\mathbf{s}}'-{\mathbf{s}}\rVert}}{4\pi\lVert{\mathbf{s}}'-{\mathbf{s}}\rVert}.
\end{align}
To separate the field into directional components, we represent the spherical wave in \eqref{Green_Function} as a superposition of plane waves. Weyl's identity \cite{chew1999waves,weyl1919ausbreitung} gives
\begin{align}\label{Green_Function_Sub2}
h_{\rm{LoS}}({\mathbf{s}}',{\mathbf{s}})=-\frac{k_0\eta_0}{8\pi^2}\int_{-\infty}^{+\infty}\int_{-\infty}^{+\infty}\frac{{\rm{e}}^{-{\rm{j}}{\bm{\kappa}}^{\mathsf{T}}({\mathbf{s}}'-{\mathbf{s}})}}{\gamma(\kappa_{\mathsf{x}},\kappa_{\mathsf{z}})}{\rm{d}}\kappa_{\mathsf{x}}{\rm{d}}\kappa_{\mathsf{z}},
\end{align}
where the wave vector is defined as ${\bm{\kappa}}=[\kappa_{\mathsf{x}},{\gamma}(\kappa_{\mathsf{x}},\kappa_{\mathsf{z}}),\kappa_{\mathsf{z}}]^{\mathsf{T}}\in{\mathbb{C}}^{3\times1}$, and the function ${\gamma}(\kappa_{\mathsf{x}},\kappa_{\mathsf{z}})$ denotes the wave number along the $\mathsf{y}$-axis (i.e., the depth direction) given by:
\begin{align}
{\gamma}(\kappa_{\mathsf{x}},\kappa_{\mathsf{z}})\triangleq\left\{\begin{matrix}
\sqrt{k_0^2-\kappa_{\mathsf{x}}^2-\kappa_{\mathsf{z}}^2},&\text{for } \kappa_{\mathsf{x}}^2+\kappa_{\mathsf{z}}^2\leq k_0^2,\\
-{\rm{j}}\sqrt{\kappa_{\mathsf{x}}^2+\kappa_{\mathsf{z}}^2-k_0^2},&\text{for } \kappa_{\mathsf{x}}^2+\kappa_{\mathsf{z}}^2> k_0^2.
\end{matrix}\right.
\end{align}
\rev{Here, $s'_{\mathsf{y}}>s_{\mathsf{y}}$, i.e., the observation point lies in front of the Tx aperture.} Equation \eqref{Green_Function_Sub2} interprets free-space propagation as an integral superposition of plane waves ${\rm{e}}^{-{\rm{j}}{\bm{\kappa}}^{\mathsf{T}}({\mathbf{s}}'-{\mathbf{s}})}$ traveling in directions $\bm{\kappa}/k_0 = \bm{\kappa}/\lVert\bm{\kappa}\rVert$, as visualized in Fig.~\ref{fig_direction1}. This expansion contains two types of components, depending on the transverse spatial frequencies $(\kappa_{\mathsf{x}}, \kappa_{\mathsf{z}})$.

\begin{remark}[\emph{Radiating vs. Evanescent Waves}] \label{remark_evanescent_wave}
    \normalfont
    When $\kappa_{\mathsf{x}}^2+\kappa_{\mathsf{z}}^2\leq k_0^2$, the factor $\gamma(\kappa_{\mathsf{x}},\kappa_{\mathsf{z}})$ is real-valued and positive. These components correspond to \textbf{radiating waves} that can propagate freely over long distances to reach the scatterers. In contrast, when $\kappa_{\mathsf{x}}^2+\kappa_{\mathsf{z}}^2> k_0^2$, the factor $\gamma(\kappa_{\mathsf{x}},\kappa_{\mathsf{z}})$ becomes purely imaginary. Consequently, the corresponding plane wave behaves as
    \begin{align}
    {\rm{e}}^{-{\rm{j}}{\bm{\kappa}}^{\mathsf{T}}({\mathbf{s}}'-{\mathbf{s}})}
    =\underbrace{{\rm{e}}^{-{\rm{j}}\kappa_{\mathsf{x}}(s_{\mathsf{x}}'-s_{\mathsf{x}})}{\rm{e}}^{-{\rm{j}}\kappa_{\mathsf{z}}(s_{\mathsf{z}}'-s_{\mathsf{z}})}}_{\text{oscillating}}
    \underbrace{{\rm{e}}^{-\sqrt{\kappa_{\mathsf{x}}^2+\kappa_{\mathsf{z}}^2-k_0^2}(s_{\mathsf{y}}'-s_{\mathsf{y}})}}_{\text{exponentially decaying}}.
    \end{align}
    This shows that the wave only oscillates along the array plane but decays exponentially along the propagation direction (the $\mathsf{y}$-axis). Such components are known as \textbf{evanescent waves}. Because their decay factor is proportional to the distance relative to the wavelength, they are strictly confined to regions very near the Tx array (typically within a few wavelengths)\rev{; their contribution to the spatial DoF will be discussed in Section~\ref{sec:fundamental_limits}}.
\end{remark}

For the multipath models considered here, the scatterers are not located in the evanescent near zone of the Tx aperture. We therefore retain only the radiating components in the disk ${\mathcal{D}}({\bm\kappa})\triangleq\{(\kappa_{\mathsf{x}},\kappa_{\mathsf{z}})\in{\mathbb{R}}^2:\kappa_{\mathsf{x}}^2+\kappa_{\mathsf{z}}^2\leq k_0^2\}$. Substituting the retained part of \eqref{Green_Function_Sub2} into \eqref{Transmit_Field_Basic} gives
\begin{align}\label{Transmit_Field_Basic_Fourier}
e_{\rm{t}}({\mathbf{s}}')=\iint_{{\mathcal{D}}({\bm\kappa})}\frac{{\rm{d}}\kappa_{\mathsf{x}}}{2\pi}\frac{{\rm{d}}\kappa_{\mathsf{z}}}{2\pi}
{\rm{e}}^{-{\rm{j}}{\bm{\kappa}}^{\mathsf{T}}{\mathbf{s}}'}{E}_{\rm{t}}(\kappa_{\mathsf{x}},\kappa_{\mathsf{z}}),
\end{align}  
where ${E}_{\rm{t}}(\kappa_{\mathsf{x}},\kappa_{\mathsf{z}})$ captures the cumulative effect of the transmit current on the specific wave direction and is defined as
\begin{align}\label{Transmit_Field_Basic_Fourier_Final}
{E}_{\rm{t}}(\kappa_{\mathsf{x}},\kappa_{\mathsf{z}})\triangleq\rev{-}\frac{k_0\eta_0}{2{\gamma}(\kappa_{\mathsf{x}},\kappa_{\mathsf{z}})}\int_{{\mathcal{S}}_{\mathrm{t}}}
{{\rm{e}}^{{\rm{j}}{\bm{\kappa}}^{\mathsf{T}}{\mathbf{s}}}}x({\mathbf{s}}) \, {\rm{d}}{\mathbf{s}}.
\end{align}
Equation \eqref{Transmit_Field_Basic_Fourier} shows that the transmitted field is a continuum of radiating plane waves, each traveling in direction ${\bm{\kappa}}/\lVert\bm{\kappa}\rVert$. The coefficient ${E}_{\rm{t}}(\kappa_{\mathsf{x}},\kappa_{\mathsf{z}})$ is the corresponding \emph{transmit plane-wave spectrum}; it specifies how strongly the aperture current excites each radiating direction.

\subsubsection{Received Field}
Unlike the transmitted field, which is generated before scattering, the \emph{received field} at position $\mathbf{r}$ is the superposition of waves arriving after interaction with the environment. It is defined as
\begin{align}\label{Correlation_Based_Model_Received_Field}
e_{\rm{r}}({\mathbf{r}})=\int_{{\mathcal{S}}_{\mathrm{t}}}h_{\rm{NLoS}}({\mathbf{r}},{\mathbf{s}})x({\mathbf{s}})\,{\rm{d}}{\mathbf{s}}.
\end{align}

Analogously to the transmitted-field expansion in \eqref{Green_Function_Sub2}, the received field can be represented as a superposition of plane waves arriving from all possible directions \cite[Sec. 6.7]{stratton2007electromagnetic}:
\begin{align}
e_{\rm{r}}({\mathbf{r}}) = &\int_{-\infty}^{+\infty}\int_{-\infty}^{+\infty}\frac{{\rm{d}}\hat{k}_{\mathsf{x}}}{2\pi}\frac{{\rm{d}}\hat{k}_{\mathsf{z}}}{2\pi} \nonumber \\
&\times\left({\rm{e}}^{-{\rm{j}}{\hat{\mathbf{k}}}_{+}^{\mathsf{T}}{\mathbf{r}}}\hat{E}_{\rm{r}}^{+}(\hat{k}_{\mathsf{x}},\hat{k}_{\mathsf{z}}) +{\rm{e}}^{-{\rm{j}}{\hat{\mathbf{k}}}_{-}^{\mathsf{T}}{\mathbf{r}}}\hat{E}_{\rm{r}}^{-}(\hat{k}_{\mathsf{x}},\hat{k}_{\mathsf{z}})\right).
\end{align}
Here, the wave vectors are ${\hat{\mathbf{k}}}_{\pm}=[\hat{k}_{\mathsf{x}},\pm{\gamma}(\hat{k}_{\mathsf{x}},\hat{k}_{\mathsf{z}}),\hat{k}_{\mathsf{z}}]^{\mathsf{T}}$. The $\pm$ superscripts distinguish plane waves arriving from the two half-spaces on either side of the $\mathsf{x}$-$\mathsf{z}$ plane, as depicted in Fig.~\ref{fig_direction1}. The function $\hat{E}_{\rm{r}}^{\pm}(\hat{k}_{\mathsf{x}},\hat{k}_{\mathsf{z}})$ is the \emph{received plane-wave spectrum}; it gives the complex amplitude of the plane wave arriving from direction ${\hat{\mathbf{k}}}_{\pm}/\lVert{\hat{\mathbf{k}}}_{\pm}\rVert$.

As in the transmitted-field case, we neglect evanescent components and restrict the integration to the radiating region. The received field then simplifies to
\begin{align}\label{Received_Field_Simplified}
e_{\rm{r}}({\mathbf{r}}) = &\iint_{{\mathcal{D}}(\hat{\mathbf{k}})}\frac{{\rm{d}}\hat{k}_{\mathsf{x}}}{2\pi}\frac{{\rm{d}}\hat{k}_{\mathsf{z}}}{2\pi} \nonumber \\
&\times\left( {\rm{e}}^{-{\rm{j}}{\hat{\mathbf{k}}}_{+}^{\mathsf{T}}{\mathbf{r}}}\hat{E}_{\rm{r}}^{+}(\hat{k}_{\mathsf{x}},\hat{k}_{\mathsf{z}}) + {\rm{e}}^{-{\rm{j}}{\hat{\mathbf{k}}}_{-}^{\mathsf{T}}{\mathbf{r}}}\hat{E}_{\rm{r}}^{-}(\hat{k}_{\mathsf{x}},\hat{k}_{\mathsf{z}}) \right),
\end{align}
where the integration domain is ${\mathcal{D}}(\hat{\mathbf{k}})=\{(\hat{k}_{\mathsf{x}},\hat{k}_{\mathsf{z}})\in{\mathbb{R}}^2 : \hat{k}_{\mathsf{x}}^2+\hat{k}_{\mathsf{z}}^2\leq k_0^2\}$. 

For later use, it is convenient to express the received field in the local coordinate system of the Rx aperture. We therefore transform the wave vectors from the global $\mathsf{x}\mathsf{y}\mathsf{z}$ frame to the local $\mathsf{x}'\mathsf{y}'\mathsf{z}'$ frame:
\begin{align}
{\hat{\mathbf{k}}}_{\pm}=\mathbf{C}{{\mathbf{k}}}_{\pm}\quad \Longleftrightarrow \quad \mathbf{C}^{\mathsf{T}}{\hat{\mathbf{k}}}_{\pm}={{\mathbf{k}}}_{\pm},
\end{align}
where ${{\mathbf{k}}}_{\pm}=[{k}_{\mathsf{x}},\pm{{\gamma}}({k}_{\mathsf{x}},{k}_{\mathsf{z}}),{k}_{\mathsf{z}}]^{\mathsf{T}}\in{\mathbb{R}}^{3\times1}$ is the wave vector expressed in the local Rx frame. Applying this invertible transformation to \eqref{Received_Field_Simplified} gives
\begin{align}\label{Received_Field_Simplified_Further}
&e_{\rm{r}}({\mathbf{r}}) = \iint_{{\mathcal{D}}({\mathbf{k}})}\frac{{\rm{d}}{k}_{\mathsf{x}}}{2\pi}\frac{{\rm{d}}{k}_{\mathsf{z}}}{2\pi} \nonumber \\
&\times\left( {\rm{e}}^{-{\rm{j}}{{\mathbf{k}}}_{+}^{\mathsf{T}}\mathbf{C}^{\mathsf{T}}{\mathbf{r}}}{E}_{\rm{r}}^{+}({k}_{\mathsf{x}},{k}_{\mathsf{z}}) + {\rm{e}}^{-{\rm{j}}{{\mathbf{k}}}_{-}^{\mathsf{T}}\mathbf{C}^{\mathsf{T}}{\mathbf{r}}}{E}_{\rm{r}}^{-}({k}_{\mathsf{x}},{k}_{\mathsf{z}}) \right),
\end{align}
where ${\mathcal{D}}({\mathbf{k}})=\{({k}_{\mathsf{x}},{k}_{\mathsf{z}})\in{\mathbb{R}}^2 \mid {k}_{\mathsf{x}}^2+{k}_{\mathsf{z}}^2\leq k_0^2\}$. Here, ${E}_{\rm{r}}^{\pm}({k}_{\mathsf{x}},{k}_{\mathsf{z}})$ is the representation of $\hat{E}_{\mathsf{r}}^{\pm}(\hat{k}_{\mathsf{x}},\hat{k}_{\mathsf{z}})$ in the local coordinate frame. A closed-form expression for this transformed spectrum is not needed; the channel model only requires the mapping from the transmit spectrum ${E}_{\rm{t}}(\kappa_{\mathsf{x}},\kappa_{\mathsf{z}})$ to the receive spectrum ${E}_{\rm{r}}^{\pm}({k}_{\mathsf{x}},{k}_{\mathsf{z}})$.

Finally, the two terms in \eqref{Received_Field_Simplified_Further} represent waves arriving from the two sides of the Rx aperture plane, as illustrated in Fig.~\ref{fig_direction2}. If the model focuses on illumination from the front side of the Rx aperture, the opposite component can be dropped. Thus, \eqref{Received_Field_Simplified_Further} reduces to
\begin{align}\label{Received_Field_Simplified_Final}
e_{\rm{r}}({\mathbf{r}})=\iint_{{\mathcal{D}}({\mathbf{k}})}\frac{{\rm{d}}{k}_{\mathsf{x}}}{2\pi}\frac{{\rm{d}}{k}_{\mathsf{z}}}{2\pi}
{\rm{e}}^{-{\rm{j}}{{\mathbf{k}}}^{\mathsf{T}}\mathbf{C}^{\mathsf{T}}{\mathbf{r}}}{E}_{\rm{r}}({k}_{\mathsf{x}},{k}_{\mathsf{z}}),
\end{align}
where we define ${\mathbf{k}}\triangleq{\mathbf{k}}_{+}$ and ${E}_{\rm{r}}({k}_{\mathsf{x}},{k}_{\mathsf{z}})\triangleq{E}_{\rm{r}}^{+}({k}_{\mathsf{x}},{k}_{\mathsf{z}})$ for brevity.

\subsubsection{Linear Scattering Operator}
According to EM scattering theory \cite{saxon1955tensor,gerjuoy1954variational,kerns1976plane,nieto1986generalized}, the scattering environment acts as a linear operator that maps the \emph{transmit plane-wave spectrum} ${E}_{\rm{t}}(\kappa_{\mathsf{x}},\kappa_{\mathsf{z}})$ into the \emph{receive plane-wave spectrum} ${E}_{\rm{r}}({k}_{\mathsf{x}},{k}_{\mathsf{z}})$. This transformation is described by
\begin{align}\label{Linear_Mapping}
{E}_{\rm{r}}({k}_{\mathsf{x}},{k}_{\mathsf{z}})=\iint_{{\mathcal{D}}({\bm\kappa})}{{K}}_a({\mathbf{k}},{\bm\kappa}){E}_{\rm{t}}(\kappa_{\mathsf{x}},\kappa_{\mathsf{z}}){{\rm{d}}\kappa}_{\mathsf{x}}{{\rm{d}}\kappa}_{\mathsf{z}},
\end{align}
where ${{K}}_a({\mathbf{k}},{\bm\kappa})\in{\mathbb{C}}$ is the \emph{propagation kernel function}. It characterizes how a transmitted plane wave in direction $\bm{\kappa}$ couples into a received plane wave in direction $\mathbf{k}$ due to the presence of scatterers.

Substituting the transmit spectrum in \eqref{Transmit_Field_Basic_Fourier_Final} into \eqref{Linear_Mapping} expresses the receive spectrum directly in terms of the transmit current $x(\mathbf{s})$:
\begin{align}
{E}_{\rm{r}}({k}_{\mathsf{x}},{k}_{\mathsf{z}})=&\rev{-}\frac{k_0\eta_0}{2}\iint_{{\mathcal{D}}({\bm\kappa})}{{K}}_a({\mathbf{k}},{\bm\kappa}) \nonumber\\
&\times\int_{{\mathcal{S}}_{\mathrm{t}}}
\frac{{\rm{e}}^{{\rm{j}}{\bm{\kappa}}^{\mathsf{T}}{\mathbf{s}}}x({\mathbf{s}}){\rm{d}}{\mathbf{s}}}{{\gamma}(\kappa_{\mathsf{x}},\kappa_{\mathsf{z}})}{{\rm{d}}\kappa_{\mathsf{x}}}{{\rm{d}}\kappa_{\mathsf{z}}}.
\end{align}
Integrating this receive spectrum over all arriving directions according to \eqref{Received_Field_Simplified_Final} yields the total received field at $\mathbf{r}$:
\begin{align}\label{Correlation_Based_Model_TR_Field}
e_{\rm{r}}({\mathbf{r}})&=\rev{-}\frac{1}{(2\pi)^2}\frac{k_0\eta_0}{2}\iiiint_{{\mathcal{D}}({\mathbf{k}})\times{\mathcal{D}}({\bm\kappa})}
{\rm{e}}^{-{\rm{j}}{{\mathbf{k}}}^{\mathsf{T}}\mathbf{C}^{\mathsf{T}}{\mathbf{r}}} \nonumber \\
&\times \frac{K_a({\mathbf{k}},{\bm\kappa})}{{{\gamma}(\kappa_{\mathsf{x}},\kappa_{\mathsf{z}})}}\int_{{\mathcal{S}}_{\mathrm{t}}}
{{\rm{e}}^{{\rm{j}}{\bm{\kappa}}^{\mathsf{T}}{\mathbf{s}}}x({\mathbf{s}}){\rm{d}}{\mathbf{s}}} \, {\rm{d}}k_{\mathsf{x}}{\rm{d}}k_{\mathsf{z}}{\rm{d}}\kappa_{\mathsf{x}}{\rm{d}}\kappa_{\mathsf{z}}.
\end{align}
Comparing \eqref{Correlation_Based_Model_TR_Field} with the input-output relation in \eqref{Correlation_Based_Model_Received_Field} identifies the NLoS channel kernel as
\begin{align}\label{4FPWD_Model}
h_{\rm{NLoS}}({\mathbf{r}},{\mathbf{s}})&=\rev{-}\frac{1}{(2\pi)^2}\frac{k_0\eta_0}{2}\iiiint_{{\mathcal{D}}({\mathbf{k}})\times{\mathcal{D}}({\bm\kappa})}
{\rm{e}}^{-{\rm{j}}{{\mathbf{k}}}^{\mathsf{T}}\mathbf{C}^{\mathsf{T}}{\mathbf{r}}} \nonumber \\
&\times \frac{K_a({\mathbf{k}},{\bm\kappa})}{{{\gamma}(\kappa_{\mathsf{x}},\kappa_{\mathsf{z}})}}{\rm{e}}^{{\rm{j}}{\bm{\kappa}}^{\mathsf{T}}{\mathbf{s}}}{\rm{d}}k_{\mathsf{x}}{\rm{d}}k_{\mathsf{z}}{\rm{d}}\kappa_{\mathsf{x}}{\rm{d}}\kappa_{\mathsf{z}}.
\end{align}
Using the coordinate transformation ${\mathbf{r}}=\mathbf{r}_o+\mathbf{C}{\mathbf{r}}'$ in \eqref{Transformation_Coordinate} and the orthogonality of $\mathbf{C}$, \eqref{4FPWD_Model} can be written in the local aperture coordinates as
\begin{align}\label{4FPWD_Model_Final}
h_{\rm{NLoS}}({\mathbf{r}},{\mathbf{s}})&=\frac{1}{(2\pi)^2}\iiiint_{{\mathcal{D}}({\mathbf{k}})\times{\mathcal{D}}({\bm\kappa})}
{\rm{e}}^{-{\rm{j}}{{\mathbf{k}}}^{\mathsf{T}}{\mathbf{r}}'}\nonumber\\
&\times{{H}}_a({\mathbf{k}},{\bm\kappa}){\rm{e}}^{{\rm{j}}{\bm{\kappa}}^{\mathsf{T}}{\mathbf{s}}}{\rm{d}}k_{\mathsf{x}}{\rm{d}}k_{\mathsf{z}}{\rm{d}}\kappa_{\mathsf{x}}{\rm{d}}\kappa_{\mathsf{z}},
\end{align}
where ${{H}}_a({\mathbf{k}},{\bm\kappa})$ is the \emph{angular} or \emph{wavenumber-domain} response and is given by
\begin{align}
{{H}}_a({\mathbf{k}},{\bm\kappa})\triangleq\rev{-}\frac{k_0\eta_0}{2}\frac{{K}_a({\mathbf{k}},{\bm\kappa})}{\rev{\gamma(\kappa_{\mathsf{x}},\kappa_{\mathsf{z}})}}{\rm{e}}^{-{\rm{j}}{{\mathbf{k}}}^{\mathsf{T}}\mathbf{C}^{\mathsf{T}}{\mathbf{r}}_{o}}.
\end{align} 
This angular response describes how each transmit direction $\bm\kappa/\lVert{\bm\kappa}\rVert$ is coupled by the environment into each receive direction $\mathbf{k}/\lVert\mathbf{k}\rVert$. In a multipath fading environment, $H_a({\mathbf{k}},{\bm\kappa})$ is typically modeled as a random process. Since $\mathbf{s}$ and $\mathbf{r}'$ lie on the local aperture planes, i.e., ${\mathbf{s}}=[s_{\mathsf{x}},0,s_{\mathsf{z}}]^{\mathsf{T}}$ and ${\mathbf{r}}'=[r_{\mathsf{x}}',0,r_{\mathsf{z}}']^{\mathsf{T}}$, \eqref{4FPWD_Model_Final} becomes the continuous spatial Fourier representation
\begin{align}\label{4FPWD_Model_Standard}
h_{\rm{NLoS}}({\mathbf{r}},{\mathbf{s}})&=\frac{1}{(2\pi)^2}\iiiint_{{\mathcal{D}}({\mathbf{k}})\times{\mathcal{D}}({\bm\kappa})}
{\rm{e}}^{-{\rm{j}}(r_{\mathsf{x}}'k_{\mathsf{x}}+r_{\mathsf{z}}'k_{\mathsf{z}})}\nonumber\\
&\times{{H}}_a({\mathbf{k}},{\bm\kappa}){\rm{e}}^{{\rm{j}}(s_{\mathsf{x}}\kappa_{\mathsf{x}}+s_{\mathsf{z}}\kappa_{\mathsf{z}})}{\rm{d}}k_{\mathsf{x}}{\rm{d}}k_{\mathsf{z}}{\rm{d}}\kappa_{\mathsf{x}}{\rm{d}}\kappa_{\mathsf{z}}.
\end{align}

If the scattering environment induces \emph{Rayleigh fading}, the NLoS channel response $h_{\rm{NLoS}}({\mathbf{r}},{\mathbf{s}})$ is modeled as a spatially stationary, circularly symmetric complex Gaussian (CSCG) random field. The angular response then adopts the statistical structure \cite{pizzo2022spatial}
\begin{align}\label{Angular_Response_Version_1}
{{H}}_a({\mathbf{k}},{\bm\kappa})=\frac{{{A}}_a({\mathbf{k}},{\bm\kappa}){{W}}_a({\mathbf{k}},{\bm\kappa})}{\sqrt{\gamma(k_{\mathsf{x}},k_{\mathsf{z}})\gamma(\kappa_{\mathsf{x}},\kappa_{\mathsf{z}})}},
\end{align}
where ${{A}}_a({\mathbf{k}},{\bm\kappa})\geq0$ is an arbitrary deterministic non-negative function known as the \emph{spectral factor}. The term ${{W}}_a({\mathbf{k}},{\bm\kappa})$ is a zero-mean, unit-variance CSCG random field defined over the spatial domain ${\mathcal{D}}({\mathbf{k}})\times{\mathcal{D}}({\bm\kappa})$. Specifically, ${{W}}_a({\mathbf{k}},{\bm\kappa})\sim{\mathcal{CN}}(0,1)$, satisfying the uncorrelated scattering condition such that its autocorrelation is a Dirac delta function, i.e., ${\mathbb{E}}\{W_a({\mathbf{k}},{\bm\kappa}){W}_a^{*}({\mathbf{k}}',{\bm\kappa}')\}=\delta({\mathbf{k}}-{\mathbf{k}}')\delta({\bm\kappa}-{\bm\kappa}')$. By adding a deterministic LoS component to this random NLoS response, we obtain the full correlation-based Rician fading model:
\begin{align}
h({\mathbf{r}},{\mathbf{s}})= h_{\rm{LoS}}({\mathbf{r}},{\mathbf{s}}) + h_{\rm{NLoS}}({\mathbf{r}},{\mathbf{s}}).
\end{align}

To interpret the spectral factor geometrically, we can rewrite \eqref{Angular_Response_Version_1} as 
\begin{align}\label{Angular_Response_Version_2}
{{H}}_a({\mathbf{k}},{\bm\kappa})=S^{\frac{1}{2}}({\mathbf{k}},{\bm\kappa}){{W}}_a({\mathbf{k}},{\bm\kappa}),
\end{align}
where $S({\mathbf{k}},{\bm\kappa})$ is defined as
\begin{align}\label{Angular_Domain_Power_Distribution_Isotropic_Scattering}
S({\mathbf{k}},{\bm\kappa})=\frac{{{A}}_a^2({\mathbf{k}},{\bm\kappa})}{\gamma(k_{\mathsf{x}},k_{\mathsf{z}})\gamma(\kappa_{\mathsf{x}},\kappa_{\mathsf{z}})}.
\end{align}
Substituting this statistical model into \eqref{4FPWD_Model_Final} gives the average NLoS channel power:
\begin{align}\label{Average_Power_Multipath_Correlation}
{\mathbb{E}}\{\lvert{{h}}_{\rm{NLoS}}({\mathbf{r}},{\mathbf{s}})\rvert^2\} =\iiiint_{{\mathcal{D}}({\mathbf{k}})\times{\mathcal{D}}({\bm\kappa})}
\frac{S({\mathbf{k}},{\bm\kappa})}{(2\pi)^4}
{\rm{d}}k_{\mathsf{x}}{\rm{d}}k_{\mathsf{z}}{\rm{d}}\kappa_{\mathsf{x}}{\rm{d}}\kappa_{\mathsf{z}}.
\end{align}
Therefore, $S({\mathbf{k}},{\bm\kappa})$ can be regarded as a \emph{continuous angular power distribution} of the channel.

\subsubsection{Physical Modeling of the Spectral Factor}
The representation in \eqref{4FPWD_Model_Final} becomes a concrete channel model once the angular power distribution is specified. This distribution determines which departure and arrival directions carry significant power, and therefore controls the spatial correlation seen by the CAPAs. To connect the abstract wavenumber variables to measurable quantities, we next express the spectral factor ${{A}}_a({\mathbf{k}},{\bm\kappa})$ in terms of physical elevation and azimuth angles.

We transform the integration variables in \eqref{Average_Power_Multipath_Correlation} from wavenumbers $({\bm\kappa}, {\mathbf{k}})$ to elevation and azimuth angles $(\theta,\phi)\in[0,\pi]\times[0,\pi)$ at the Tx and Rx. The corresponding directional unit vectors are
\begin{align}
\frac{{\bm{\kappa}}}{\lVert{\bm{\kappa}}\rVert}
&=\frac{{\bm{\kappa}}}{k_0}=\hat{\bm{\mathsf{x}}}\frac{\kappa_{\mathsf{x}}}{k_0}+\hat{\bm{\mathsf{y}}}\frac{{\gamma}(\kappa_{\mathsf{x}},\kappa_{\mathsf{z}})}{k_0}+\hat{\bm{\mathsf{z}}}\frac{\kappa_{\mathsf{z}}}{k_0}\nonumber\\
&=\hat{\bm{\mathsf{x}}}\sin{\theta_{\rm{t}}}\cos{\phi_{\rm{t}}}+\hat{\bm{\mathsf{y}}}\sin{\theta_{\rm{t}}}\sin{\phi_{\rm{t}}}+\hat{\bm{\mathsf{z}}}\cos{\theta_{\rm{t}}}, \label{Wavenumber_Angle_Trans_TX} \\
\frac{{\mathbf{k}}}{\lVert{\mathbf{k}}\rVert}
&=\frac{{\mathbf{k}}}{k_0}=\hat{\bm{\mathsf{x}}}'\frac{k_{\mathsf{x}}}{k_0}+\hat{\bm{\mathsf{y}}}'\frac{{\gamma}(k_{\mathsf{x}},k_{\mathsf{z}})}{k_0}+\hat{\bm{\mathsf{z}}}'\frac{k_{\mathsf{z}}}{k_0}\nonumber\\
&=\hat{\bm{\mathsf{x}}}'\sin{\theta_{\rm{r}}}\cos{\phi_{\rm{r}}}+\hat{\bm{\mathsf{y}}}'\sin{\theta_{\rm{r}}}\sin{\phi_{\rm{r}}}+\hat{\bm{\mathsf{z}}}'\cos{\theta_{\rm{r}}},
\label{Wavenumber_Angle_Trans_RX}
\end{align}
where $(\theta_{\rm{t}},\phi_{\rm{t}})$ correspond to the elevation and azimuth angles in the localized Tx coordinate frame, while $(\theta_{\rm{r}},\phi_{\rm{r}})$ denote their counterparts at the Rx. 

\begin{figure}[!t]
 \centering
\includegraphics[width=0.45\textwidth]{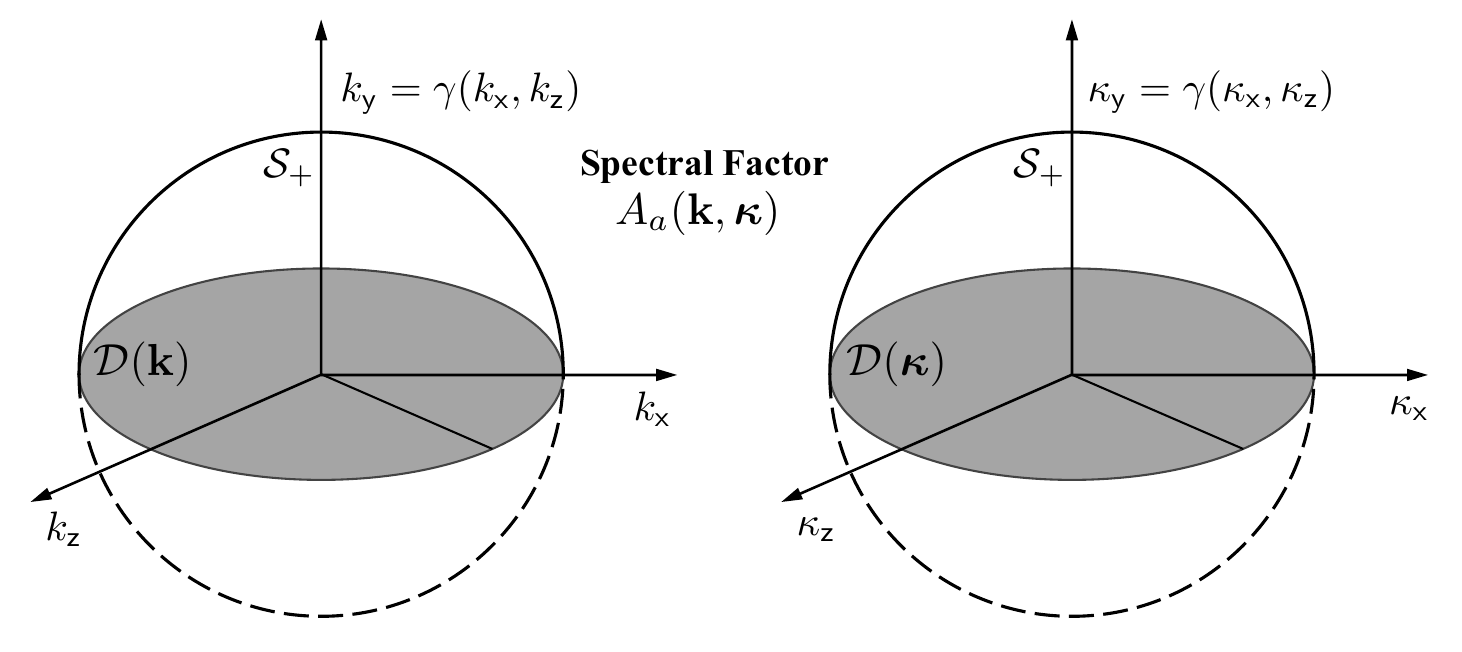}
\caption{Illustration of the power spectral density of a spatially stationary channel impulse response ${{h}}_{\rm{NLoS}}({\mathbf{r}},{\mathbf{s}})$ for the correlation-based channel model.}
\label{Figure: Multipath Channel Models: Correlation-Based: PSD_Correlation}
\end{figure}

Applying this change of variables, the average channel power in \eqref{Average_Power_Multipath_Correlation} can be rewritten as
\begin{align}\label{Average_Power_Multipath_Correlation_Transform}
&{\mathbb{E}}\{\lvert{{h}}_{\rm{NLoS}}({\mathbf{r}},{\mathbf{s}})\rvert^2\}\nonumber\\
&=\iiiint_{{\mathcal{D}}({\mathbf{k}})\times{\mathcal{D}}({\bm\kappa})}
\frac{{{A}}_a^2({\mathbf{k}},{\bm\kappa})}{(2\pi)^4}\frac{{\rm{d}}k_{\mathsf{x}}{\rm{d}}k_{\mathsf{z}}{\rm{d}}\kappa_{\mathsf{x}}{\rm{d}}\kappa_{\mathsf{z}}}{\gamma(k_{\mathsf{x}},k_{\mathsf{z}})\gamma(\kappa_{\mathsf{x}},\kappa_{\mathsf{z}})}\nonumber\\
&=\int_{0}^{\pi}\int_{0}^{\pi}\int_{0}^{\pi}\int_{0}^{\pi}
\frac{1}{(2\pi)^4}\frac{p(\theta_{\rm{r}},\phi_{\rm{r}},\theta_{\rm{t}},\phi_{\rm{t}})}{\gamma(k_{\mathsf{x}},k_{\mathsf{z}})\gamma(\kappa_{\mathsf{x}},\kappa_{\mathsf{z}})}\nonumber\\
&\times \left\lvert\frac{\partial(k_{\mathsf{x}},k_{\mathsf{z}})}{\partial(\theta_{\rm{r}},\phi_{\rm{r}})}\right\rvert\left\lvert\frac{\partial(\kappa_{\mathsf{x}},\kappa_{\mathsf{z}})}{\partial(\theta_{\rm{t}},\phi_{\rm{t}})}\right\rvert {\rm{d}}\theta_{\rm{r}}{\rm{d}}\phi_{\rm{r}}{\rm{d}}\theta_{\rm{t}}{\rm{d}}\phi_{\rm{t}},
\end{align}
where $p(\theta_{\rm{r}},\phi_{\rm{r}},\theta_{\rm{t}},\phi_{\rm{t}})\triangleq{{A}}_a^2(\eqref{Wavenumber_Angle_Trans_RX}\times k_0,\eqref{Wavenumber_Angle_Trans_TX}\times k_0)$ denotes the \emph{angular power distribution function}. The Jacobian determinants introduce the required geometric scaling:
\begin{subequations}
\begin{align}
&\left\lvert\frac{\partial(k_{\mathsf{x}},k_{\mathsf{z}})}{\partial(\theta_{\rm{r}},\phi_{\rm{r}})}\right\rvert=k_0^2\sin^2{\theta_{\rm{r}}}\sin{\phi_{\rm{r}}}=k_0\gamma(k_{\mathsf{x}},k_{\mathsf{z}})\sin{\theta_{\rm{r}}},\\
&\left\lvert\frac{\partial(\kappa_{\mathsf{x}},\kappa_{\mathsf{z}})}{\partial(\theta_{\rm{t}},\phi_{\rm{t}})}\right\rvert=k_0^2\sin^2{\theta_{\rm{t}}}\sin{\phi_{\rm{t}}}=k_0\gamma(\kappa_{\mathsf{x}},\kappa_{\mathsf{z}})\sin{\theta_{\rm{t}}}.
\end{align}
\end{subequations}
Substituting these Jacobians into \eqref{Average_Power_Multipath_Correlation_Transform} yields
\begin{align}\label{Average_Power_Multipath_Correlation_Transform_Again}
{\mathbb{E}}\{\lvert{{h}}_{\rm{NLoS}}({\mathbf{r}},{\mathbf{s}})\rvert^2\}
&=\iiiint_{{\mathcal{S}}_{+}\times {\mathcal{S}}_{+}}
\frac{k_0^2}{(2\pi)^4}\nonumber\\
&\times p(\theta_{\rm{r}},\phi_{\rm{r}},\theta_{\rm{t}},\phi_{\rm{t}}){\rm{d}}\Omega_{\rm{r}}{\rm{d}}\Omega_{\rm{t}},
\end{align}
where ${\rm{d}}\Omega_{\rm{r}}=\sin{\theta_{\rm{r}}} {\rm{d}}\theta_{\rm{r}}{\rm{d}}\phi_{\rm{r}}$ and ${\rm{d}}\Omega_{\rm{t}}=\sin{\theta_{\rm{t}}} {\rm{d}}\theta_{\rm{t}}{\rm{d}}\phi_{\rm{t}}$ are the differential solid angles for the receive and transmit directions, respectively. Here, ${\mathcal{S}}_{+}$ denotes integration over the upper unit hemisphere. This compact form, visualized in Fig.~\ref{Figure: Multipath Channel Models: Correlation-Based: PSD_Correlation}, shows that spatial correlation is induced by the angular power distribution over departure and arrival directions.

Different propagation environments correspond to different choices of $p(\theta_{\rm{r}},\phi_{\rm{r}},\theta_{\rm{t}},\phi_{\rm{t}})$.

\paragraph{Isotropic Propagation}
The simplest reference case is completely unstructured scattering, where waves depart and arrive with uniform intensity over all directions. This yields an \emph{isotropic} channel \cite{paulraj2003introduction,molisch2012wireless,marzetta2018spatially} with constant angular power distribution:
\begin{align}
p(\theta_{\rm{r}},\phi_{\rm{r}},\theta_{\rm{t}},\phi_{\rm{t}})=\left(\frac{2\pi}{k_0}\right)^2,\label{eq:isotropic_density}
\end{align}
which effectively normalizes the average NLoS channel power such that ${\mathbb{E}}\{\lvert{{h}}_{\rm{NLoS}}({\mathbf{r}},{\mathbf{s}})\rvert^2\}=1$.

\paragraph{Non-Isotropic Propagation}
In practice, blockages and finite scattering clusters often restrict multipath components to dominant angular regions, leading to \emph{non-isotropic} scattering. A common tractable approximation is to decouple the angular distributions at the Tx and Rx, which yields the Kronecker model \cite{molisch2012wireless}:
\begin{align}\label{vMF_Basic_RX}
p(\theta_{\rm{r}},\phi_{\rm{r}},\theta_{\rm{t}},\phi_{\rm{t}})=p_{\rm{r}}(\theta_{\rm{r}},\phi_{\rm{r}})p_{\rm{t}}(\theta_{\rm{t}},\phi_{\rm{t}}).
\end{align}

\begin{figure}[!t]
\centering
\setlength{\abovecaptionskip}{0pt}
    \subfigure[Single cluster (Tx) and isotropic propagation (Rx).]
    {
        \includegraphics[width=0.4\textwidth]{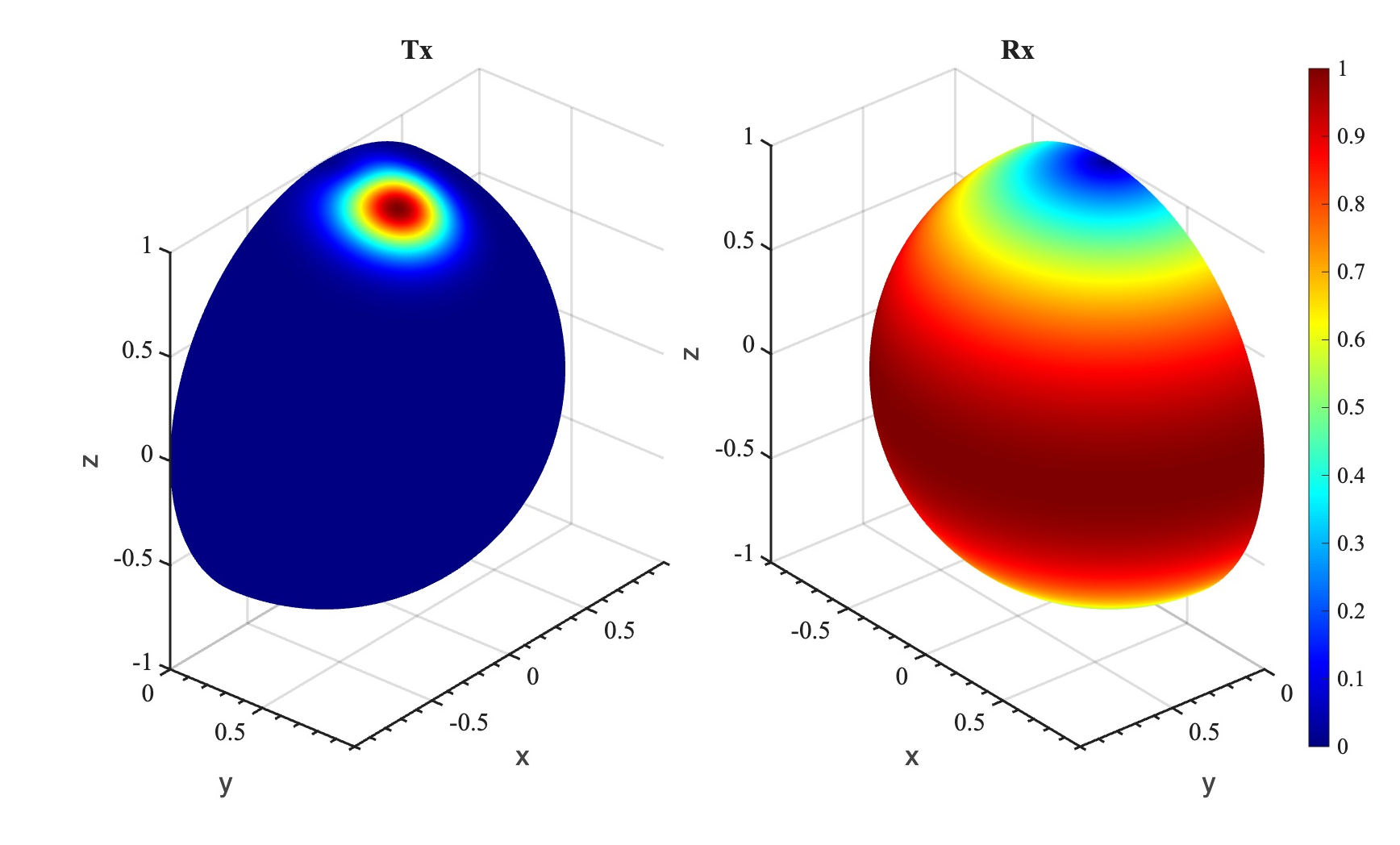}
	   \label{fig_vMF_Single}	
    }
    \subfigure[Multiple clusters (Tx and Rx).]
    {
        \includegraphics[width=0.4\textwidth]{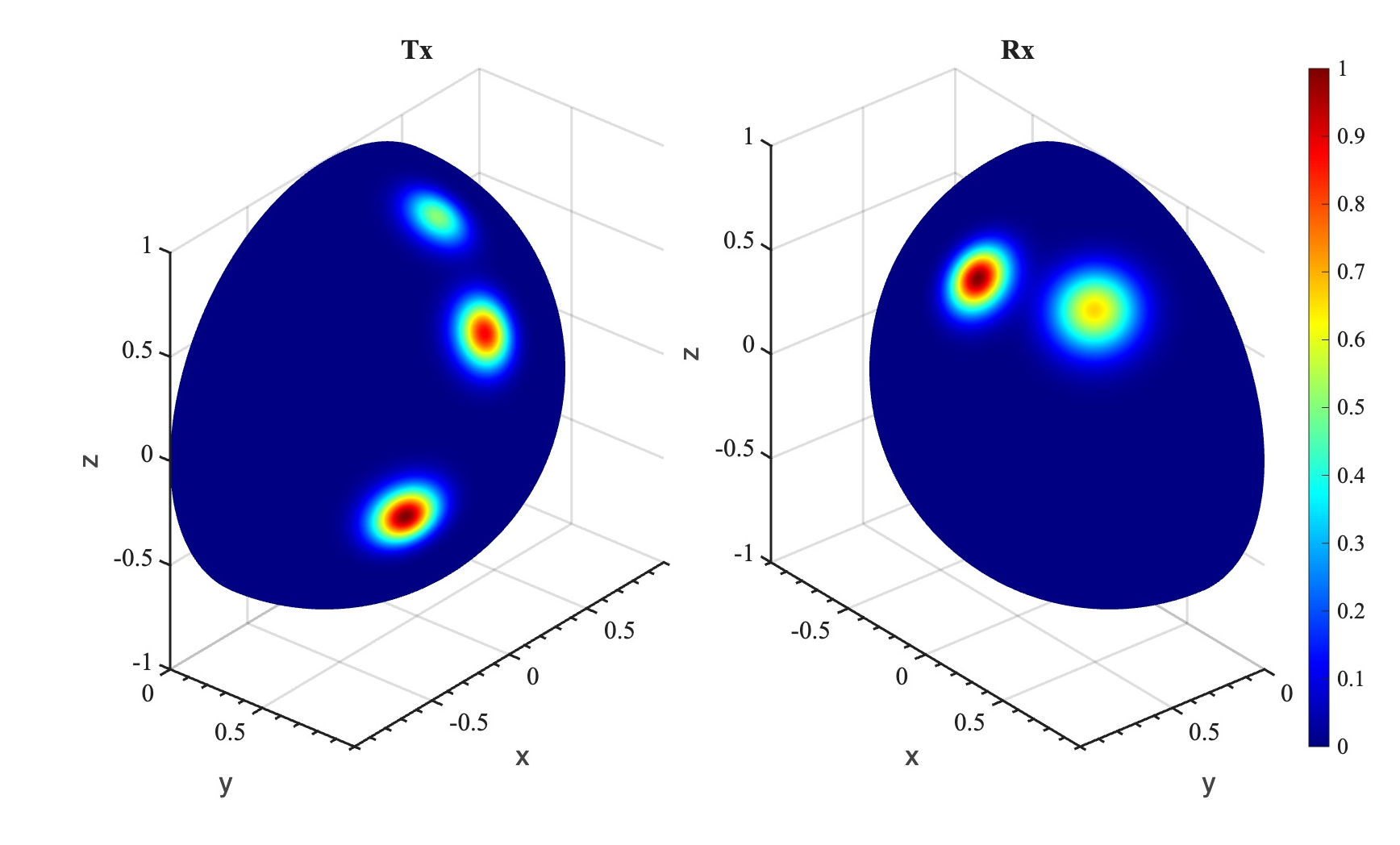}
	   \label{fig_vMF_Multiple}	
    }
\caption{Illustration of the angular power distribution function $p(\theta_{\rm{r}},\phi_{\rm{r}},\theta_{\rm{t}},\phi_{\rm{t}})$ generated by three-dimensional von Mises-Fisher (vMF) distributions. In (a), the Tx-side power is concentrated around a single dominant angular cluster, while the Rx side is isotropic. In (b), both the Tx and Rx sides contain multiple clusters with different modal directions and angular spreads. The modal angles determine the centers of the visible lobes, whereas the vMF concentration parameters control how tightly the power is localized around each cluster direction.}
\label{Figure_vMF}
\end{figure}

A useful tradeoff between tractability and accuracy is obtained by modeling the angular density as a mixture of three-dimensional von Mises-Fisher (vMF) distributions. For example, the Rx angular density can be constructed as a weighted sum over the scattering clusters:
\begin{align}
p_{\rm{r}}(\theta_{\rm{r}},\phi_{\rm{r}})=\sum_{i=1}^{N_{\rm{c}}}w_ip_{{\rm{r}},i}(\theta_{\rm{r}},\phi_{\rm{r}}),
\end{align}
where $N_{\rm{c}}$ is the number of clusters and the positive weights $\{w_i\}$ represent fractional cluster powers, satisfying $\sum_{i=1}^{N_{\rm{c}}}w_i=1$. 

Furthermore, the angular density characterizing each individual cluster $i$ is governed by the vMF distribution:
\begin{align}\label{vMF_Basic}
p_{{\rm{r}},i}(\theta,\phi)=c(\alpha_i){\rm{e}}^{\alpha_i(\sin{\theta}\sin{\mu_{\theta,i}}\cos{(\phi-\mu_{\phi,i})}+\cos{\theta}\cos{\mu_{\theta,i}})}.
\end{align}
The parameters have direct geometric interpretations:
\begin{itemize}
    \item $({\mu_{\theta,i}, \mu_{\phi,i}})$ are the \emph{modal angles}, specifying the central direction around which the cluster power is concentrated.
    \item $\alpha_i \geq 0$ is the \emph{concentration parameter}, which controls angular spread. When $\alpha_i = 0$, the distribution is isotropic; as $\alpha_i \to \infty$, it approaches an impulsive distribution around the modal direction.
    \item $c(\alpha_i)=\rev{\frac{\pi}{k_0}}\frac{\alpha_i}{\sinh{\alpha_i}}$ is a deterministic normalization constant. \rev{For the isotropic case $\alpha_i=0$, the density $p_{{\rm{r}},i}=2\pi/k_0$, consistent with \eqref{eq:isotropic_density}, is used instead, since $c(\alpha_i)$ normalizes the vMF distribution over the full sphere.}
\end{itemize}
The same construction can be applied on the Tx side, allowing different non-isotropic angular spreads at the two ends of the link. Further details are provided in \cite[Section VI-B]{pizzo2022spatial}.

Fig.~\ref{Figure_vMF} illustrates example vMF angular distributions through $p(\theta,\phi)\sin\theta$. In Fig.~\ref{fig_vMF_Single}, the Tx has one concentrated cluster with $\alpha_1=39.5$ pointing toward $({\mu_{\theta,1}, \mu_{\phi,1}}) = (20^{\circ},90^{\circ})$, while the Rx side is isotropic with $\alpha_1=0$. Fig.~\ref{fig_vMF_Multiple} shows a richer case with multiple clusters at both ends of the link. The Tx observes three clusters pointing toward $({\mu_{\theta,1}, \mu_{\phi,1}}) = (60^{\circ}, 90^{\circ})$, $({\mu_{\theta,2}, \mu_{\phi,2}}) = (30^{\circ}, 75^{\circ})$, and $({\mu_{\theta,3}, \mu_{\phi,3}}) = (100^{\circ}, 120^{\circ})$, with concentration parameters $\alpha_1=100$, $\alpha_2=99.5$, and $\alpha_3=100$, respectively. The Rx observes two concentrated incoming clusters from $({\mu_{\theta,1}, \mu_{\phi,1}}) = (45^{\circ}, 45^{\circ})$ and $({\mu_{\theta,2}, \mu_{\phi,2}}) = (45^{\circ}, 90^{\circ})$, with concentration parameters $\alpha_1=66.2$ and $\alpha_2=100$.

\rev{The correlation-based models above rely on spatial stationarity, Gaussian fading, uncorrelated scattering, and, in the Kronecker case, separable Tx and Rx angular distributions. Spatial stationarity requires all parts of the aperture to observe the same scatterers from approximately the same directions, which may not hold when the scatterers are close to the aperture compared with its physical size or are partially blocked. In such cases, spatially non-stationary models based on visibility regions~\cite{de2020non} or near-field scattering~\cite{wan2024near} can be adopted instead.}

\subsection{Time-Domain and Doubly-Dispersive Channel Models} \label{sec:time_domain}
The preceding channel models are frequency-domain models and therefore correspond to time-harmonic fields at a fixed frequency. Practical wireless signals occupy non-negligible bandwidth and may experience both frequency selectivity and time selectivity. Since waveform design and receiver processing are often expressed in the time domain, we next revisit the EM formulation from a space-time perspective.

\subsubsection{Time-Domain Electromagnetic Wave Equation}
From the time-domain macroscopic Maxwell’s equations, Ampère’s law and Faraday’s law in a linear, isotropic, and non-dispersive medium are given by
\begin{subequations}\label{eq:maxwell_time_domain}
\begin{align}
\nabla \times \pmb{\mathcal{E}}(\mathbf{r}, t) &= -\mu_0 \frac{\partial \pmb{\mathcal{H}}(\mathbf{r}, t)}{\partial t},\\
\nabla \times \pmb{\mathcal{H}}(\mathbf{r}, t) &= \pmb{\mathcal{J}}(\mathbf{r}, t) + \epsilon_0 \frac{\partial \pmb{\mathcal{E}}(\mathbf{r}, t)}{\partial t}.
\end{align}
\end{subequations}
By taking the curl of Faraday's law and substituting Ampère's law, we obtain the time-domain \rev{inhomogeneous vector wave equation}:
\begin{equation}\label{eq:time_domain_wave_equation}
\nabla \times \nabla \times \pmb{\mathcal{E}}(\mathbf{r}, t) + \mu_0 \epsilon_0 \frac{\partial^2 \pmb{\mathcal{E}}(\mathbf{r}, t)}{\partial t^2} = -\mu_0 \frac{\partial \pmb{\mathcal{J}}(\mathbf{r}, t)}{\partial t}.
\end{equation}
Unlike the frequency-domain equation in \eqref{wave_equation}, the time-domain formulation explicitly contains the \emph{time derivative} of the source current. This derivative is important when relating physical passband currents to equivalent baseband signals.

\subsubsection{Time-Domain Green's Function and Channel Response}
The solution of \eqref{eq:time_domain_wave_equation} can be expressed through the \emph{time-domain dyadic Green's function} $\pmb{\mathcal{G}}(\mathbf{r}, t; \mathbf{s}, t')$, which gives the electric field at $\mathbf{r}$ and time $t$ generated by a unit impulsive current source at $\mathbf{s}$ and time $t'$. The generated electric field is therefore a space-time convolution:
\begin{equation}\label{eq:time_domain_electric_field_solution}
\pmb{\mathcal{E}}(\mathbf{r}, t) = -\mu_0 \int_{-\infty}^{t} \int_{\mathcal{S}_{\mathrm{t}}} \pmb{\mathcal{G}}(\mathbf{r}, t; \mathbf{s}, t') \frac{\partial \pmb{\mathcal{J}}(\mathbf{s}, t')}{\partial t'} \, {\rm{d}}\mathbf{s} \, {\rm{d}}t'.
\end{equation}

In a free-space LoS environment, the time-domain dyadic Green's function acts as a delay operator determined by the finite speed of light $c = 1/\sqrt{\mu_0 \epsilon_0}$. A wave radiated from $\mathbf{s}$ reaches $\mathbf{r}$ after the retarded delay $\tau = \lVert\mathbf{r} - \mathbf{s}\rVert / c$. In a general multipath environment, these propagation effects can be summarized by a time-varying channel impulse response $h(\mathbf{r}, \mathbf{s}, t, \tau)$, where $\tau = t-t'$ is the relative delay. The received baseband signal is then modeled as
\begin{equation}\label{eq:time_domain_baseband_received}
y(\mathbf{r}, t) = \int_{0}^{\infty} \int_{\mathcal{S}_{\mathrm{t}}} h(\mathbf{r}, \mathbf{s}, t, \tau) x(\mathbf{s}, t - \tau) \, {\rm{d}}\tau \, {\rm{d}}\mathbf{s} + n(\mathbf{r}, t),
\end{equation}
where $x(\mathbf{s}, t)$ represents the time-varying baseband equivalent of the source excitation signal across the CAPA.

\begin{remark}[\emph{Impact of the time derivative}]
    \normalfont
    To understand how $x(\mathbf{s}, t)$ relates to the time derivative of the physical passband source current $\pmb{\mathcal{J}}(\mathbf{s}, t)$, recall that the current in a wireless transmitter is typically a modulated passband signal, which can be expressed as $\pmb{\mathcal{J}}(\mathbf{s}, t) = \Re\left\{\pmb{\mathcal{J}}_{\mathrm{BB}}(\mathbf{s}, t) {\rm{e}}^{{\rm{j}}\omega_c t}\right\}$, where $\omega_c$ is the carrier frequency and $\pmb{\mathcal{J}}_{\mathrm{BB}}(\mathbf{s}, t)$ is the spatial baseband envelope. The time derivative governing radiation in \eqref{eq:time_domain_electric_field_solution} becomes
    \begin{equation}
        \frac{\partial \pmb{\mathcal{J}}(\mathbf{s}, t)}{\partial t} = \Re\left\{ \left( \frac{\partial \pmb{\mathcal{J}}_{\mathrm{BB}}(\mathbf{s}, t)}{\partial t} + {\rm{j}}\omega_c \pmb{\mathcal{J}}_{\mathrm{BB}}(\mathbf{s}, t) \right) {\rm{e}}^{{\rm{j}}\omega_c t} \right\},
    \end{equation}
    For typical narrowband baseband signals, the envelope varies much more slowly than the carrier, i.e., $\lvert\frac{\partial \pmb{\mathcal{J}}_{\mathrm{BB}}(\mathbf{s}, t)}{\partial t}\rvert \ll \omega_c\lvert\pmb{\mathcal{J}}_{\mathrm{BB}}(\mathbf{s}, t)\rvert$. The derivative is then dominated by the carrier term, so $\frac{\partial \pmb{\mathcal{J}}(\mathbf{s}, t)}{\partial t} \approx \Re\left\{ {\rm{j}}\omega_c \pmb{\mathcal{J}}_{\mathrm{BB}}(\mathbf{s}, t) {\rm{e}}^{{\rm{j}}\omega_c t} \right\}$. After downconversion and RF filtering, the equivalent baseband signal observes an effective transmit signal $x(\mathbf{s}, t) \propto {\rm{j}}\omega_c \pmb{\mathcal{J}}_{\mathrm{BB}}(\mathbf{s}, t)$. Thus, $x(\mathbf{s}, t)$ absorbs both the baseband current envelope and the derivative effect introduced by EM radiation.
\end{remark}

\subsubsection{Doubly-Dispersive Channel Modeling}
In practical environments, the wireless channel is rarely flat in both time and frequency. It often exhibits \emph{doubly dispersive} behavior, meaning that the transmitted signal experiences dispersion in two domains simultaneously:
\begin{itemize}
    \item \emph{Time dispersion (delay spread):} Multipath components arrive at the Rx through paths of different lengths, causing frequency-selective fading.
    \item \emph{Frequency dispersion (Doppler spread):} Relative motion among the Tx, Rx, and scatterers makes the path delays time-dependent, producing Doppler shifts and time-selective fading.
\end{itemize}
To capture these effects jointly, the spatial CAPA channel must be extended to include both delay and Doppler variables. We next outline physics-based and correlation-based doubly dispersive models.

\paragraph{Physics-based Model}

For the physics-based approach, each scattering path is assigned a propagation delay and a Doppler shift \cite{tse2005fundamentals}. Recalling the time-harmonic multipath model in \eqref{eq:physics_based_multipath_model}, the corresponding time-varying impulse response can be written as
\begin{align}\label{eq:physics_based_doubly_dispersive_model}
h(\mathbf{r}, {\mathbf{s}}, t, \tau) &= h_{\rm{LoS}}({\mathbf{r}}, {\mathbf{s}}, t)\delta(\tau - \tau_{\rm{LoS}}) \nonumber \\
&+ \sum_{l=1}^{N_{l}}\sum_{j\in\varOmega_{l}}\Gamma_{l,j}(t)
h_{\rm{LoS}}(\mathbf{r}, \mathbf{q}_{l,j}, t) \nonumber \\
& \hspace{1.5cm} \times h_{\rm{LoS}}(\mathbf{q}_{l,j}, \mathbf{s}, t)\delta(\tau - \tau_{l,j}).
\end{align}
Here, $\tau_{\rm{LoS}}$ and $\tau_{l,j}$ are the delays of the LoS path and the $(l,j)$-th NLoS path, respectively. Mobility or dynamic scatterers make the path lengths vary over time, producing a time-varying phase that is equivalent to a Doppler shift $\nu_{l,j}$. Over a short coherence interval with approximately constant velocities, the path gain can be modeled as $\Gamma_{l,j}(t) = \Gamma_{l,j} {\rm{e}}^{{\rm{j}}2\pi \nu_{l,j}t}$. A detailed treatment of this model is given in \cite{ranasinghe2025doubly}. \rev{The model in \eqref{eq:physics_based_doubly_dispersive_model} assigns a single delay to each path across the entire aperture. This is accurate when the delay differences across the aperture are much smaller than the symbol duration $1/B$, i.e., when the spatial length of a symbol, $c/B$, is much larger than the aperture size $L$. For large apertures and wide bandwidths, however, $\tau_{\rm{LoS}}$ and $\tau_{l,j}$ should be replaced by the point-dependent delays $\lVert\mathbf{r}-\mathbf{s}\rVert/c$ and $(\lVert\mathbf{r}-\mathbf{q}_{l,j}\rVert+\lVert\mathbf{q}_{l,j}-\mathbf{s}\rVert)/c$, respectively. This spatial-wideband effect causes beam squint when the aperture current is kept fixed over the band \cite{do2024hybrid}. Apart from this propagation effect, the hardware also responds differently at different frequencies. Specifically, the surface impedance, mutual coupling, and circuit response in Section~\ref{sec:circuit_to_field} all vary with frequency, e.g., due to material dispersion. Hence, the same port excitation produces different current distributions and beam patterns at different frequencies, which can be captured by evaluating the circuit-to-field model at each frequency.}

\paragraph{Correlation-based Model}
For the correlation-based approach, the spatial spectrum $S({\mathbf{k}}, {\bm\kappa})$ in \eqref{Angular_Domain_Power_Distribution_Isotropic_Scattering} is extended to include delay and Doppler. Under the wide-sense stationary uncorrelated scattering (WSSUS) assumption \cite{bello1963characterization}, define the delay-Doppler angular response
\begin{align}\label{eq:delay_doppler_angular_response}
\widetilde{H}_a({\mathbf{k}}, {\bm\kappa}, \tau, \nu) = S^{\frac{1}{2}}({\mathbf{k}}, {\bm\kappa}, \tau, \nu) {{W}}_a({\mathbf{k}}, {\bm\kappa}, \tau, \nu).
\end{align}
Here, $S({\mathbf{k}}, {\bm\kappa}, \tau, \nu)$ is the joint space-delay-Doppler power spectral density. It specifies the average power density of components associated with transmit direction ${\bm\kappa}$, receive direction ${\mathbf{k}}$, delay $\tau$, and Doppler shift $\nu$. The term ${{W}}_a({\mathbf{k}}, {\bm\kappa}, \tau, \nu)$ is a zero-mean, unit-variance CSCG random field that models uncorrelated scattering across these domains:
\begin{align}
\mathbb{E}\Big\{{{W}}_a({\mathbf{k}}, & {\bm\kappa}, \tau, \nu) {{W}}_a^*({\mathbf{k}}', {\bm\kappa}', \tau', \nu')\Big\} \nonumber \\
&= \delta({\mathbf{k}} - {\mathbf{k}}') \delta({\bm\kappa} - {\bm\kappa}') \delta(\tau - \tau') \delta(\nu - \nu').
\end{align}
Based on this, the time-varying, delay-dispersive angular response ${{H}}_a({\mathbf{k}},{\bm\kappa}, t, \tau)$ is obtained by taking the inverse Fourier transform of $\widetilde{H}_a({\mathbf{k}}, {\bm\kappa}, \tau, \nu)$ over the Doppler domain as
\begin{equation}\label{eq:time_varying_angular_response}
{{H}}_a({\mathbf{k}}, {\bm\kappa}, t, \tau) = \int_{-\infty}^{\infty} \widetilde{H}_a({\mathbf{k}}, {\bm\kappa}, \tau, \nu) {\rm{e}}^{{\rm{j}}2\pi\nu t} \, {\rm{d}}\nu.
\end{equation}
Physically, this integral superposes scattered components with different Doppler shifts, each contributing a phase rotation ${\rm{e}}^{{\rm{j}}2\pi\nu t}$.
Integrating this angular response over all radiating plane-wave directions gives the doubly dispersive space-time NLoS channel
\begin{align}\label{Doubly_Dispersive_Correlation_Model_New}
h_{\rm{NLoS}}({\mathbf{r}}, {\mathbf{s}}, t, \tau) &= \frac{1}{(2\pi)^2}\iiiint_{{\mathcal{D}}({\mathbf{k}})\times{\mathcal{D}}({\bm\kappa})}
{\rm{e}}^{-{\rm{j}}{{\mathbf{k}}}^{\mathsf{T}}{\mathbf{r}}'} \nonumber \\
&\times {{H}}_a({\mathbf{k}},{\bm\kappa}, t, \tau){\rm{e}}^{{\rm{j}}{\bm{\kappa}}^{\mathsf{T}}{\mathbf{s}}}{\rm{d}}k_{\mathsf{x}}{\rm{d}}k_{\mathsf{z}}{\rm{d}}\kappa_{\mathsf{x}}{\rm{d}}\kappa_{\mathsf{z}}.
\end{align}
\rev{Then, under the WSSUS assumption, the NLoS channel is uncorrelated across different delays, i.e.,
\begin{align}
    {\mathbb{E}}\left\{ h_{\rm{NLoS}}({\mathbf{r}}, {\mathbf{s}}, t, \tau)\, h_{\rm{NLoS}}^*({\mathbf{r}}, {\mathbf{s}}, t, \tau') \right\} = P_{\rm{NLoS}}(\tau)\,\delta(\tau-\tau'),
\end{align}
where the power delay profile of the NLoS channel is given by}
\begin{align}\label{eq:Average_Power_Doubly_Dispersive_NLoS}
    \rev{P_{\rm{NLoS}}(\tau)}
    = &\int_{-\infty}^{\infty} \iiiint_{{\mathcal{D}}({\mathbf{k}})\times{\mathcal{D}}({\bm\kappa})} \frac{S({\mathbf{k}}, {\bm\kappa}, \tau, \nu)}{(2\pi)^4} \nonumber\\
    &\qquad {\rm{d}}k_{\mathsf{x}}{\rm{d}}k_{\mathsf{z}}{\rm{d}}\kappa_{\mathsf{x}}{\rm{d}}\kappa_{\mathsf{z}} \, {\rm{d}}\nu.
\end{align}

If the scattering environment allows the joint spectrum to be approximated by separable spatial, delay, and Doppler profiles, i.e., $S({\mathbf{k}}, {\bm\kappa}, \tau, \nu) \approx S_{\mathrm{space}}({\mathbf{k}}, {\bm\kappa}) S_{\mathrm{delay}}(\tau) S_{\mathrm{Doppler}}(\nu)$, then the four-dimensional fading process ${{H}}_a({\mathbf{k}},{\bm\kappa}, t, \tau)$ can be generated and evaluated efficiently in separated domains. \rev{The DoF of wideband CAPA systems will be discussed in Section~\ref{sec:dof}.}

\begin{table*}[!t]
\centering
\setlength{\abovecaptionskip}{0pt}
\caption{Summary of the Optimization Methods for CAPAs in Section~\ref{sec:beamforming_Optimization}.}
\label{tab:beamforming_optimization_summary}
\footnotesize
\setlength{\tabcolsep}{2.5pt}
\renewcommand{\arraystretch}{1.2}
\begin{tabular}{|>{\centering\arraybackslash}m{2.1cm}|>{\centering\arraybackslash}m{2.5cm}|>{\raggedright\arraybackslash}m{2.8cm}|>{\raggedright\arraybackslash}m{3cm}|>{\raggedright\arraybackslash}m{2.8cm}|>{\raggedright\arraybackslash}m{2.8cm}|}
\hline
\textbf{Method} & \textbf{Mathematical Tool} & \textbf{Main Idea} & \textbf{Approximation} & \textbf{Pros} & \textbf{Cons} \\
\hline
\textbf{Wavenumber-domain optimization} &
Fourier analysis &
Convert the functional problem into a finite-dimensional MIMO-like model. &
Yes; discretizes and truncates the spectrum to the radiating region. &
Compatible with standard discrete beamforming design. &
Accuracy depends on discretization and truncation; high computational complexity. \\
\hline
\textbf{Calculus of variations} &
Functional analysis &
Derive functional derivatives and solve the resulting Fredholm integral equations. &
No model approximation; only numerical quadrature may be used in implementation. &
Retains the exact continuous model; reveals analytical beam structure. &
More mathematically involved; integral equations can be difficult to solve. \\
\hline
\end{tabular}
\end{table*}

\section{Design and Optimization of CAPA Systems} \label{sec:design_optimization}

The preceding sections established continuous-space channel models for CAPAs. In this section, we turn these models into practical design tools with a focus on beamforming optimization and channel estimation. The central challenge is that the optimization variables and the unknown channels are functions defined over continuous apertures, rather than finite-dimensional vectors. We therefore develop tractable finite-dimensional representations that preserve the essential continuous-aperture structure while enabling practical analysis and algorithm design.

\subsection{Beamforming Optimization} \label{sec:beamforming_Optimization}

As discussed above, CAPA systems are naturally described by continuous-space operators rather than finite-dimensional vectors and matrices. Consequently, conventional optimization methods developed for discrete wireless systems, such as those described in \cite{bjornson2013optimal}, cannot be applied directly without first introducing an appropriate finite-dimensional representation. This motivates beamforming frameworks specifically tailored to continuous apertures.

Recall the signal model for time-harmonic, uni-polarized CAPAs in \eqref{basic_CAPA_model}:
\begin{align} \label{sec_IV_signal_model}
    y(\mathbf{r}) = \int_{{\mathcal{S}}_{\mathrm{t}}} h(\mathbf{r},{\mathbf{s}}) x({\mathbf{s}}) \,\mathrm{d}{\mathbf{s}}+ n(\mathbf{r}),\qquad \forall {\mathbf{r}}\in{\mathcal{S}}_{\mathrm{r}}.
\end{align}
A typical CAPA design problem is to determine the source current distribution $x({\mathbf{s}})$ so as to optimize a given utility function $\mathcal{U}$. Depending on the application, the utility may represent the received signal-to-noise ratio (SNR), signal-to-interference-plus-noise ratio (SINR), achievable rate, or power efficiency, while the constraint set may account for transmit power, aperture loss, mutual coupling, or feasible hardware excitations. A generic formulation is
\begin{align}
    \underset{{x({\mathbf{s}})}}{\mathrm{maximize}} \,\, \mathcal{U}\big(x(\mathbf{s})\big) \quad \text{subject to} \quad x(\mathbf{s}) \in \mathcal{C},
\end{align}
where $\mathcal{C}$ denotes the feasible set. This is a functional optimization problem, since the optimization variable is the function $x({\mathbf{s}})$ defined over the continuous Tx aperture, rather than a finite-dimensional vector.

In the following, we introduce two representative solution strategies: \emph{wavenumber-domain optimization} and \emph{calculus-of-variations-based optimization}. Their main ideas, advantages, and tradeoffs are summarized in Table~\ref{tab:beamforming_optimization_summary}.

\subsubsection{Wavenumber-Domain Optimization}

A direct way to solve the functional optimization problem is to discretize the aperture in space. However, this approach can be inefficient, because accurately approximating the aperture current may require a very dense spatial grid, especially when the aperture is electrically large. Wavenumber-domain discretization offers a more structured alternative \cite{pizzo2022fourier, 9906802, 10158997}. Its key idea is to represent the aperture current and channel response in terms of spatial-frequency components, and then retain only the radiating components that are relevant to far-field or radiative near-field propagation. In this way, the original continuous problem is converted into a finite-dimensional model that resembles conventional MIMO and can therefore be handled using standard matrix-based optimization tools.

\paragraph{Fourier Transform to the Wavenumber Domain}

The core idea is to transform spatial signals and channel responses into their spectral representations. For a planar aperture ${\mathcal{S}}_{\mathrm{t}}$ with dimensions $L_{{\rm{t}},\mathsf{x}}$ and $L_{{\rm{t}},\mathsf{z}}$, the continuous source current $x(\mathbf{s})$ can be represented by its two-dimensional spatial Fourier transform, which defines the wavenumber-domain spectrum $X(\boldsymbol{\kappa})$ as follows:
\begin{align}
    X(\boldsymbol{\kappa})
    =
    \int_{\frac{-L_{{\rm{t}},\mathsf{z}}}{2}}^{\frac{L_{{\rm{t}},\mathsf{z}}}{2}}
    \int_{\frac{-L_{{\rm{t}},\mathsf{x}}}{2}}^{\frac{L_{{\rm{t}},\mathsf{x}}}{2}}
    x(\mathbf{s})
    {\rm{e}}^{-{\rm{j}}(\kappa_{\mathsf{x}} s_{\mathsf{x}} + \kappa_{\mathsf{z}} s_{\mathsf{z}})}
    \, {\rm{d}}s_{\mathsf{x}} {\rm{d}}s_{\mathsf{z}},
\end{align}
and the inverse transform is
\begin{align}
    \label{wavenumber_transform_x}
    x(\mathbf{s})
    =
    \frac{1}{(2\pi)^2}
    \int_{-\infty}^{\infty}
    \int_{-\infty}^{\infty}
    X(\boldsymbol{\kappa})
    {\rm{e}}^{{\rm{j}}(\kappa_{\mathsf{x}} s_{\mathsf{x}} + \kappa_{\mathsf{z}} s_{\mathsf{z}})}
    \, {\rm{d}}\kappa_{\mathsf{x}} {\rm{d}}\kappa_{\mathsf{z}}.
\end{align}
Here, $\boldsymbol{\kappa}=(\kappa_{\mathsf{x}},\kappa_{\mathsf{z}})$ denotes the transverse wavenumber vector. Physically, $X(\boldsymbol{\kappa})$ describes how strongly the aperture current excites each spatial-frequency component.

Similarly, for the received signal, its wavenumber-domain spectrum $Y(\mathbf{k})$ is defined by
\begin{align}
    \label{wavenumber_transform_y}
    Y(\mathbf{k})
    =
    \int_{\frac{-L_{{\rm{r}},\mathsf{z}}}{2}}^{\frac{L_{{\rm{r}},\mathsf{z}}}{2}}
    \int_{\frac{-L_{{\rm{r}},\mathsf{x}}}{2}}^{\frac{L_{{\rm{r}},\mathsf{x}}}{2}}
    y(\mathbf{r})
    {\rm{e}}^{-{\rm{j}}(k_{\mathsf{x}} r_{\mathsf{x}} + k_{\mathsf{z}} r_{\mathsf{z}})}
    \, {\rm{d}}r_{\mathsf{x}} {\rm{d}}r_{\mathsf{z}},
\end{align}
with inverse transform
\begin{align}
    y(\mathbf{r})
    =
    \frac{1}{(2\pi)^2}
    \int_{-\infty}^{\infty}
    \int_{-\infty}^{\infty}
    Y(\mathbf{k})
    {\rm{e}}^{{\rm{j}}(k_{\mathsf{x}} r_{\mathsf{x}} + k_{\mathsf{z}} r_{\mathsf{z}})}
    \, {\rm{d}}k_{\mathsf{x}} {\rm{d}}k_{\mathsf{z}},
\end{align}
where $\mathbf{k}=(k_{\mathsf{x}},k_{\mathsf{z}})$ is the receive-side transverse wavenumber vector.
Substituting \eqref{wavenumber_transform_x} into \eqref{sec_IV_signal_model} yields
\begin{align} \label{wavenumber_transform_y_2}
    y(\mathbf{r})
    =
    \frac{1}{(2\pi)^2}
    \int_{-\infty}^{\infty}
    \int_{-\infty}^{\infty}
    \int_{\frac{-L_{{\rm{t}},\mathsf{z}}}{2}}^{\frac{L_{{\rm{t}},\mathsf{z}}}{2}}
    \int_{\frac{-L_{{\rm{t}},\mathsf{x}}}{2}}^{\frac{L_{{\rm{t}},\mathsf{x}}}{2}}
    h(\mathbf{r}, \mathbf{s}) X(\boldsymbol{\kappa})
    \nonumber\\
    \times
    {\rm{e}}^{{\rm{j}}(\kappa_{\mathsf{x}} s_{\mathsf{x}} + \kappa_{\mathsf{z}} s_{\mathsf{z}})}
    \, {\rm{d}}\kappa_{\mathsf{x}} {\rm{d}}\kappa_{\mathsf{z}} {\rm{d}}s_{\mathsf{x}} {\rm{d}}s_{\mathsf{z}}
    + n(\mathbf{r}).
\end{align}
Combining \eqref{wavenumber_transform_y_2} with \eqref{wavenumber_transform_y} gives the wavenumber-domain input-output relation
\begin{align} \label{wavenumber_transform_y_3}
    Y(\mathbf{k})
    =
    \frac{1}{(2\pi)^2}
    \int_{-\infty}^{\infty}
    \int_{-\infty}^{\infty}
    H_a(\mathbf{k}, \boldsymbol{\kappa}) X(\boldsymbol{\kappa})
    \, {\rm{d}}\kappa_{\mathsf{x}} {\rm{d}}\kappa_{\mathsf{z}}
    + N(\mathbf{k}),
\end{align}
where $H_a(\mathbf{k}, \boldsymbol{\kappa})$ is the wavenumber-domain channel response and $N(\mathbf{k})$ is the wavenumber-domain noise spectrum. More specifically,
\begin{align} \label{wavenumber_transform_y_4}
    H_a(\mathbf{k}, \boldsymbol{\kappa})
    = &
    \int_{\frac{-L_{{\rm{t}},\mathsf{z}}}{2}}^{\frac{L_{{\rm{t}},\mathsf{z}}}{2}}
    \int_{\frac{-L_{{\rm{t}},\mathsf{x}}}{2}}^{\frac{L_{{\rm{t}},\mathsf{x}}}{2}}
    \int_{\frac{-L_{{\rm{r}},\mathsf{z}}}{2}}^{\frac{L_{{\rm{r}},\mathsf{z}}}{2}}
    \int_{\frac{-L_{{\rm{r}},\mathsf{x}}}{2}}^{\frac{L_{{\rm{r}},\mathsf{x}}}{2}}
    {\rm e}^{-\mathrm{j} (k_{\mathsf{x}} r_{\mathsf{x}} + k_{\mathsf{z}} r_{\mathsf{z}})}
    \nonumber\\
    & \times
    h(\mathbf{r}, \mathbf{s})
    {\rm{e}}^{{\rm{j}}(\kappa_{\mathsf{x}} s_{\mathsf{x}} + \kappa_{\mathsf{z}} s_{\mathsf{z}})}
    \, {\rm{d}}s_{\mathsf{x}} {\rm{d}}s_{\mathsf{z}} {\rm{d}}r_{\mathsf{x}} {\rm{d}}r_{\mathsf{z}}.
\end{align}

\paragraph{Discretization of the Wavenumber-Domain Model} \label{sec:discretization_wavenumber_domain}

Since both the Tx and Rx apertures, ${\mathcal{S}}_{\mathrm{t}}$ and ${\mathcal{S}}_{\mathrm{r}}$, are finite, their corresponding wavenumber representations can be sampled on regular grids whose spacings are determined by the inverse aperture dimensions following the Nyquist-Shannon sampling theorem \cite{pizzo2022fourier}. Specifically, define
\begin{align}
    \boldsymbol{\kappa}_{m,n}
    &\triangleq
    (\kappa_{\mathsf{x},m}, \kappa_{\mathsf{z},n})
    =
    (m \Delta \kappa_{\mathsf{x}}, n \Delta \kappa_{\mathsf{z}}), \\
    \mathbf{k}_{p,q}
    &\triangleq
    (k_{\mathsf{x},p}, k_{\mathsf{z},q})
    =
    (p \Delta k_{\mathsf{x}}, q \Delta k_{\mathsf{z}}),
\end{align}
where the sampling intervals are $\Delta \kappa_{\mathsf{x}} = 2\pi/L_{{\rm{t}},\mathsf{x}}$, $\Delta \kappa_{\mathsf{z}} = 2\pi/L_{{\rm{t}},\mathsf{z}}$, $\Delta k_{\mathsf{x}} = 2\pi/L_{{\rm{r}},\mathsf{x}}$, and $\Delta k_{\mathsf{z}} = 2\pi/L_{{\rm{r}},\mathsf{z}}$.
These sampling intervals reflect the standard Fourier duality between aperture extent and spectral resolution, i.e., a larger physical aperture leads to a finer sampling grid in the wavenumber domain. Evaluating $Y(\mathbf{k})$ on this grid transforms the continuous spectral-domain model into the discrete summation
\begin{align} \label{WD_discrete_model}
    Y(\mathbf{k}_{p,q})
    =
    \frac{\Delta \kappa_{\mathsf{x}} \Delta \kappa_{\mathsf{z}}}{(2\pi)^2}
    & \sum_{m \in \mathbb{Z}} \sum_{n \in \mathbb{Z}}
    H_a(\mathbf{k}_{p,q}, \boldsymbol{\kappa}_{m,n})
    X(\boldsymbol{\kappa}_{m,n})
    \nonumber\\
    & + N(\mathbf{k}_{p,q}),
    \quad p\in \mathbb{Z},\ q \in \mathbb{Z}.
\end{align}

\begin{figure}[!t]
 \centering
\includegraphics[width=0.47\textwidth]{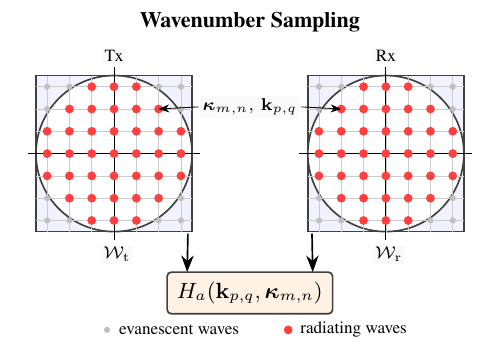}
\caption{Illustration of wavenumber sampling and truncation for the wavenumber-domain optimization method. The original continuous spectrum is discretized onto a regular grid, and only the samples within the radiating region are retained for optimization.}
\label{Figure: waveform_optimization}
\end{figure}

The summation in \eqref{WD_discrete_model} is still infinite. To obtain a finite model, we exploit the radiating support of the EM channel. As highlighted in \textbf{Remark~\ref{remark_evanescent_wave}}, only the wavenumber components satisfying $\kappa_{\mathsf{x}}^2 + \kappa_{\mathsf{z}}^2 \le k_0^2$ and $k_{\mathsf{x}}^2 + k_{\mathsf{z}}^2 \le k_0^2$ correspond to radiating plane waves. Components outside these regions are evanescent and decay rapidly with distance. Therefore, for propagation beyond the evanescent near zone, the channel effectively acts as a spatial low-pass filter with cutoff wavenumber $k_0 = 2\pi/\lambda$. \rev{Owing to the finite aperture size, a small amount of energy leaks into the spectral samples just outside the radiating region, and this leakage becomes negligible when the apertures are much larger than the wavelength.} Under these conditions, \eqref{WD_discrete_model} can be accurately approximated as
\begin{align} \label{WD_discrete_model_approx}
    Y(\mathbf{k}_{p,q})
    \approx
    \frac{\Delta \kappa_{\mathsf{x}} \Delta \kappa_{\mathsf{z}}}{(2\pi)^2}
    & \sum_{(m, n) \in \mathcal{W}_{\mathrm{t}}}
    H_a(\mathbf{k}_{p,q}, \boldsymbol{\kappa}_{m,n})
    X(\boldsymbol{\kappa}_{m,n})
    \nonumber\\
    & + N(\mathbf{k}_{p,q}),
    \quad (p, q) \in \mathcal{W}_{\mathrm{r}},
\end{align}
where
\begin{align}
    \mathcal{W}_{\mathrm{t}}
    &=
    \Big\{(m,n) \in \mathbb{Z}^2 :
    (m \Delta \kappa_{\mathsf{x}})^2 + (n \Delta \kappa_{\mathsf{z}})^2 \le k_0^2 \Big\}, \\
    \mathcal{W}_{\mathrm{r}}
    &=
    \Big\{(p,q) \in \mathbb{Z}^2 :
    (p \Delta k_{\mathsf{x}})^2 + (q \Delta k_{\mathsf{z}})^2 \le k_0^2 \Big\}.
\end{align}
That is, only the spectral samples inside the radiating disks are retained. Consequently, the design problem reduces to selecting a finite set of complex Fourier coefficients $X(\boldsymbol{\kappa}_{m,n})$ associated with the channel samples $H_a(\mathbf{k}_{p,q}, \boldsymbol{\kappa}_{m,n})$ inside the radiating region, as illustrated in Fig.~\ref{Figure: waveform_optimization}. The original functional optimization problem is thereby converted into a finite-dimensional matrix optimization problem.

We now stack the received samples $Y(\mathbf{k}_{p,q})$ for $(p,q) \in \mathcal{W}_{\mathrm{r}}$ into $\mathbf{y}_a \in \mathbb{C}^{N_{\mathrm{r}} \times 1}$, where $N_{\mathrm{r}} = |\mathcal{W}_{\mathrm{r}}|$, and stack the transmit coefficients $X(\boldsymbol{\kappa}_{m,n})$ for $(m,n) \in \mathcal{W}_{\mathrm{t}}$ into $\mathbf{x}_a \in \mathbb{C}^{N_{\mathrm{t}} \times 1}$, where $N_{\mathrm{t}} = |\mathcal{W}_{\mathrm{t}}|$. Then \eqref{WD_discrete_model_approx} can be written in the discrete MIMO-like form 
\begin{equation}\label{WD_MIMO_Model}
    \mathbf{y}_a = \mathbf{H}_a \mathbf{x}_a + \mathbf{n}_a,
\end{equation}
where $\mathbf{n}_a \in \mathbb{C}^{N_{\mathrm{r}} \times 1}$ collects the samples $N(\mathbf{k}_{p,q})$, and $\mathbf{H}_a \in \mathbb{C}^{N_{\mathrm{r}} \times N_{\mathrm{t}}}$ is the equivalent wavenumber-domain channel matrix with entries given by $\frac{\Delta \kappa_{\mathsf{x}} \Delta \kappa_{\mathsf{z}}}{(2\pi)^2}
H_a(\mathbf{k}_{p,q}, \boldsymbol{\kappa}_{m,n})$. This finite-dimensional representation enables the use of mature matrix-based tools for power allocation, precoding, and beamforming design. Once the optimal wavenumber-domain signal $\mathbf{x}_a$ is obtained, the continuous current profile can be reconstructed from \eqref{wavenumber_transform_x} as
\begin{equation}
    x(\mathbf{s})
    \approx
    \frac{\Delta \kappa_{\mathsf{x}} \Delta \kappa_{\mathsf{z}}}{(2\pi)^2}
    \sum_{(m,n) \in \mathcal{W}_{\mathrm{t}}}
    X(\boldsymbol{\kappa}_{m,n})
    \mathrm{e}^{\mathrm{j}\left(m \Delta \kappa_{\mathsf{x}} s_{\mathsf{x}} + n \Delta \kappa_{\mathsf{z}} s_{\mathsf{z}}\right)}.
\end{equation}

\paragraph{Discussion}

The main advantage of the wavenumber-domain method is its broad applicability to standard wireless optimization problems. Once the continuous CAPA design is transformed into the finite-dimensional model in \eqref{WD_MIMO_Model}, many mature optimization frameworks originally developed for conventional SPDA systems can be directly reused. As a result, common formulations such as SNR/SINR maximization, power minimization, and rate maximization can all be handled in a familiar matrix form.

Its main drawback is the computational cost of constructing $\mathbf{H}_a$. Although only the radiating modes in $\mathcal{W}_{\mathrm{t}}$ and $\mathcal{W}_{\mathrm{r}}$ are retained, the number of samples can still be very large for electrically large apertures\rev{, e.g., at} high carrier frequencies. For example, if both the Tx and Rx use planar $1\,\mathrm{m}\times 1\,\mathrm{m}$ apertures at $30\,\mathrm{GHz}$, then $\lambda=0.01\,\mathrm{m}$ and the number of radiating wavenumber samples on each side is approximately $N_{\mathrm{t}} = N_{\mathrm{r}} \approx \pi (L/\lambda)^2 \approx 3.14\times 10^4$. Hence, $\mathbf{H}_a$ contains entries on the scale of $N_{\mathrm{r}}N_{\mathrm{t}} \approx 9.9\times 10^8$. Moreover, each entry is obtained from the four-dimensional integral in \eqref{wavenumber_transform_y_4}, which generally \rev{needs to} be evaluated numerically. Therefore, although the optimization step becomes standard after discretization, the channel-construction step can dominate the overall complexity. \rev{It is worth noting that $N_{\mathrm{t}}$ is roughly $\pi/4$ of the $(2L/\lambda)^2$ samples required by conventional half-wavelength spatial sampling, indicating that the wavenumber-domain representation alone does not significantly decrease the variable count. However, under LoS or clustered-scattering conditions, the channel power is predominantly concentrated in a small subset of $\mathbf{H}_a$. Consequently, storage and computational overhead can be drastically reduced by retaining only these dominant elements.}

\subsubsection{Calculus of Variations}
An alternative is the calculus of variations. While the wavenumber-domain method first maps the problem into a finite-dimensional matrix space, the calculus of variations works directly with the continuous problem by finding the function $x(\mathbf{s})$ that makes the objective functional stationary. It is therefore the infinite-dimensional analogue of standard vector calculus and does not introduce a model approximation through discretization.

\paragraph{Preliminaries of Calculus of Variations}

Before formulating the optimization problem, we briefly review the basic tools from the calculus of variations, namely the \emph{functional derivative} and the \emph{fundamental lemma of calculus of variations}.

In ordinary calculus, extrema are characterized by derivatives with respect to scalar variables. In the calculus of variations, the optimization variable is a function. Specifically, we seek a function $x(\mathbf{s})$ that minimizes or maximizes a functional $\mathcal{U}\big(x(\mathbf{s})\big)$. By perturbing $x(\mathbf{s})$ with an arbitrary test function $\eta(\mathbf{s})$ scaled by a small real parameter $\epsilon$, the first-order change is defined as the \emph{first variation}
\begin{align}\label{eq:first_variation_def}
    \delta \mathcal{U}(x, \eta) = \left. \frac{{\rm d}}{{\rm d}\epsilon} \mathcal{U}\big(x(\mathbf{s}) + \epsilon \eta(\mathbf{s})\big) \right|_{\epsilon=0}.
\end{align}
If $x(\mathbf{s})$ is indeed an extremal function, this first variation must vanish for any permissible test function $\eta(\mathbf{s})$, yielding $\delta \mathcal{U}(x, \eta)=0$.

In CAPA systems, $x(\mathbf{s})$ is generally complex-valued, while $\mathcal{U}\big(x(\mathbf{s})\big)$ is real-valued. Therefore, consider a general form of \rev{$\mathcal{U}\big(x(\mathbf{s})\big) = f\big(I_1, \dots, I_K, I_1^*, \dots, I_K^*\big)$, where $I_k(x,x^*)=\int_{\mathcal{S}_{\mathrm{t}}}F_k\big(x(\mathbf{s}),x^*(\mathbf{s})\big)\,{\rm d}\mathbf{s}$ may be complex-valued and the real-valued differentiable function $f$ is regarded as a function of both $\{I_k\}$ and $\{I_k^*\}$}. Applying \eqref{eq:first_variation_def} and the chain rule gives
\begin{align}\label{eq:first_variation_expand}
    \delta \mathcal{U}(x, \eta)
    &= \rev{\sum_{k=1}^K
    \int_{\mathcal{S}_{\mathrm{t}}}
    \bigg[
    \frac{\partial f}{\partial I_k}
    \left(
    \frac{\partial F_k}{\partial x}\eta(\mathbf{s})
    +\frac{\partial F_k}{\partial x^*}\eta^*(\mathbf{s})
    \right)} \nonumber\\
    &\quad \rev{+\frac{\partial f}{\partial I_k^*}
    \left(
    \frac{\partial F_k^*}{\partial x}\eta(\mathbf{s})
    +\frac{\partial F_k^*}{\partial x^*}\eta^*(\mathbf{s})
    \right)
    \bigg]{\rm d}\mathbf{s}} \nonumber\\
    &=2\Re\left\{
    \int_{\mathcal{S}_{\mathrm{t}}}
    \frac{\delta \mathcal{U}}{\delta x^*(\mathbf{s})}
    \eta^*(\mathbf{s})\,{\rm d}\mathbf{s}
    \right\},
\end{align}
where \rev{
\begin{equation}
    \frac{\delta \mathcal{U}}{\delta x^*(\mathbf{s})}
    =\sum_{k=1}^K \left(\frac{\partial f}{\partial I_k}\frac{\partial F_k}{\partial x^*}
    +\frac{\partial f}{\partial I_k^*}\frac{\partial F_k^*}{\partial x^*}\right)
\end{equation}
is} the \emph{functional derivative} of $\mathcal{U}$ with respect to $x^*(\mathbf{s})$\rev{, and the last step uses the fact that $\mathcal{U}$ is real-valued. When all $I_k$ are real-valued, $f$ depends on $\{I_k\}$ only, and the functional derivative reduces to $\sum_{k}\frac{\partial f}{\partial I_k}\frac{\partial F_k}{\partial x^*}$. As an example, for $\mathcal{U}=\big|\int_{\mathcal{S}_{\mathrm{t}}}h(\mathbf{s})x(\mathbf{s})\,{\rm d}\mathbf{s}\big|^2=I I^*$ with $I=\int_{\mathcal{S}_{\mathrm{t}}}h(\mathbf{s})x(\mathbf{s})\,{\rm d}\mathbf{s}$, we obtain $\delta\mathcal{U}/\delta x^*(\mathbf{s})=h^*(\mathbf{s})\int_{\mathcal{S}_{\mathrm{t}}}h(\mathbf{s}')x(\mathbf{s}')\,{\rm d}\mathbf{s}'$}.

The key step for finding $x(\mathbf{s})$ such that $\delta \mathcal{U}(x, \eta)=0$ is the \emph{fundamental lemma of the calculus of variations}. In particular, if a continuous function $f(\mathbf{s})$ satisfies
\begin{align}\label{eq:fundamental_lemma}
    \Re\left\{\int_{\mathcal{S}} f(\mathbf{s})\eta^*(\mathbf{s})\,{\rm d}\mathbf{s}\right\}=0
\end{align}
for every smooth, compactly supported $\eta(\mathbf{s})$, then it must be true that $f(\mathbf{s})=0$ for all $\mathbf{s}\in\mathcal{S}$. \rev{Indeed, if $f(\mathbf{s}_0)\neq0$ at some point $\mathbf{s}_0\in\mathcal{S}$, choosing $\eta(\mathbf{s})={\rm{e}}^{{\rm{j}}\varphi_0}w(\mathbf{s})$ with $\varphi_0=\arg f(\mathbf{s}_0)$ and a smooth nonnegative function $w(\mathbf{s})$ concentrated around $\mathbf{s}_0$ makes the integral in \eqref{eq:fundamental_lemma} strictly positive by continuity, yielding a contradiction.} Therefore, the optimality condition $\delta\mathcal{U}(x,\eta)=0$ implies the following equivalent condition
\begin{align}
    \frac{\delta \mathcal{U}\big(x(\mathbf{s})\big)}{\delta x^*(\mathbf{s})}
    =\rev{\sum_{k=1}^K \left(\frac{\partial f}{\partial I_k}\frac{\partial F_k}{\partial x^*}+\frac{\partial f}{\partial I_k^*}\frac{\partial F_k^*}{\partial x^*}\right)}=0,
    \quad \forall \mathbf{s}\in\mathcal{S}_{\mathrm{t}}.
\end{align}

In practice, $x(\mathbf{s})$ is also subject to physical constraints, such as the total power constraint $\int_{\mathcal{S}_{\mathrm{t}}}|x(\mathbf{s})|^2{\rm d}\mathbf{s}\le P_{\rm t}$\rev{, i.e., the simplified power metric in \eqref{eq:simplified_power}}. This can be incorporated through the Lagrangian as follows:
\begin{align}
    \mathcal{L}\big(x(\mathbf{s}),\mu\big)
    =\mathcal{U}\big(x(\mathbf{s})\big)
    -\mu\left(
    \int_{\mathcal{S}_{\mathrm{t}}}|x(\mathbf{s})|^2{\rm d}\mathbf{s}-P_{\rm t}
    \right),
\end{align}
where $\mu\ge 0$ is the Lagrange multiplier. The optimal current is then obtained from
\begin{align}
    \frac{\delta \mathcal{L}\big(x(\mathbf{s}),\mu\big)}{\delta x^*(\mathbf{s})}=0,
\end{align}
together with the complementary slackness conditions. Taking the functional derivative typically yields an integral equation, often a Fredholm integral equation, which will be further discussed in the following sections. 

\paragraph{Generalizing the Calculus of Variations to Parametrizable Surfaces}

The above discussion focuses on optimizing a complex-valued current over a fixed surface. Recent studies have shown that the surface geometry itself can also be optimized to improve communication performance, giving rise to so-called flexible surface techniques \cite{10922153, ranasinghe2025flexible}. We now extend the formulation to this more general case, where the surface is also an optimization variable. Specifically, the surface is parameterized by a real-valued vector function $\mathbf{s}:\mathcal{D}\to\mathbb{R}^3$, defined over a fixed reference domain $\mathcal{D}\subset\mathbb{R}^n$. The objective functional then takes the general form
\begin{align}\label{eq:shape_functional}
\mathcal{V}(\mathbf{s})
=\int_{\mathcal{D}} F\big(x^\star(\mathbf{s}),\mathbf{s},\nabla\mathbf{s}\big) \, {\rm d}\mathcal{D},
\end{align}
where $\nabla\mathbf{s}\in\mathbb{R}^{3\times n}$ collects the partial derivatives of $\mathbf{s}$, and $x^\star(\mathbf{s})$ denotes a secondary function that is optimally adapted to $\mathbf{s}$.

Compared with the previous case, the key difference is that the integrand now depends not only on $\mathbf{s}$ but also on its gradient $\nabla\mathbf{s}$. Since the optimal solution $x^\star(\mathbf{s})$ satisfies the stationarity condition
\begin{align}\label{eq:current_stationarity_general}
    \frac{\delta \mathcal{V}(\mathbf{s})}{\delta x^*(\mathbf{s})}=0,
    \quad \forall\,\mathbf{s}\in\mathcal{D},
\end{align}
the first-order implicit dependence of $\mathcal{V}$ on $\mathbf{s}$ through $x^\star$ vanishes by the \emph{envelope theorem} \cite{ranasinghe2025flexible}, i.e., $x^\star(\mathbf{s})$ can be treated as a known function when differentiating $\mathcal{V}$ with respect to $\mathbf{s}$, and \eqref{eq:shape_functional} reduces to a functional with only explicit dependence on $\mathbf{s}$ and $\nabla\mathbf{s}$, i.e., $F\big(x^\star(\mathbf{s}), \mathbf{s}, \nabla\mathbf{s}\big) \equiv F\big(\mathbf{s}, \nabla\mathbf{s}\big)$.

Then, perturbing $\mathbf{s}$ by an arbitrary real-valued test function $\boldsymbol{\eta}:\mathcal{D}\to\mathbb{R}^3$ with $\boldsymbol{\eta}=0$ on $\partial\mathcal{D}$, the first variation is given by
\begin{align}\label{eq:shape_first_variation}
    \delta\mathcal{V}(\mathbf{s},\boldsymbol{\eta})
    =\left.\frac{{\rm d}}{{\rm d}\epsilon}\mathcal{V}\big(\mathbf{s}+\epsilon\boldsymbol{\eta}\big)\right|_{\epsilon=0}.
\end{align}
Applying the chain rule gives
\begin{align}\label{eq:shape_variation_expand}
    \delta\mathcal{V}(\mathbf{s},\boldsymbol{\eta})
    =\int_{\mathcal{D}}
    \left[
    \left(\frac{\partial F}{\partial \mathbf{s}}\right)^{\!\mathsf{T}}\boldsymbol{\eta}
    +
    \rev{\operatorname{tr}\!\left(\left(\frac{\partial F}{\partial (\nabla\mathbf{s})}\right)^{\!\mathsf{T}}\nabla\boldsymbol{\eta}\right)}
    \right]{\rm d}\mathcal{D}.
\end{align}
The second term involves $\nabla\boldsymbol{\eta}$ and therefore cannot be handled directly by the fundamental lemma. To address this, we can apply integration by parts to the second term and use $\boldsymbol{\eta}=0$ on $\partial\mathcal{D}$, which yields
\begin{align}\label{eq:ibp}
    \int_{\mathcal{D}}
    \rev{\operatorname{tr}\!\left(\left(\frac{\partial F}{\partial(\nabla\mathbf{s})}\right)^{\!\mathsf{T}}
    \nabla\boldsymbol{\eta}\right)}\,{\rm d}\mathcal{D}
    =
    -\int_{\mathcal{D}}
    \left(\nabla\cdot\frac{\partial F}{\partial(\nabla\mathbf{s})}\right)^{\!\mathsf{T}}
    \boldsymbol{\eta}\,{\rm d}\mathcal{D}.
\end{align}
Substituting this back gives
\begin{align}\label{eq:shape_variation_compact}
    \delta\mathcal{V}(\mathbf{s},\boldsymbol{\eta})
    =
    \int_{\mathcal{D}}
    \underbrace{
    \left(
    \frac{\partial F}{\partial \mathbf{s}}
    -\nabla\cdot\frac{\partial F}{\partial(\nabla\mathbf{s})}
    \right)^{\!\mathsf{T}}
    }_{\triangleq\left(\frac{\delta\mathcal{V}(\mathbf{s})}{\delta\mathbf{s}}\right)^{\!\mathsf{T}}}
    \boldsymbol{\eta}\,{\rm d}\mathcal{D}.
\end{align}
Thus, the functional derivative of $\mathcal{V}$ with respect to $\mathbf{s}$ is
\begin{equation}
    \frac{\delta\mathcal{V}(\mathbf{s})}{\delta\mathbf{s}}
=
\frac{\partial F}{\partial \mathbf{s}}
-\nabla\cdot\frac{\partial F}{\partial(\nabla\mathbf{s})}.
\end{equation}

Since both $\mathbf{s}$ and $\boldsymbol{\eta}$ are real-valued, the fundamental lemma applies componentwise, and the stationary condition becomes the Euler-Lagrange equation
\begin{align}\label{eq:EL_general}
    \frac{\delta\mathcal{V}(\mathbf{s})}{\delta\mathbf{s}}
    =
    \frac{\partial F}{\partial \mathbf{s}}
    -\nabla\cdot\frac{\partial F}{\partial(\nabla\mathbf{s})}
    =\mathbf{0}_{3\times 1},
    \quad \forall\,\mathbf{s}\in\mathcal{D}.
\end{align}
This is generally a system of second-order PDEs in $\mathbf{s}$. Closed-form solutions are usually unavailable, so iterative numerical methods are required, such as the method proposed in \cite{ranasinghe2025flexible}.

\paragraph{Fredholm Integral Equation}

As discussed above, taking the functional derivative typically yields a Fredholm integral equation. There are two kinds of such integral equations. The equation of the first kind takes the form:
\begin{equation} \label{Fredholm_equation_first}
    g(\mathbf{s}) = \int_{\mathcal{S}_{\mathrm{t}}} K(\mathbf{s}, \mathbf{s}') x(\mathbf{s}') \, \mathrm{d} \mathbf{s}', \quad \forall \mathbf{s} \in \mathcal{S}_{\mathrm{t}}.
\end{equation}
Given the kernel $K(\mathbf{s}, \mathbf{s}')$ and the function $g(\mathbf{s})$, the task is to recover the unknown function $x(\mathbf{s})$. This equation is the continuous analogue of the discrete linear system $\mathbf{g} = \mathbf{K}\mathbf{x}$, whose solution is $\mathbf{x} = \mathbf{K}^{-1}\mathbf{g}$ when $\mathbf{K}$ is invertible. Similarly, solving \eqref{Fredholm_equation_first} requires the inverse kernel $K^{-1}(\mathbf{s}, \mathbf{s}')$ satisfying~\cite{10938678}
\begin{equation}
    \int_{\mathcal{S}_{\mathrm{t}}} K^{-1}(\mathbf{s}, \mathbf{s}') K(\mathbf{s}', \mathbf{s}'') \, \mathrm{d} \mathbf{s}' = \delta(\mathbf{s} - \mathbf{s}'').
\end{equation}
For a general kernel $K(\mathbf{s}, \mathbf{s}')$, however, the inverse kernel may not have a closed-form expression and can be difficult to evaluate numerically. Approximation methods can therefore be useful. \rev{One} example is the mutual-coupling kernel approximation method in \cite{wang2025mutual}. \rev{Moreover, when $K(\mathbf{s}, \mathbf{s}')$ is smooth, as is the case for radiation kernels and for channel kernels between separated apertures, its eigenvalues decay to zero. Hence, \eqref{Fredholm_equation_first} is ill-posed, i.e., small perturbations of $g(\mathbf{s})$ may cause arbitrarily large changes of $x(\mathbf{s})$. In practice, it is therefore solved with regularization, e.g., by replacing $K(\mathbf{s}, \mathbf{s}')$ with $K(\mathbf{s}, \mathbf{s}')+\varrho\,\delta(\mathbf{s}-\mathbf{s}')$ for some $\varrho>0$, which is the continuous counterpart of replacing $\mathbf{K}^{-1}$ with $(\mathbf{K}+\varrho\mathbf{I})^{-1}$ in the discrete case. This turns \eqref{Fredholm_equation_first} into an equation of the second kind. In CAPA systems, such a regularization term naturally arises from ohmic loss or the simplified power metric.}

The equation of the second kind takes the form:
\begin{equation} \label{Fredholm_equation_second}
    x(\mathbf{s}) = g(\mathbf{s}) + \int_{\mathcal{S}_{\mathrm{t}}} K(\mathbf{s}, \mathbf{s}') x(\mathbf{s}') \, \mathrm{d} \mathbf{s}', \quad \forall \mathbf{s} \in \mathcal{S}_{\mathrm{t}}.
\end{equation}
In contrast to the equation of the first kind, the second-kind equation \rev{can be solved iteratively using the} Neumann series\rev{, which converges when the eigenvalues of the kernel are smaller than one in magnitude}. In many CAPA problems, the kernel is separable, i.e., $K(\mathbf{s}, \mathbf{s}') = \sum_{k=1}^K U_k(\mathbf{s}) V_k(\mathbf{s}')$\rev{, in which case the solution can be obtained in closed form. Specifically,} \eqref{Fredholm_equation_second} becomes
\begin{align} \label{Fredholm_equation_second_separable}
    x(\mathbf{s}) = & g(\mathbf{s}) + \sum_{k=1}^K U_k(\mathbf{s}) \int_{\mathcal{S}_{\mathrm{t}}} V_k(\mathbf{s}') x(\mathbf{s}') \, \mathrm{d} \mathbf{s}' \nonumber \\
    = & g(\mathbf{s}) + \sum_{k=1}^K a_k U_k(\mathbf{s}),
\end{align}
where $a_k = \int_{\mathcal{S}_{\mathrm{t}}} V_k(\mathbf{s}') x(\mathbf{s}') \, \mathrm{d} \mathbf{s}'$. To solve this separable equation, we can multiply both sides by $V_j(\mathbf{s})$ and integrate over $\mathcal{S}_{\mathrm{t}}$ for $j=1, \dots, K$. Then, we have 
\begin{align} \label{Fredholm_equation_second_separable_2}
    a_j = b_j + \sum_{k=1}^K a_k c_{j,k} & \, \Rightarrow \, \mathbf{a} = \mathbf{b} + \mathbf{C} \mathbf{a} \nonumber \\
    & \, \Rightarrow \, \mathbf{a} = (\mathbf{I} - \mathbf{C})^{-1} \mathbf{b},
\end{align}
where $b_j = \int_{\mathcal{S}_{\mathrm{t}}} V_j(\mathbf{s}) g(\mathbf{s}) \, \mathrm{d} \mathbf{s}$, $c_{j,k} = \int_{\mathcal{S}_{\mathrm{t}}} V_j(\mathbf{s}) U_k(\mathbf{s}) \, \mathrm{d} \mathbf{s}$, \rev{$\mathbf{a}=[a_1,\ldots,a_K]^{\mathsf{T}}$, $\mathbf{b}=[b_1,\ldots,b_K]^{\mathsf{T}}$, and $[\mathbf{C}]_{j,k}=c_{j,k}$. Provided that $\mathbf{I}-\mathbf{C}$ is invertible, substituting $\mathbf{a}$ into \eqref{Fredholm_equation_second_separable} yields the solution} $x(\mathbf{s})$.

\paragraph{Gauss-Legendre Quadrature}
The preceding derivations show that efficient numerical integration is essential in CAPA optimization, for example, when computing $\mathbf{C}$ and $\mathbf{b}$ in \eqref{Fredholm_equation_second_separable_2}. A standard high-accuracy method for bounded continuous integrals is Gauss-Legendre quadrature. Unlike uniform sampling, which evaluates the integrand at equally spaced points, Gauss-Legendre quadrature selects non-uniform nodes and weights to maximize algebraic accuracy.

Specifically, a 1D integral over a general interval $[a, b]$ can be numerically approximated as a weighted sum of function values:
\begin{align}
    \int_{a}^{b} f(s) \, \mathrm{d} s \approx \frac{b-a}{2} \sum_{m=1}^{M} w_m f\left( \frac{b-a}{2} s_m + \frac{b+a}{2} \right),
\end{align}
where $M$ is the number of sample points, $s_m$ are the roots of the $M$-th order Legendre polynomial $P_M(s)$, and $w_m$ are the associated Gauss-Legendre weights given by $w_m = \frac{2}{(1-s_m^2)[P_M'(s_m)]^2}$, with $P_M'(s_m)$ denoting the $M$-th order Legendre polynomial derivative at $s_m$. 
This quadrature rule is highly precise, yielding theoretically exact results for polynomials of degree up to $2M-1$. 

\begin{figure}[!t]
 \centering
\includegraphics[width=0.5\textwidth]{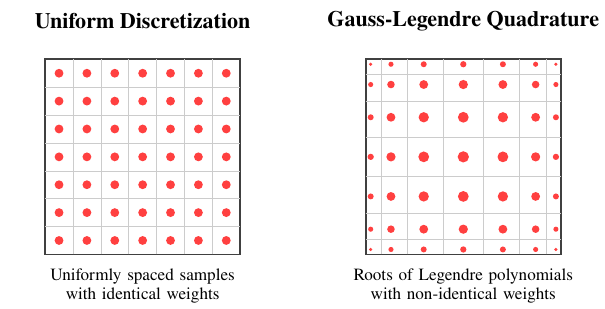}
\caption{Comparison of 2D numerical integration strategies. Left: Uniform discretization with uniformly spaced samples and identical weights. Right: Gauss-Legendre quadrature, which samples at the roots of Legendre polynomials. The area of each red dot represents its non-identical 2D quadrature weight $w_m w_n$.}
\label{Figure: GL_quadrature}
\end{figure}

Applying this numerical integration to the 2D continuous planar CAPA aperture region $\mathcal{S}_{\mathrm{t}} = [-L_{{\rm{t}},\mathsf{x}}/2, L_{{\rm{t}},\mathsf{x}}/2] \times [-L_{{\rm{t}},\mathsf{z}}/2, L_{{\rm{t}},\mathsf{z}}/2]$ yields
\begin{align}
    &\int_{\frac{-L_{{\rm{t}},\mathsf{z}}}{2}}^{\frac{L_{{\rm{t}},\mathsf{z}}}{2}}\int_{\frac{-L_{{\rm{t}},\mathsf{x}}}{2}}^{\frac{L_{{\rm{t}},\mathsf{x}}}{2}} f(s_x, s_z) \, \mathrm{d} s_x \, \mathrm{d} s_z \nonumber \\
    &\approx \frac{L_{{\rm{t}},\mathsf{x}}}{2} \frac{L_{{\rm{t}},\mathsf{z}}}{2} \sum_{m=1}^{M} \sum_{n=1}^{M} w_m w_n f\left( \frac{L_{{\rm{t}},\mathsf{x}}}{2} s_{m,\mathsf{x}}, \frac{L_{{\rm{t}},\mathsf{z}}}{2} s_{n,\mathsf{z}} \right),
\end{align}
where $s_{m,\mathsf{x}}$ and $s_{n,\mathsf{z}}$ are the roots of the $M$-th order Legendre polynomials $P_M(s_x)$ and $P_M(s_z)$, respectively. An illustration of the 2D Gauss-Legendre quadrature is shown in Fig. \ref{Figure: GL_quadrature}.

\begin{remark}[\emph{Difference from the Wavenumber-Domain Approximation}]

    \normalfont
    Although both approaches yield finite-dimensional computations, they serve different purposes. Wavenumber-domain discretization approximates the \emph{model} by retaining \rev{the} sampled radiating spectral components, \rev{which leads} to the MIMO-like representation in \eqref{WD_MIMO_Model}. In contrast, Gauss-Legendre quadrature is \rev{applied} after the continuous variational model has been derived, \rev{e.g.,} to compute the coefficients $b_j$ and $c_{j,k}$ in \eqref{Fredholm_equation_second_separable_2}\rev{, and its only approximation is} the numerical evaluation of integrals. \rev{Hence,} the number of quadrature points $M$ is chosen \rev{according to the required integration accuracy} \rev{rather than a spatial sampling rule for the model. Since the integrands oscillate on the wavelength scale, the required $M$ grows approximately linearly with the aperture size normalized by the wavelength in each dimension. Nevertheless, since these integrands are smooth, Gauss-Legendre quadrature typically requires far fewer points than uniform sampling to achieve the same accuracy.}
\end{remark}

\paragraph{A Case Study of Multi-User Beamforming}
To illustrate how the calculus of variations can be applied to CAPA design, consider a multi-user multiple-input single-output (MISO) downlink where a base station (BS) equipped with a planar CAPA ${\mathcal{S}}_{\mathrm{t}}$ serves $K$ single-antenna users. Under the simplified power metric adopted in this example, the objective is to determine the continuous transmit currents that minimize the total transmit power while guaranteeing a target SINR for every user. The key point is that, although the optimization variable is an infinite-dimensional function, the optimal solution admits a finite-dimensional structure.

For user $k$, the received signal is
\begin{align}
    y_k = \int_{{\mathcal{S}}_{\mathrm{t}}} h_k({\mathbf{s}}) x({\mathbf{s}}) \, {\rm{d}}{\mathbf{s}} + n_k,
\end{align}
where $h_k({\mathbf{s}})$ is the channel from the surface point ${\mathbf{s}}$ to user $k$, $x({\mathbf{s}})$ is the total transmit current over the aperture, and $n_k \sim {\mathcal{CN}}(0,\sigma^2)$ is the additive noise. To simultaneously serve all $K$ users, the BS superposes $K$ data streams as
\begin{align}
    x({\mathbf{s}}) = \sum_{k=1}^{K} w_k({\mathbf{s}}) c_k,
\end{align}
where $c_k \sim {\mathcal{CN}}(0,1)$ is the information symbol for user $k$, and $w_k({\mathbf{s}})$ is the corresponding beamforming current distribution. If user $k$ has channel state information and treats interference as noise, an achievable spectral efficiency is $\log_2(1+\gamma_k)$, where the SINR is
\begin{align}\label{MU_SINR}
    \gamma_k =
    \frac{\left\lvert \int_{{\mathcal{S}}_{\mathrm{t}}} h_k({\mathbf{s}}) w_k({\mathbf{s}}) \, {\rm{d}}{\mathbf{s}} \right\rvert^2}
    {\sum_{j \neq k} \left\lvert \int_{{\mathcal{S}}_{\mathrm{t}}} h_k({\mathbf{s}}) w_j({\mathbf{s}}) \, {\rm{d}}{\mathbf{s}} \right\rvert^2 + \sigma^2}.
\end{align}

Suppose the users have specific SINR requirements, $\bar{\gamma}_1,\ldots,\bar{\gamma}_K$.
We now minimize the total transmit power subject to these SINR requirements: $\gamma_k \ge \bar{\gamma}_k$. Using \eqref{MU_SINR}, the constraint for user $k$ can be written as
\begin{align}\label{MU_SINR_Constraint}
    \frac{1}{\bar{\gamma}_k}\left\lvert \int_{{\mathcal{S}}_{\mathrm{t}}} h_k({\mathbf{s}}) w_k({\mathbf{s}}) \, {\rm{d}}{\mathbf{s}} \right\rvert^2
    -
    \sum_{j \neq k} \left\lvert \int_{{\mathcal{S}}_{\mathrm{t}}} h_k({\mathbf{s}}) w_j({\mathbf{s}}) \, {\rm{d}}{\mathbf{s}} \right\rvert^2
    \ge \sigma^2.
\end{align}
The resulting functional optimization problem is
\begin{subequations}\label{MU_Power_Min}
\begin{align}
    \underset{{\{w_k({\mathbf{s}})\}}}{\mathrm{minimize}} \ \sum_{k=1}^{K} \int_{{\mathcal{S}}_{\mathrm{t}}} \lvert w_k({\mathbf{s}})\rvert^2 \, {\rm{d}}{\mathbf{s}} \quad \text{s.t.} \quad  \eqref{MU_SINR_Constraint}, \quad k = 1, \dots, K.
\end{align}
\end{subequations}
To solve \eqref{MU_Power_Min}, we form the Lagrangian
\begin{align}\label{MU_Lagrangian}
    \mathcal{L}
    =
    \sum_{k=1}^{K} &\Bigg(
    \int_{{\mathcal{S}}_{\mathrm{t}}} \lvert w_k({\mathbf{s}})\rvert^2 \, {\rm{d}}{\mathbf{s}}
    - \frac{\mu_k}{\bar{\gamma}_k \sigma^2}
    \left\lvert \int_{{\mathcal{S}}_{\mathrm{t}}} h_k({\mathbf{s}}) w_k({\mathbf{s}}) \, {\rm{d}}{\mathbf{s}} \right\rvert^2
    \nonumber\\
    &+ \frac{\mu_k}{\sigma^2}
    \sum_{j \neq k}
    \left\lvert \int_{{\mathcal{S}}_{\mathrm{t}}} h_k({\mathbf{s}}) w_j({\mathbf{s}}) \, {\rm{d}}{\mathbf{s}} \right\rvert^2
    + \mu_k
    \Bigg),
\end{align}
where $\mu_k \ge 0$ is the multiplier associated with the SINR constraint of user $k$. To derive the stationarity condition, we perturb each beamforming current as $w_k({\mathbf{s}})\rightarrow w_k({\mathbf{s}})+\epsilon\eta_k({\mathbf{s}})$, where $\eta_k({\mathbf{s}})$ is an arbitrary complex test function and $\epsilon$ is a small real scalar. Following \eqref{eq:first_variation_expand}, the first variation with respect to $w_k$ becomes
\begin{align}\label{MU_First_Variation}
    &\delta \mathcal{L}(w_k, \eta_k)
    =
    \left. \frac{{\rm d}}{{\rm d}\epsilon} \mathcal{L}\big(\{w_k + \epsilon \eta_k\}\big) \right|_{\epsilon=0}
    \nonumber\\
    &=
    2\Re\Bigg\{
    \int_{{\mathcal{S}}_{\mathrm{t}}}
    \Bigg[
    w_k({\mathbf{s}})
    -
    \frac{\mu_k}{\bar{\gamma}_k \sigma^2}
    h_k^*({\mathbf{s}})
    \int_{{\mathcal{S}}_{\mathrm{t}}} h_k({\mathbf{s}}') w_k({\mathbf{s}}') \, {\rm{d}}{\mathbf{s}}'
    \nonumber\\
    &\quad
    +
    \sum_{j \neq k}
    \frac{\mu_j}{\sigma^2}
    h_j^*({\mathbf{s}})
    \int_{{\mathcal{S}}_{\mathrm{t}}} h_j({\mathbf{s}}') w_k({\mathbf{s}}') \, {\rm{d}}{\mathbf{s}}'
    \Bigg]
    \eta_k^*({\mathbf{s}}) \, {\rm{d}}{\mathbf{s}}
    \Bigg\}.
\end{align}
By comparing \eqref{MU_First_Variation} with \eqref{eq:fundamental_lemma} and applying the fundamental lemma of the calculus of variations, the optimal beamforming current must satisfy the Fredholm integral equation as follows:
\begin{align}\label{MU_Fredholm}
    w_k({\mathbf{s}})
    = &
    \frac{\mu_k}{\bar{\gamma}_k \sigma^2}
    h_k^*({\mathbf{s}})
    \int_{{\mathcal{S}}_{\mathrm{t}}} h_k({\mathbf{s}}') w_k({\mathbf{s}}') \, {\rm{d}}{\mathbf{s}}'
    \nonumber\\
    &
    -
    \sum_{j \neq k}
    \frac{\mu_j}{\sigma^2}
    h_j^*({\mathbf{s}})
    \int_{{\mathcal{S}}_{\mathrm{t}}} h_j({\mathbf{s}}') w_k({\mathbf{s}}') \, {\rm{d}}{\mathbf{s}}'.
\end{align}

At first glance, \eqref{MU_Fredholm} is still an infinite-dimensional equation. However, its kernel is separable, so the method in \eqref{Fredholm_equation_second_separable}-\eqref{Fredholm_equation_second_separable_2} can be applied. Defining $a_{k,j} \triangleq \int_{\mathcal{S}_{\mathrm{t}}} h_j(\mathbf{s}') w_k(\mathbf{s}') \, \mathrm{d}\mathbf{s}'$, we can rewrite \eqref{MU_Fredholm} as
\begin{align}\label{MU_Fredholm_Rearranged}
    w_k(\mathbf{s})
    =
    \frac{\mu_k}{\bar{\gamma}_k \sigma^2} a_{k,k} h_k^*(\mathbf{s})
    -
    \sum_{j \neq k} \frac{\mu_j}{\sigma^2} a_{k,j} h_j^*(\mathbf{s}).
\end{align}
Equation \eqref{MU_Fredholm_Rearranged} immediately reveals the key structural result: \emph{the optimal beamforming current $w_k(\mathbf{s})$ lies in the linear span of the conjugate user channels $\{h_1^*(\mathbf{s}),\dots,h_K^*(\mathbf{s})\}$}, which is a continuous generalization of \cite[Prop.~2]{4558045}. Hence, we can \rev{reformulate} \eqref{MU_Fredholm_Rearranged} into the following equivalent form:
\begin{align}\label{MU_span_expansion}
    w_k(\mathbf{s}) = \sum_{j=1}^K \beta_{k,j} h_j^*(\mathbf{s}),
\end{align}
where linear coefficients are given by
\begin{align}\label{MU_Fredholm_Coefficients}
    \beta_{k,j} =
    \begin{cases}
        \dfrac{\mu_k}{\bar{\gamma}_k \sigma^2} a_{k,k}, &\text{for } j = k, \\[6pt]
        -\dfrac{\mu_j}{\sigma^2} a_{k,j}, & \text{for } j \neq k.
    \end{cases}
\end{align}
Therefore, the original functional optimization problem has been reduced to determining the finite set of coefficients $\{\beta_{k,j}\}$.

To determine these coefficients, we substitute \eqref{MU_span_expansion} back into the definition of $a_{k,j}$, which yields
\begin{align}\label{MU_Fredholm_Inner_Product}
    a_{k,j}
    =
    \sum_{l=1}^K \beta_{k,l}
    \int_{\mathcal{S}_{\mathrm{t}}} h_j(\mathbf{s}') h_l^*(\mathbf{s}') \, \mathrm{d}\mathbf{s}'.
\end{align}
Define $R_{j,l} \triangleq \int_{\mathcal{S}_{\mathrm{t}}} h_j(\mathbf{s}') h_l^*(\mathbf{s}') \, \mathrm{d}\mathbf{s}'$, which is the spatial inner product between the channels of users $j$ and $l$. These coefficients form a $K\times K$ positive semidefinite \rev{channel correlation matrix} $\mathbf{R}$, whose entries can be computed efficiently using the Gauss-Legendre quadrature discussed earlier. In vector form, let $\mathbf{a}_k = [a_{k,1}, \dots, a_{k,K}]^{\mathsf{T}}$ and $\boldsymbol{\beta}_k = [\beta_{k,1}, \dots, \beta_{k,K}]^{\mathsf{T}}$. Then \eqref{MU_Fredholm_Inner_Product} becomes
\begin{align}
    \mathbf{a}_k = \mathbf{R}\boldsymbol{\beta}_k.
\end{align}
Combining this relation with \eqref{MU_Fredholm_Coefficients} yields the following finite-dimensional linear system:
\begin{align}\label{MU_Fredholm_Linear_System}
    \left(\mathbf{I}_K + \frac{1}{\sigma^2} \boldsymbol{\Lambda}_{\mu} \mathbf{R} \right) \boldsymbol{\beta}_k
    =
    \rev{\frac{\mu_k (1+\bar{\gamma}_k)}{\bar{\gamma}_k \sigma^2}} a_{k,k} \, \mathbf{e}_k,
\end{align}
where $\boldsymbol{\Lambda}_{\mu} = \mathrm{diag}(\mu_1, \dots, \mu_K)$ and $\mathbf{e}_k$ is the $k$-th column of $\mathbf{I}_K$. \rev{Here, the $k$-th entry of the right-hand side collects the term $\frac{\mu_k}{\bar{\gamma}_k\sigma^2}a_{k,k}$ from \eqref{MU_Fredholm_Coefficients} and the term $\frac{\mu_k}{\sigma^2}a_{k,k}$ arising from the $k$-th entry of $\frac{1}{\sigma^2}\boldsymbol{\Lambda}_{\mu}\mathbf{R}\boldsymbol{\beta}_k=\frac{1}{\sigma^2}\boldsymbol{\Lambda}_{\mu}\mathbf{a}_k$.} Solving \eqref{MU_Fredholm_Linear_System} gives
\begin{align}
    \boldsymbol{\beta}_k
    =
    \rev{\frac{\mu_k (1+\bar{\gamma}_k)}{\bar{\gamma}_k \sigma^2}} a_{k,k}
    \left(\mathbf{I}_K + \frac{1}{\sigma^2} \boldsymbol{\Lambda}_{\mu} \mathbf{R} \right)^{-1}
    \mathbf{e}_k.
\end{align}

Finally, substituting $\boldsymbol{\beta}_k$ back into \eqref{MU_span_expansion} gives the optimal continuous beam. More compactly, define $\mathbf{w}(\mathbf{s}) = [w_1(\mathbf{s}),\dots,w_K(\mathbf{s})]$, $\mathbf{h}(\mathbf{s}) = [h_1(\mathbf{s}),\dots,h_K(\mathbf{s})]$, and $\mathbf{P} = \mathrm{diag}\!\left(\rev{\frac{\mu_1 (1+\bar{\gamma}_1) a_{1,1}}{\bar{\gamma}_1 \sigma^2}, \dots, \frac{\mu_K (1+\bar{\gamma}_K) a_{K,K}}{\bar{\gamma}_K \sigma^2}}\right)$.
Hence, the beamformer can be written as
\begin{align}\label{MU_RZF_Continuous}
    \mathbf{w}(\mathbf{s})
    =
    \mathbf{h}^*(\mathbf{s})
    \left(\mathbf{I}_K + \frac{1}{\sigma^2} \boldsymbol{\Lambda}_{\mu} \mathbf{R} \right)^{-1}
    \mathbf{P},
\end{align}
which is the continuous-space analog of the classical minimum mean-squared error (MMSE) beamformer in discrete MIMO systems \cite{bjornson2013optimal, 10938678}, also known as the regularized zero-forcing (RZF) beamformer. In particular, the matrix $\mathbf{P}$ essentially plays the role of a diagonal power-allocation matrix across users. Under this formulation, the infinite-dimensional beamforming problem over the CAPA surface is reduced to matrix operations on the \rev{channel correlation matrix} $\mathbf{R}$, while the resulting beamformer itself remains a continuous current distribution over the aperture.

It is important to note that, although \eqref{MU_RZF_Continuous} provides a closed-form expression for the continuous beamformer, the Lagrange multipliers $\{\mu_k\}$ and the coefficients $\{a_{k,k}\}$ are still unknown and must be determined iteratively. We refer to \cite{10910020} and \cite{10938678} for more details on the iterative algorithm design, where a suboptimal alternating-optimization method and an optimal monotonic-optimization method were proposed, respectively.

Based on the MMSE beamformer in \eqref{MU_RZF_Continuous}, the maximum-ratio transmission (MRT) and zero-forcing (ZF) beamformers can also be obtained as special cases by adjusting the regularization term $ \left(\mathbf{I}_K + \frac{1}{\sigma^2} \boldsymbol{\Lambda}_{\mu} \mathbf{R}\right)$, which controls the tradeoff between beamforming gain and interference suppression. In particular, the MRT beamformer is obtained by retaining only the identity term $\mathbf{I}_K$, which corresponds to the low-SNR regime:
\begin{align}\label{MU_MRT_Continuous}
    \mathbf{w}_{\mathrm{MRT}}(\mathbf{s})
    =
    \mathbf{h}^*(\mathbf{s}) \mathbf{P}.
\end{align}
Thus, each user beam is matched to its own conjugate channel, i.e., $w_k(\mathbf{s}) \propto h_k^*(\mathbf{s})$, which maximizes the desired received signal power but does not explicitly suppress inter-user interference.

In contrast, the ZF beamformer is obtained in the high-SNR regime by neglecting the identity term relative to the interference term, which yields
\begin{align}\label{MU_ZF_Continuous}
    \mathbf{w}_{\mathrm{ZF}}(\mathbf{s})
    =
    \mathbf{h}^*(\mathbf{s}) \mathbf{R}^{-1}\mathbf{P},
\end{align}
where the factor $\frac{1}{\sigma^2} \boldsymbol{\Lambda}_{\mu}$ is absorbed into $\mathbf{P}$. This expression assumes that $\mathbf{R}$ is invertible. In this case, the inter-user interference is completely eliminated, since
\begin{align}
    \int_{\mathcal{S}_{\mathrm{t}}}
    \mathbf{h}^{\mathsf{T}}(\mathbf{s})
    \mathbf{w}_{\mathrm{ZF}}(\mathbf{s}) \, \mathrm{d}\mathbf{s}
    =
    \mathbf{R}\mathbf{R}^{-1}\mathbf{P}
    =
    \mathbf{P},
\end{align}
where the off-diagonal entries of $\mathbf{P}$ are zero, indicating that each user receives only its own signal without interference from the others.

\rev{The above derivation adopts the simplified power metric in \eqref{eq:simplified_power}. If the physical radiated power in \eqref{eq:power_radiation_integral} is used instead, the power term in the Lagrangian becomes a quadratic form with kernel $-\frac{1}{2}\Re\{\mathcal{G}_{\mathrm{p}}(\mathbf{s},\mathbf{s}')\}$, and the stationarity condition becomes a Fredholm integral equation of the first kind with this kernel. As discussed for \eqref{Fredholm_equation_first}, this equation is ill-posed, and the ohmic loss provides the required regularization. In this case, the optimal beamformers lie in the span of the conjugate user channels filtered by the inverse of the regularized kernel \cite{wang2025mutual}. The finite-dimensional structure in \eqref{MU_RZF_Continuous} is therefore unchanged, with the entries of $\mathbf{R}$ replaced by the inner products of the user channels weighted by this inverse kernel.}

\begin{figure}[!t]
    \centering
    \subfigure[Spectral efficiency.]{
        \includegraphics[width=0.4\textwidth]{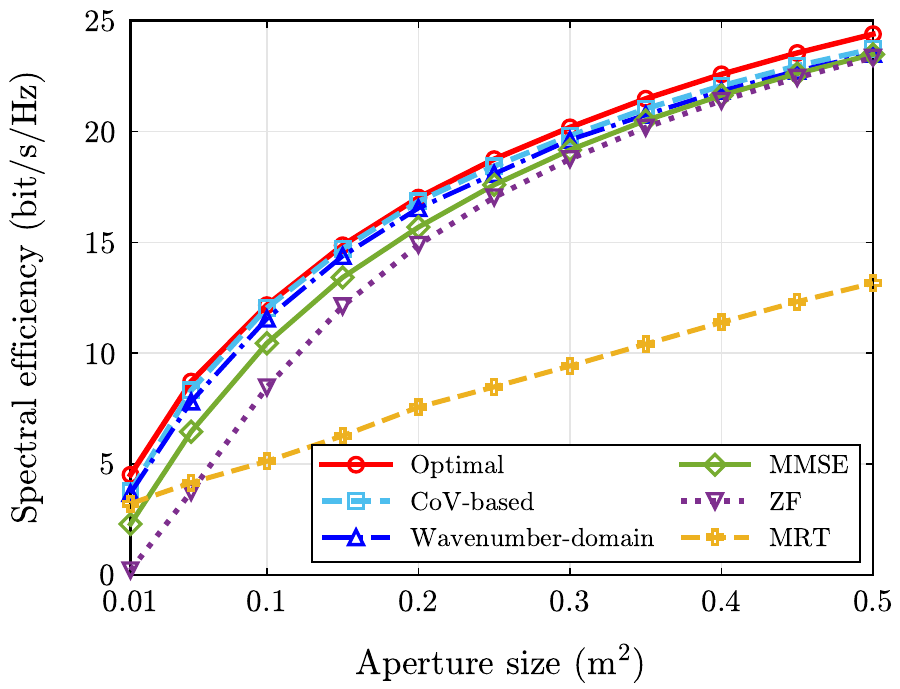}
        \label{fig:mag_rate_vs_aperture}
    }
    \subfigure[Computational complexity.]{
        \includegraphics[width=0.4\textwidth]{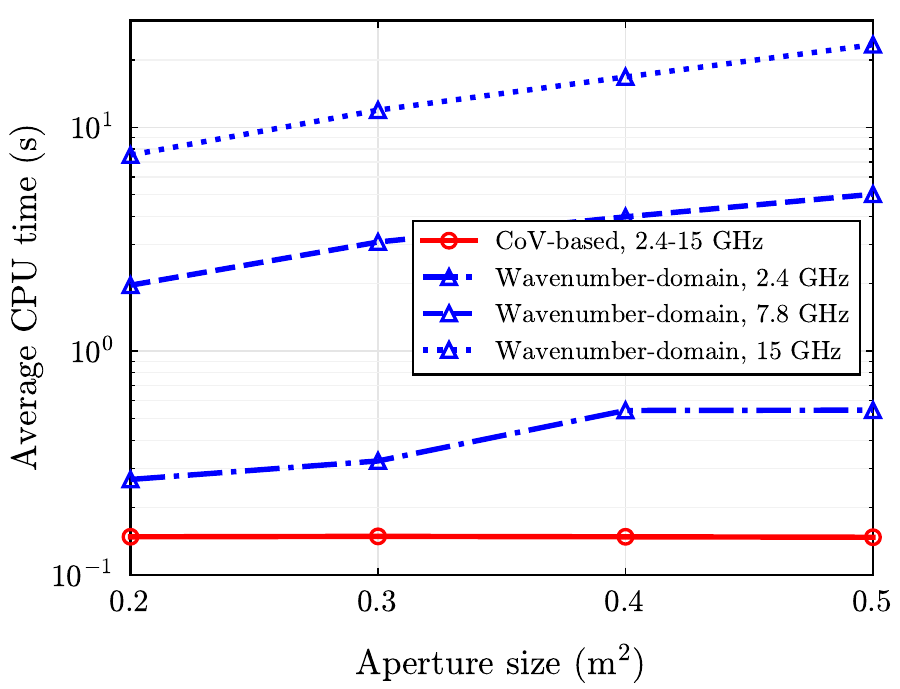}
        \label{fig:CPU_time}
    }
    \subfigure[\rev{Comparison with SPDAs.}]{
        \includegraphics[width=0.4\textwidth]{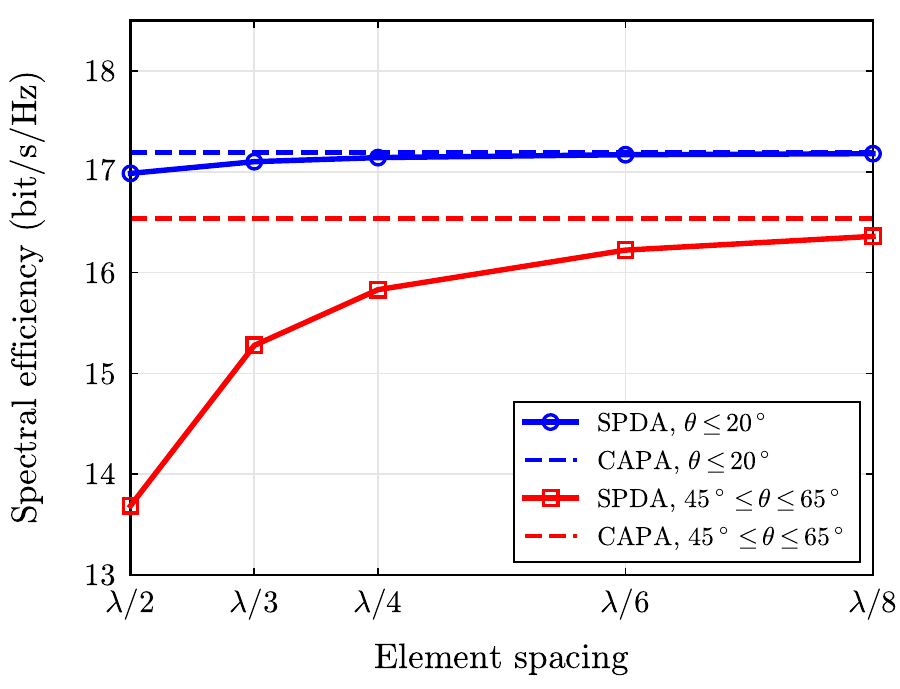}
        \label{fig:spda_convergence}
    }
    \caption{Numerical comparison of continuous beamforming designs. (a) Spectral efficiency versus aperture size for different CAPA beamforming designs at $2.4$ GHz. (b) CPU time comparison for the calculus-of-variations (CoV)-based and wavenumber-domain beamforming designs. \rev{(c) Spectral efficiency of the CAPA and of SPDAs with element spacings from $\lambda/2$ to $\lambda/8$, i.e., from $8^2$ to $32^2$ elements, under MMSE beamforming for $|\mathcal{S}_{\mathrm{t}}|=0.25$~m$^2$, where $\theta$ is the angle between the user direction and the aperture normal.} The detailed simulation can be found in \cite{10910020} and \cite{10938678}.}
    \label{fig:beamforming_numerical_examples}
\end{figure}

\paragraph{Discussion}
The main advantage of the calculus-of-variations method is that it operates directly on the original continuous optimization problem, without introducing an intermediate model approximation such as wavenumber truncation. Hence, it can reveal the exact structure of the optimal current distribution. When the resulting stationarity condition has exploitable structure, the infinite-dimensional problem may collapse into a much smaller finite-dimensional one. In the multi-user example above, for instance, the optimal beamformer is shown to lie in the span of the $K$ conjugate user channels.

Its drawback, however, is that the method is less universal and more mathematically involved than the wavenumber-domain approach. For each new utility function and constraint set, one must rederive the functional derivative, establish the stationarity condition, and analyze the resulting integral equation. If the kernel is not separable or does not admit a convenient structure, a closed-form solution may not exist. Therefore, although the calculus-of-variations method can be more exact and sometimes lower-dimensional, it is generally more problem-dependent and harder to develop into a plug-and-play optimization framework.

\rev{\begin{remark}[\emph{From Optimal Currents to Realizable Excitations}]\label{remark:realizable}
    \normalfont
    The beamformers derived above are optimal over all square-integrable currents. In a practical CAPA described by the circuit-to-field model in Section~\ref{sec:circuit_to_field}, however, only currents in the span of the basis modes $\{\psi_n(\mathbf{s})\}_{n=1}^N$ can be generated. A realizable beamformer can be obtained by projecting $w_k(\mathbf{s})$ onto this span. A better approach is to substitute $w_k(\mathbf{s})=\boldsymbol{\psi}^{\mathsf{T}}(\mathbf{s})\mathbf{Y}\mathbf{w}_k$ from \eqref{eq:effective_beamformer_rewrite} into the design problem and optimize the digital precoders $\{\mathbf{w}_k\}$ directly, which yields the best beamformer within this span. This reduces to a conventional MISO problem in the port currents $\mathbf{Y}\mathbf{w}_k$, with $N$-dimensional effective channels $\mathbf{g}_k=\int_{\mathcal{S}_{\mathrm{t}}}h_k(\mathbf{s})\boldsymbol{\psi}(\mathbf{s})\,\mathrm{d}\mathbf{s}$ and a power constraint weighted by the Gram matrix $\int_{\mathcal{S}_{\mathrm{t}}}\boldsymbol{\psi}^*(\mathbf{s})\boldsymbol{\psi}^{\mathsf{T}}(\mathbf{s})\,\mathrm{d}\mathbf{s}$. The performance gap between the continuous and the realizable beamformers quantifies the loss caused by restricting the current to the span of the basis modes. For example, consider a pixel-based CAPA with square pixels of side length $\Delta$, each carrying a uniform current. Each pixel then has the space factor of a uniformly excited aperture, so that a beam steered by an angle $\theta$ from the aperture normal within a principal plane is attenuated by the factor $\mathrm{sinc}^2(k_0\Delta\sin\theta/2)$, where $\mathrm{sinc}(x)=\sin(x)/x$ \cite{balanis2016antenna}. This corresponds to a gain loss of, e.g., $2.9$~dB at $\theta=60^\circ$ for $\Delta=\lambda/2$, but only $0.7$~dB for $\Delta=\lambda/4$. In addition, preserving the spatial multiplexing capability requires at least as many modes as the spatial DoF, which will be discussed in Section~\ref{sec:dof}. Moreover, quantizing the phases of the excitation coefficients with $b$ bits reduces the beamforming gain approximately by the factor $\big(\frac{2^b}{\pi}\sin\frac{\pi}{2^b}\big)^2$ when the number of modes is large and the quantization errors are approximately independent, e.g., by $3.9$, $0.9$, and $0.2$~dB for $b=1$, $2$, and $3$, respectively \cite{wu2020beamforming}.
\end{remark}}

\subsubsection{Numerical Examples}

To illustrate the performance of the continuous beamforming designs discussed above, Fig.~\ref{fig:mag_rate_vs_aperture} compares their spectral efficiency as the transmit aperture size $|\mathcal{S}_{\mathrm{t}}|$ increases. The detailed simulation setup is provided in \cite{10910020} and \cite{10938678}. More specifically, the calculus-of-variations (CoV)-based beamformers are implemented using the iterative algorithm proposed in \cite{10910020}, while the performance benchmark labeled as the optimal solution is obtained via the monotonic optimization method in \cite{10938678}. \rev{As shown in Fig.~\ref{fig:mag_rate_vs_aperture}, the CoV-based and wavenumber-domain designs both approach the optimal solution, and the gap between the linear beamformers and the optimal solution decreases as the aperture size increases.}

Moreover, although the CoV-based approach provides only a slight performance improvement over the wavenumber-domain method, its main advantage lies in a substantial reduction in computational complexity, as illustrated in Fig.~\ref{fig:CPU_time}. Specifically, the wavenumber-domain method requires a large number of spectral samples to accurately represent the channel, which leads to high-dimensional matrix operations and a considerable computational burden. In contrast, the CoV-based method directly exploits the structure of the optimal solution to reduce the problem dimension, while also using fast-converging Gauss-Legendre quadrature to efficiently evaluate the required integrals. As a result, the CoV-based beamformers can achieve near-optimal performance with much lower computational complexity than the wavenumber-domain method, making them more attractive for practical implementation in large-scale CAPA systems.

\rev{Fig.~\ref{fig:spda_convergence} further compares the CAPA with SPDAs of the same aperture size{ under the same power metric and MMSE beamformer, where each SPDA element is modeled as a pixel that carries a uniform current, as in Remark~\ref{remark:realizable}, with the side length $\Delta$ equal to the element spacing.} Two setups are considered, in which the users are randomly located at distances between $15$ and $30$~m, either within $20^\circ$ of the aperture normal or between $45^\circ$ and $65^\circ$ from it. In the first setup, the half-wavelength SPDA achieves about $99\%$ of the CAPA spectral efficiency, since its element sums accurately approximate the aperture integrals in the channel correlation matrix $\mathbf{R}$ of \eqref{MU_RZF_Continuous}. In the second setup, however, it loses about $17\%$ due to the pixel space factor discussed in Remark~\ref{remark:realizable}, and this gap closes only as the element spacing decreases toward $\lambda/8$, which requires $16$ times as many elements and RF chains. Hence, the CAPA can be viewed as the limit of increasingly dense SPDAs{, and its advantage over a half-wavelength SPDA lies in stronger effective user channels at large steering angles rather than in a larger number of separable users.}}

\subsection{Channel Estimation} \label{sec:ce}

Channel estimation for CAPA systems is fundamentally different from that in conventional SPDA-based systems. In classical MIMO systems, the channel is represented by a finite-dimensional matrix, so training reduces to estimating a finite number of coefficients. In contrast, the CAPA channel is a continuous-space operator described by the field response $h(\mathbf{r},\mathbf{s})$, which maps the transmit current over ${\mathcal{S}}_{\mathrm{t}}$ to the received field over ${\mathcal{S}}_{\mathrm{r}}$. Hence, the unknown object is infinite-dimensional. Directly estimating $h(\mathbf{r},\mathbf{s})$ pointwise is therefore impractical, and conventional least-squares (LS) or linear minimum mean-squared error (LMMSE) estimators, which are designed to recover all entries of a finite-dimensional channel matrix, \rev{cannot be applied directly before a suitable finite-dimensional representation of the channel is established}.

To make this explicit, consider uplink training over $\tau_{\mathrm{p}}$ pilot uses. Let $w_{l}(\mathbf{s})$ denote the known pilot current transmitted in slot $l$. The corresponding received field is
\begin{align}\label{CE_continuous_training}
    y_{l}(\mathbf{r}) = \int_{{\mathcal{S}}_{\mathrm{t}}} h(\mathbf{r},\mathbf{s}) w_{l}(\mathbf{s}) \, \mathrm{d}\mathbf{s} + n_{l}(\mathbf{r}), \quad l = 1,\dots,\tau_{\mathrm{p}},
\end{align}
where $n_{l}(\mathbf{r})$ is the noise field. The unknown quantity in \eqref{CE_continuous_training} is the bivariate function $h(\mathbf{r},\mathbf{s})$, rather than a finite-dimensional vector. Therefore, channel estimation must first transform the original infinite-dimensional problem into a finite-dimensional one, while still preserving the continuous EM structure of the channel.

\subsubsection{\rev{Finite-Dimensional Channel Representations}}
A natural first step is to project the operator onto finite transmit and receive codebooks. Let $\mathcal{C}_{\mathrm{t}}=\{u_n(\mathbf{s})\}_{n=1}^{N_{\mathrm{t}}}$ and $\mathcal{C}_{\mathrm{r}}=\{v_m(\mathbf{r})\}_{m=1}^{N_{\mathrm{r}}}$ denote finite sets of continuous transmit and receive patterns. The projected channel coefficients are defined as
\begin{align}\label{CE_projected_channel}
    g_{m,n} \triangleq \int_{{\mathcal{S}}_{\mathrm{r}}} \int_{{\mathcal{S}}_{\mathrm{t}}} v_m^*(\mathbf{r}) h(\mathbf{r},\mathbf{s}) u_n(\mathbf{s}) \, \mathrm{d}\mathbf{s} \, \mathrm{d}\mathbf{r},
\end{align}
for $m=1,\dots,N_{\mathrm{r}}$ and $n=1,\dots,N_{\mathrm{t}}$. This transforms the continuous operator into a finite matrix $\mathbf{G}\in\mathbb{C}^{N_{\mathrm{r}}\times N_{\mathrm{t}}}$, and the channel can be approximated as
\begin{align}\label{CE_channel_reconstruction_basis}
    h(\mathbf{r},\mathbf{s}) \approx \sum_{m=1}^{N_{\mathrm{r}}}\sum_{n=1}^{N_{\mathrm{t}}} g_{m,n} v_m(\mathbf{r}) u_n^*(\mathbf{s}).
\end{align}
\rev{Here, we consider orthonormal codebooks, i.e., $\int_{\mathcal{S}_{\mathrm{t}}}u_n^*(\mathbf{s})u_{n'}(\mathbf{s})\,\mathrm{d}\mathbf{s}=\delta_{n,n'}$ and $\int_{\mathcal{S}_{\mathrm{r}}}v_m^*(\mathbf{r})v_{m'}(\mathbf{r})\,\mathrm{d}\mathbf{r}=\delta_{m,m'}$, so that \eqref{CE_channel_reconstruction_basis} is the orthogonal projection of $h(\mathbf{r},\mathbf{s})$ onto the span of the codebooks.}
This representation is important because it preserves the continuous nature of the channel without resorting to pointwise spatial sampling. However, it is still not the most suitable form for estimation, since recovering the full matrix $\mathbf{G}$ requires estimating $N_{\mathrm{r}}N_{\mathrm{t}}$ coefficients. When the codebooks are large, this leads to high pilot overhead and computational complexity. Moreover, under a generic pair of projection bases, $\mathbf{G}$ is typically dense, so the representation in \eqref{CE_channel_reconstruction_basis} does not explicitly exploit the physical sparsity of practical channels.

To obtain a more efficient model, we next introduce a dictionary aligned with the dominant propagation mechanisms. Specifically, suppose the continuous channel admits the parametric representation
\begin{align}\label{CE_parametric_family}
    h(\mathbf{r},\mathbf{s}) = \int_{\Theta} \beta(\boldsymbol{\theta}) \, \phi(\mathbf{r},\mathbf{s};\boldsymbol{\theta}) \, \mathrm{d}\boldsymbol{\theta},
\end{align}
where $\boldsymbol{\theta}\in\Theta$ is a continuous parameter vector describing one propagation mode, and $\beta(\boldsymbol{\theta})$ is its complex gain density. A finite dictionary is then obtained by discretizing $\Theta$ into a codebook $\{\boldsymbol{\theta}_i\}_{i=1}^{N_{\mathrm{c}}}$ and defining $\varphi_i(\mathbf{r},\mathbf{s}) \triangleq \phi(\mathbf{r},\mathbf{s};\boldsymbol{\theta}_i), \quad i=1,\dots,N_{\mathrm{c}}$. This gives the sparse approximation
\begin{align}\label{CE_sparse_operator}
    h(\mathbf{r},\mathbf{s}) \approx \sum_{i=1}^{N_{\mathrm{c}}} \alpha_i \varphi_i(\mathbf{r},\mathbf{s}),
\end{align}
where only a small number of coefficients in $\boldsymbol{\alpha}=[\alpha_1,\dots,\alpha_{N_{\mathrm{c}}}]^{\mathsf{T}}$ are significant. In other words, the channel is first mapped to a finite-dimensional model and then further structured so that the unknown coefficient vector is sparse or approximately sparse.

The choice of the dictionary functions $\varphi_i(\mathbf{r},\mathbf{s})$ should follow the underlying physics\rev{, as illustrated by the following two examples.

\paragraph{Far-Field Dictionary} In a far-field angular-domain model}, one propagation mode is characterized by a departure direction $\boldsymbol{\kappa}$ at the Tx and an arrival direction $\mathbf{k}$ at the Rx. A natural basis function is
\begin{align}\label{CE_plane_wave_atom}
    \phi(\mathbf{r},\mathbf{s};\mathbf{k},\boldsymbol{\kappa}) = a_{\mathrm{r}}(\mathbf{r};\mathbf{k}) a_{\mathrm{t}}^*(\mathbf{s};\boldsymbol{\kappa}),
\end{align}
where $a_{\mathrm{r}}(\mathbf{r};\mathbf{k}) = {\rm e}^{{\rm j}(k_{\mathsf{x}}r_{\mathsf{x}}+k_{\mathsf{z}}r_{\mathsf{z}})}$ and $a_{\mathrm{t}}(\mathbf{s};\boldsymbol{\kappa}) = {\rm e}^{{\rm j}(\kappa_{\mathsf{x}}s_{\mathsf{x}}+\kappa_{\mathsf{z}}s_{\mathsf{z}})}$. If the angular domain is discretized into finite grids $\mathbf{k}\in\mathcal{K}_{\mathrm{r}}$ and $\boldsymbol{\kappa}\in\mathcal{K}_{\mathrm{t}}$, then each pair $(\mathbf{k},\boldsymbol{\kappa})$ defines one dictionary function, i.e.,
\begin{align}\label{CE_dictionary_angular_grid}
    \varphi_i(\mathbf{r},\mathbf{s}) = a_{\mathrm{r}}(\mathbf{r};\mathbf{k}_{p,q}) a_{\mathrm{t}}^*(\mathbf{s};\boldsymbol{\kappa}_{m,n}),
\end{align}
where the index $i$ is a one-to-one mapping of the tuple $(p,q,m,n)$. In this case, sparse channel estimation amounts to identifying a small number of dominant angular departure-arrival pairs.

\paragraph{\rev{Near-Field Dictionary}} For near-field propagation, the relevant parameters include not only direction but also distance or focal location. In that case, $\boldsymbol{\theta}$ can include a scatterer or focal point $\mathbf{q}$, and the \rev{dictionary} function can be chosen as a spherical-wave kernel, for example
\begin{align}\label{CE_spherical_atom}
    \phi(\mathbf{r},\mathbf{s};\mathbf{q}) = \frac{{\rm e}^{-{\rm j}k_0\|\mathbf{r}-\mathbf{q}\|}}{\|\mathbf{r}-\mathbf{q}\|} \frac{{\rm e}^{-{\rm j}k_0\|\mathbf{q}-\mathbf{s}\|}}{\|\mathbf{q}-\mathbf{s}\|}.
\end{align}
Sampling the candidate scatterer region with a finite set $\{\mathbf{q}_i\}$ then gives $\varphi_i(\mathbf{r},\mathbf{s})=\phi(\mathbf{r},\mathbf{s};\mathbf{q}_i)$. Compared with the far-field plane-wave dictionary, this near-field dictionary is able to capture the spherical-wave structure and focal behavior of CAPA channels more accurately.

\paragraph{\rev{Approximation Accuracy}} From an approximation viewpoint, the quality of the dictionary can be measured by the residual
\begin{align}\label{CE_dictionary_residual}
    \left\| h(\mathbf{r},\mathbf{s}) - \sum_{i=1}^{N_{\mathrm{c}}} \alpha_i \varphi_i(\mathbf{r},\mathbf{s}) \right\|_{L^2({\mathcal{S}}_{\mathrm{r}} \times {\mathcal{S}}_{\mathrm{t}})}^2,
\end{align}
where, for any bivariate function $f(\mathbf{r},\mathbf{s})$, the norm is defined as
\begin{align}
    \|f(\mathbf{r},\mathbf{s})\|_{L^2({\mathcal{S}}_{\mathrm{r}} \times {\mathcal{S}}_{\mathrm{t}})}^2
    \triangleq \int_{{\mathcal{S}}_{\mathrm{r}}}\int_{{\mathcal{S}}_{\mathrm{t}}} |f(\mathbf{r},\mathbf{s})|^2 \, \mathrm{d}\mathbf{s}\,\mathrm{d}\mathbf{r}.
\end{align}
Hence, \eqref{CE_dictionary_residual} measures the total squared approximation error of the channel kernel over all Tx-Rx point pairs on the two apertures. Ideally, this residual should be small while $\boldsymbol{\alpha}$ remains sparse. If the parametric family in \eqref{CE_parametric_family} is physically accurate, then refining the sampling grid $\{\boldsymbol{\theta}_i\}$ generally improves the approximation. In practice, however, the grid resolution and the codebook size $N_{\mathrm{c}}$ must be chosen by balancing approximation accuracy, pilot overhead, and sparse-recovery complexity.

\subsubsection{\rev{Sparse Channel Estimation}}
With the sparse model in hand, channel training can now be written in a standard finite-dimensional form. Specifically, the Rx correlates the observed field with a receive combining pattern $b_{l}(\mathbf{r})$ and obtains the scalar measurement
\begin{align}\label{CE_scalar_measurement}
    v_{l}
    &= \int_{{\mathcal{S}}_{\mathrm{r}}} b_{l}^*(\mathbf{r}) y_{l}(\mathbf{r}) \, \mathrm{d}\mathbf{r} \nonumber \\
    &\approx \sum_{i=1}^{N_{\mathrm{c}}} \alpha_i \int_{{\mathcal{S}}_{\mathrm{r}}} \int_{{\mathcal{S}}_{\mathrm{t}}} b_{l}^*(\mathbf{r}) \varphi_i(\mathbf{r},\mathbf{s}) w_{l}(\mathbf{s}) \, \mathrm{d}\mathbf{s} \, \mathrm{d}\mathbf{r} + z_{l},
\end{align}
where $z_{l}$ is the effective projected noise. By stacking $\{v_{l}\}_{l=1}^{\tau_{\mathrm{p}}}$, we obtain the finite-dimensional linear model
\begin{align}\label{CE_CS_model}
    \mathbf{v} = \mathbf{A}\boldsymbol{\alpha} + \boldsymbol{z},
\end{align}
where the sensing matrix $\mathbf{A} \in \mathbb{C}^{\tau_{\mathrm{p}} \times N_{\mathrm{c}}}$ has entries
\begin{align}\label{CE_sensing_matrix}
    [\mathbf{A}]_{l,i}
    =
    \int_{{\mathcal{S}}_{\mathrm{r}}} \int_{{\mathcal{S}}_{\mathrm{t}}}
    b_{l}^*(\mathbf{r}) \varphi_i(\mathbf{r},\mathbf{s}) w_{l}(\mathbf{s})
    \, \mathrm{d}\mathbf{s} \, \mathrm{d}\mathbf{r}.
\end{align}
Here, the $l$-th row of $\mathbf{A}$ is determined by the pair $\rev{(w_{l}, b_{l})}$. Therefore, the probing patterns and receive combining patterns should vary with $l$ so as to generate sufficiently diverse rows in $\mathbf{A}$. Otherwise, repeatedly using the same pair would produce identical or highly correlated measurements and thus provide insufficient information for sparse recovery. It is also important to note that $\mathbf{A}$ is known at the Rx a priori, since it depends only on the designed pilot currents $\{w_{l}(\mathbf{s})\}$, the receive combiners $\{b_{l}(\mathbf{r})\}$, and the predefined dictionary functions $\{\varphi_i(\mathbf{r},\mathbf{s})\}$. Hence, the only unknown quantity in \eqref{CE_CS_model} is the sparse coefficient vector $\boldsymbol{\alpha}$.

Equation \eqref{CE_CS_model} is now a standard compressive-sensing problem. Although $N_{\mathrm{c}}$ can be large, the coefficient vector $\boldsymbol{\alpha}$ is sparse or approximately sparse, so it can be recovered from far fewer measurements than $N_{\mathrm{c}}$. Accordingly, the channel estimation problem can be solved via sparse recovery methods such as
\begin{align}\label{CE_sparse_recovery}
   \underset{{\boldsymbol{\alpha}}}{\mathrm{minimize}} \ \|\boldsymbol{\alpha}\|_{0}
    \quad \text{s.t.} \quad
    \|\rev{\mathbf{v}} - \mathbf{A}\boldsymbol{\alpha}\|_2^2 \le \epsilon,
\end{align}
or, in its convex relaxation form,
\begin{align}\label{CE_sparse_recovery_convex}
    \underset{{\boldsymbol{\alpha}}}{\mathrm{minimize}} \ \|\boldsymbol{\alpha}\|_{1}
    \quad \text{s.t.} \quad
    \|\rev{\mathbf{v}} - \mathbf{A}\boldsymbol{\alpha}\|_2^2 \le \epsilon,
\end{align}
where $\epsilon$ is determined by the noise power and the degree of model mismatch. Problem \eqref{CE_sparse_recovery} is non-convex and is typically handled by greedy or iterative sparse-pursuit algorithms, such as orthogonal matching pursuit (OMP) \cite{tropp2007signal} and subspace pursuit \cite{dai2009subspace}. In contrast, the convex relaxation in \eqref{CE_sparse_recovery_convex} can be solved by standard convex-optimization methods, such as basis pursuit denoising \cite{chen2001atomic}, the alternating direction method of multipliers (ADMM) \cite{boyd2011distributed}, and proximal-gradient algorithms such as the fast iterative shrinkage-thresholding algorithm (FISTA) \cite{beck2009fast}. Once the estimates $\hat{\boldsymbol{\alpha}}$ are obtained, the continuous-space channel can be reconstructed as
\begin{align}\label{CE_final_reconstruction}
    \hat{h}(\mathbf{r},\mathbf{s}) = \sum_{i=1}^{N_{\mathrm{c}}} \hat{\alpha}_i \varphi_i(\mathbf{r},\mathbf{s}).
\end{align}

\rev{\subsubsection{Bayesian Channel Estimation}
The sparse-recovery formulation above treats $\boldsymbol{\alpha}$ as deterministic. When the second-order statistics of the channel are available, e.g., through the correlation-based model in Section~\ref{Section: Multipath Fading Channel Models: Correlation-Based}, a Bayesian estimator can be used instead. Specifically, suppose that the angular dictionary in \eqref{CE_dictionary_angular_grid} uses the same wavenumber grids as the wavenumber-domain beamforming in Section~\ref{sec:beamforming_Optimization}. When the apertures are much larger than the wavelength, the NLoS coefficients $\{\alpha_i\}$ are then approximately uncorrelated zero-mean Gaussian variables, whose variances are proportional to the integrals of the angular power distribution $S(\mathbf{k},\boldsymbol{\kappa})$ over the corresponding wavenumber cells \cite{pizzo2022fourier,pizzo2022spatial}. Collecting these variances in a diagonal matrix $\boldsymbol{\Lambda}_{S}$ and denoting the covariance matrix of $\boldsymbol{z}$ by $\boldsymbol{\Sigma}_{z}$, the LMMSE estimate is
\begin{equation}\label{CE_LMMSE}
    \hat{\boldsymbol{\alpha}}_{\mathrm{LMMSE}}=\boldsymbol{\Lambda}_{S}\mathbf{A}^{\mathsf{H}}\left(\mathbf{A}\boldsymbol{\Lambda}_{S}\mathbf{A}^{\mathsf{H}}+\boldsymbol{\Sigma}_{z}\right)^{-1}\mathbf{v}.
\end{equation}
This estimator exploits the angular sparsity, since the variances of the coefficients outside the effective angular supports are close to zero. Moreover, the number of significant entries in $\boldsymbol{\Lambda}_{S}$ is determined by the effective angular supports at the Tx and Rx rather than by the dictionary size $N_{\mathrm{c}}$. Hence, the pilot overhead required for reliable estimation can be much smaller than $N_{\mathrm{c}}$.}

\rev{Both the sparse-recovery and Bayesian estimators realize} the desired infinite-to-finite dimensional transformation. The unknown object remains a continuous EM channel, but estimation is carried out through a finite-size dictionary that spans the dominant radiative subspace of the channel. In this way, the estimator preserves the essential continuous-space structure of CAPA propagation while keeping the pilot overhead and computational complexity manageable. Moreover, by enlarging or refining the codebook, the approximation in \eqref{CE_sparse_operator} can be made increasingly accurate, thereby approaching the original infinite-dimensional channel model as closely as needed. \rev{However, refining the codebook also increases the correlation between the columns of $\mathbf{A}$, which makes sparse recovery less reliable, while paths whose parameters lie between grid points still leak into several coefficients. These off-grid effects can be mitigated by off-grid sparse Bayesian inference~\cite{yang2013off} or by gridless methods based on atomic norm minimization~\cite{tang2013compressed}.} 

\section{Fundamental Limits of CAPA Systems} \label{sec:fundamental_limits}

The continuous-space modeling framework developed in the previous sections provides a basis for characterizing the ultimate performance of CAPA systems. Since a CAPA interacts with EM fields over a finite physical aperture, its performance is jointly constrained by aperture size, wavelength, propagation geometry, scattering richness, transmit power, and receiver noise. This section studies these constraints from two complementary perspectives. We first characterize the spatial DoFs, which determine how many independent spatial modes can be supported by the channel, and then discuss capacity, which quantifies how much information can be transmitted over these modes under practical power and noise models.

\subsection{Degrees of Freedom and Spatial Multiplexing} \label{sec:dof}

In conventional SPDA systems, the spatial DoF is often directly associated with the rank of the finite-dimensional channel matrix obtained via singular value decomposition (SVD), which quantifies the maximum number of independent data streams that can be transmitted to the receiver simultaneously. However, in CAPA systems, the spatial DoF needs to be reinterpreted from an operator-theoretic perspective. Although the transmit and receive signals reside in infinite-dimensional function spaces, a physically realizable aperture cannot support infinitely many independent communication modes. The finite aperture size and the limited spatial-frequency content that can be radiated and observed over a finite wavelength channel collectively constrain the system to an effectively finite-dimensional subspace. Therefore, the relevant quantity for characterizing DoF in CAPA systems is the number of \rev{significant} singular values of the continuous channel operator, which determines the number of spatial streams that can be multiplexed reliably~\cite{bucci1989spatial,poon2005degrees,franceschetti2017wave}.

We focus on the uni-polarized model in \eqref{Uni_Polarized_Signal_Model_Expression}. Since the Tx and Rx apertures are bounded and spatially separated, the channel kernel $h(\mathbf{r},\mathbf{s})$ is square-integrable. Consequently, the associated channel operator is Hilbert-Schmidt and hence bounded and compact, admitting the following singular-function decomposition:
\begin{align}\label{eq:LoS_HS_decomposition}
h(\mathbf{r},\mathbf{s})=\sum_{n=1}^{\infty}\sigma_n \zeta_n(\mathbf{r})\xi_n^*(\mathbf{s}),
\end{align}
where $\{\zeta_n\}$ and $\{\xi_n\}$ are orthonormal sets in $L^2(\mathcal{S}_{\mathrm{r}})$ and $L^2(\mathcal{S}_{\mathrm{t}})$, respectively, and $\sigma_n\ge 0$ are the singular values arranged in descending order. Substituting \eqref{eq:LoS_HS_decomposition} into \eqref{Uni_Polarized_Signal_Model_Expression} yields
\begin{align}
y(\mathbf{r})=\sum_{n=1}^{\infty}\sigma_n \zeta_n(\mathbf{r}) \int_{\mathcal{S}_{\mathrm{t}}} \xi_n^*(\mathbf{s})x(\mathbf{s})\,{\rm{d}}\mathbf{s}+n(\mathbf{r}).
\end{align}
This decomposition is the continuous-aperture counterpart of the matrix SVD in conventional SPDA systems. Each pair $(\zeta_n,\xi_n)$ represents an orthogonal transmit-receive eigenmode, with $\sigma_n$ characterizing its channel gain. \rev{Equivalently, the transmit eigenfunctions $\xi_n$ satisfy the eigenvalue problem}
\begin{align}\label{eq:LoS_eigen_problem}
\int_{\mathcal{S}_{\mathrm{t}}}K_{\mathrm{t}}(\mathbf{s},\mathbf{s}')\xi_n(\mathbf{s}')\,\mathrm{d}\mathbf{s}'=\sigma_n^2\xi_n(\mathbf{s}),
\end{align}
\rev{where $K_{\mathrm{t}}(\mathbf{s},\mathbf{s}')\triangleq\int_{\mathcal{S}_{\mathrm{r}}}h^*(\mathbf{r},\mathbf{s})h(\mathbf{r},\mathbf{s}')\,\mathrm{d}\mathbf{r}$ is the transmit correlation kernel. This follows directly from \eqref{eq:LoS_HS_decomposition} and the orthonormality of $\{\zeta_n\}$, and is the continuous counterpart of $\mathbf{H}^{\mathsf{H}}\mathbf{H}\mathbf{v}_n=\sigma_n^2\mathbf{v}_n$.} Accordingly, the spatial DoF is characterized by the number of significant singular values. \rev{Specifically, we regard a mode as significant if its normalized eigenvalue $\mu_n\triangleq\sigma_n^2/\sigma_1^2\in[0,1]$ satisfies $\mu_n\ge0.5$.}

While the Hilbert-Schmidt decomposition provides a rigorous mathematical framework for characterizing the modal structure of the CAPA channel, it does not yield explicit insights into the physical nature of the eigenmodes or their corresponding singular values. Furthermore, there is typically no closed-form solution for the singular values and eigenfunctions of the \rev{exact} LoS operator, and therefore the related spatial DoF can only be evaluated numerically\rev{, e.g., by discretizing the kernel with the Gauss-Legendre quadrature discussed in Section~\ref{sec:beamforming_Optimization}, which converges rapidly since the kernel between separated apertures is smooth}. To obtain analytical insights, in the following, we first analyze the DoF of LoS channels, which are fundamental to understanding the spatial multiplexing capabilities of CAPA systems. Then, we discuss how multipath scattering can further enrich the spatial DoF and enhance multiplexing performance.

\subsubsection{LoS Channels}

Consider the LoS channel in \eqref{dyadic Green's function_Standard_Scalar_Often_Used}. Its spatial DoF can be characterized using Landau’s eigenvalue theorem through the following procedure.

\paragraph{Fresnel Approximation and Fourier Structure}

To obtain analytical insights, let us first consider the representative broadside configuration where the two planar CAPAs are parallel, i.e., $\mathbf{C}=\mathbf{I}_3$, and the Rx center is located at $\mathbf{r}_o=[0,D,0]^{\mathsf{T}}$. Let $\mathbf{s}=[s_{\mathsf{x}},0,s_{\mathsf{z}}]^{\mathsf{T}}$ and $\mathbf{r}=[r_{\mathsf{x}},D,r_{\mathsf{z}}]^{\mathsf{T}}$, and define the in-plane coordinates $\mathbf{s}_{\paral}\triangleq[s_{\mathsf{x}},s_{\mathsf{z}}]^{\mathsf{T}}$ and $\mathbf{r}_{\paral}\triangleq[r_{\mathsf{x}},r_{\mathsf{z}}]^{\mathsf{T}}$. Under the Fresnel approximation, the propagation distance becomes
\begin{align}\label{eq:LoS_Fresnel_distance}
\lVert\mathbf{r}-\mathbf{s}\rVert \approx D+\frac{\lVert\mathbf{r}_{\paral}\rVert^2+\lVert\mathbf{s}_{\paral}\rVert^2-2\mathbf{r}_{\paral}^{\mathsf{T}}\mathbf{s}_{\paral}}{2D}.
\end{align}
Substituting \eqref{eq:LoS_Fresnel_distance} into \eqref{dyadic Green's function_Standard_Scalar_Often_Used} and replacing the slowly varying amplitude factor $1/\lVert\mathbf{r}-\mathbf{s}\rVert$ by $1/D$ result in
\begin{align}\label{eq:LoS_Fresnel_kernel}
h_{\rm{LoS}}(\mathbf{r},\mathbf{s}) \approx c_0
{\rm{e}}^{-{\rm{j}}\frac{k_0}{2D}\lVert\mathbf{r}_{\paral}\rVert^2}
{\rm{e}}^{-{\rm{j}}\frac{k_0}{2D}\lVert\mathbf{s}_{\paral}\rVert^2}
{\rm{e}}^{{\rm{j}}\frac{k_0}{D}\mathbf{r}_{\paral}^{\mathsf{T}}\mathbf{s}_{\paral}},
\end{align}
where $c_0=-{\rm{j}}\eta_0k_0{\rm{e}}^{-{\rm{j}}k_0D}/(4\pi D)$ is a constant. The first two factors are quadratic phase terms that depend only on the local coordinates at the receive and transmit apertures, respectively. Since these factors have unit magnitude, they only rotate the phase of the input and output fields and therefore do not change the singular values of the operator. After factoring out these aperture-dependent phase rotations, the remaining kernel has the form of a truncated two-dimensional Fourier transform:
\begin{align}\label{eq:LoS_truncated_FT}
\widetilde{y}(\mathbf{r}_{\paral})=\int_{\mathcal{A}_{\mathrm{t}}}
{\rm{e}}^{{\rm{j}}\bm{\kappa}^{\mathsf{T}}(\mathbf{r}_{\paral})\mathbf{s}_{\paral}}
\widetilde{x}(\mathbf{s}_{\paral})\,{\rm{d}}\mathbf{s}_{\paral},
\quad
\bm{\kappa}(\mathbf{r}_{\paral})\triangleq \frac{k_0}{D}\mathbf{r}_{\paral},
\end{align}
where
\begin{align}
\mathcal{A}_{\mathrm{t}}&\triangleq \left[-\frac{L_{{\rm{t}},\mathsf{x}}}{2},\frac{L_{{\rm{t}},\mathsf{x}}}{2}\right]\times\left[-\frac{L_{{\rm{t}},\mathsf{z}}}{2},\frac{L_{{\rm{t}},\mathsf{z}}}{2}\right],\\
\mathcal{A}_{\mathrm{r}}&\triangleq \left[-\frac{L_{{\rm{r}},\mathsf{x}}}{2},\frac{L_{{\rm{r}},\mathsf{x}}}{2}\right]\times\left[-\frac{L_{{\rm{r}},\mathsf{z}}}{2},\frac{L_{{\rm{r}},\mathsf{z}}}{2}\right].
\end{align}

Equation \eqref{eq:LoS_truncated_FT} has a clear physical interpretation, i.e., the receive aperture does not observe the entire Fourier transform of the transmit current, but only the portion corresponding to the following wavenumber window:
\begin{align}
\mathcal{K}_{\mathrm{r}}
&=\left\{\bm{\kappa}=\frac{k_0}{D}\mathbf{r}_{\paral}:\mathbf{r}_{\paral}\in\mathcal{A}_{\mathrm{r}}\right\}\nonumber\\
&=\left[-\frac{k_0L_{{\rm{r}},\mathsf{x}}}{2D},\frac{k_0L_{{\rm{r}},\mathsf{x}}}{2D}\right]
\times
\left[-\frac{k_0L_{{\rm{r}},\mathsf{z}}}{2D},\frac{k_0L_{{\rm{r}},\mathsf{z}}}{2D}\right].\label{eq:LoS_window}
\end{align}
Therefore, the LoS channel is simultaneously limited in two domains. It is space-limited by the finite Tx aperture and band-limited by the finite Rx aperture projected into the spatial-frequency domain. In other words, the Tx aperture restricts where the source distribution can exist, while the Rx aperture restricts which spatial frequencies can be observed. This is precisely the setting where Landau's eigenvalue theorem becomes applicable \cite{landau1980eigenvalue,franceschetti2015landau}.

\paragraph{Landau's Eigenvalue Theorem and DoF Scaling Law}
\rev{Shannon observed that a signal of duration $T$ and bandwidth $W$ has approximately $2WT$ DoFs~\cite{shannon1949communication}, a result later made rigorous in~\cite{slepian1961prolate,landau1980eigenvalue}.} Landau's classical eigenvalue theorem, together with its multidimensional extensions, states that for an operator that is jointly limited in space and spatial frequency, the number of eigenvalues that remain close to one is asymptotically equal to the corresponding space-bandwidth product divided by $(2\pi)^d$ \cite{landau1980eigenvalue,franceschetti2015landau}, \rev{where $d$ is the number of dimensions}. In the present 2D case in \eqref{eq:LoS_truncated_FT}, the effective number of significant LoS eigenmodes is thus given by the area of the Tx aperture in the spatial domain multiplied by the area of the Rx aperture in the spatial-frequency domain, normalized by $(2\pi)^2$:
\begin{align}\label{eq:LoS_DoF_scaling}
N_{\rm{DoF}}^{\rm{LoS}}
\approx \frac{|\mathcal{A}_{\mathrm{t}}||\mathcal{K}_{\mathrm{r}}|}{(2\pi)^2}
=\frac{L_{{\rm{t}},\mathsf{x}}L_{{\rm{t}},\mathsf{z}}}{(2\pi)^2}
\cdot
\frac{k_0^2L_{{\rm{r}},\mathsf{x}}L_{{\rm{r}},\mathsf{z}}}{D^2}
=\frac{A_{\mathrm{t}}A_{\mathrm{r}}}{\lambda^2D^2},
\end{align}
where $A_{\mathrm{t}}\triangleq L_{{\rm{t}},\mathsf{x}}L_{{\rm{t}},\mathsf{z}}$ and $A_{\mathrm{r}}\triangleq L_{{\rm{r}},\mathsf{x}}L_{{\rm{r}},\mathsf{z}}$ are the physical aperture areas. This scaling law has been widely reported in the literature \cite{miller2000communicating, 9139337, pizzo2022landau}, which reveals that the LoS DoF scales quadratically with the aperture sizes and inversely with the square of the Tx-Rx distance. Physically, this scaling can be understood as follows. As the apertures become electrically larger or the Tx-Rx distance decreases, more spatial modes become resolvable within the Fresnel region, allowing for increased spatial multiplexing. Conversely, if the apertures are small or the distance is large, only a few modes remain significant, and the channel behaves more like a rank-one link.

\rev{The scaling law in \eqref{eq:LoS_DoF_scaling} holds within the validity range of the Fresnel approximation, which requires the Tx-Rx distance to be large compared with the aperture sizes. As $D$ decreases, \eqref{eq:LoS_DoF_scaling} grows without bound, whereas the actual DoF saturates at the aperture-limited value of approximately $\pi\min\{A_{\mathrm{t}},A_{\mathrm{r}}\}/\lambda^2$, which will be derived in the multipath analysis below. This is because only the part of the observed wavenumber window $\bm{\kappa}=(k_0/D)\mathbf{r}_{\paral}$ inside the radiating disk corresponds to propagating waves. Hence, \eqref{eq:LoS_DoF_scaling} also requires this window to lie inside the radiating disk, i.e., $\sqrt{L_{{\rm{r}},\mathsf{x}}^2+L_{{\rm{r}},\mathsf{z}}^2}/(2D)<1$, which gives $D>L/\sqrt{2}\approx0.71L$ for two equal square apertures with side length $L$.}

\rev{Furthermore, the above analysis assumes} the parallel broadside case with $\mathbf{C}=\mathbf{I}_3$. If the Rx-CAPA is still centered at $\mathbf{r}_o=[0,D,0]^{\mathsf{T}}$ but is no longer parallel to the Tx-CAPA, the Fresnel approximation can be extended by keeping the full orientation matrix $\mathbf{C}$ in the phase term. Following the derivation in \cite{ouyang2025concise}, let $\mathbf{r}'_{\paral}\triangleq[r'_{\mathsf{x}},r'_{\mathsf{z}}]^{\mathsf{T}}$ and define the in-plane projection matrix
\begin{align}
\mathbf{C}_{\paral}\triangleq
\begin{bmatrix}
C_{11} & C_{13}\\
C_{31} & C_{33}
\end{bmatrix},
\end{align}
where $C_{ij}$ is the $(i,j)$-th element of the orientation matrix $\mathbf{C}$.  
Then, after separating the aperture-dependent quadratic phase terms as before, the effective Fourier variable is no longer $\bm{\kappa}=(k_0/D)\mathbf{r}_{\paral}$, but instead $\bm{\kappa}=(k_0/D)\mathbf{C}_{\paral}\mathbf{r}'_{\paral}$. Therefore, the receive-side wavenumber window is the linearly transformed set $\mathcal{K}_{\mathrm{r}}=(k_0/D)\mathbf{C}_{\paral}\mathcal{A}_{\mathrm{r}}$, whose area is
\begin{align}
|\mathcal{K}_{\mathrm{r}}|=\frac{k_0^2}{D^2}\left|\det(\mathbf{C}_{\paral})\right|A_{\mathrm{r}}.
\end{align}
Applying Landau's theorem again yields the orientation-dependent LoS DoF scaling
\begin{align}\label{eq:LoS_DoF_scaling_rotated}
N_{\rm{DoF}}^{\rm{LoS}}
\approx \frac{A_{\mathrm{t}}A_{\mathrm{r}}}{\lambda^2D^2}\left|\det(\mathbf{C}_{\paral})\right|.
\end{align}
This expression reduces to \eqref{eq:LoS_DoF_scaling} when $\mathbf{C}_{\paral}=\mathbf{I}_2$. The factor $\left|\det(\mathbf{C}_{\paral})\right|$ essentially describes the area reduction caused by projecting the Rx aperture onto the Tx aperture plane, and therefore quantifies the LoS DoF reduction due to aperture misalignment. If the relative orientation is parameterized by Euler angles $(\alpha,\beta,\gamma)$, then \cite{ouyang2025concise}
\begin{align}
\left|\det(\mathbf{C}_{\paral})\right|
=\left|\cos\alpha\cos\gamma+\sin\alpha\sin\gamma\sin\beta\right|\le 1.
\end{align}
Therefore, the LoS spatial DoF is maximized when the two planar CAPAs are parallel and decreases as their relative orientation departs from this setup.

\begin{figure}[!t]
    \centering
    \subfigure[Spatial DoF versus Tx-Rx distance.]{
        \includegraphics[width=0.4\textwidth]{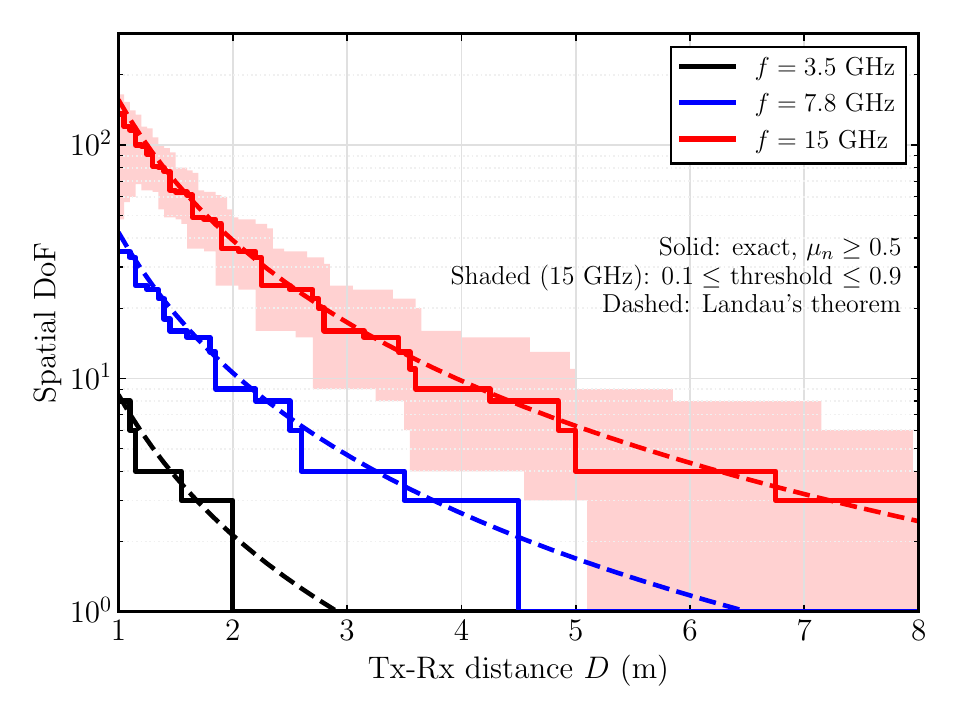}
        \label{fig:los_dof_vs_distance_landau}
    }
    \subfigure[Normalized eigenvalue spectra.]{
        \includegraphics[width=0.4\textwidth]{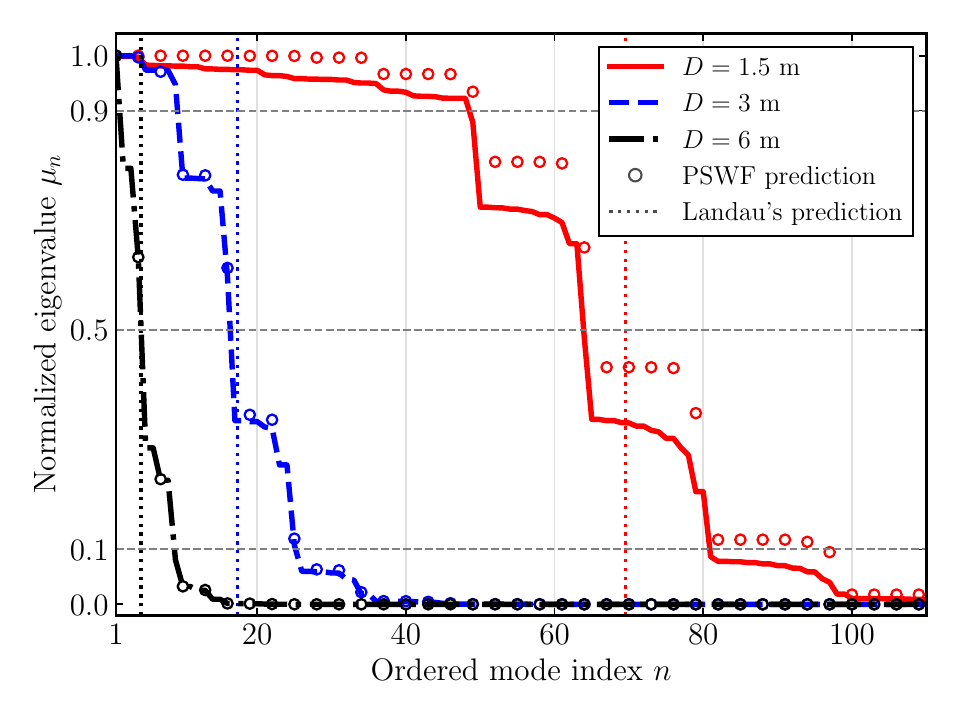}
        \label{fig:los_normalized_eigenvalues_landau}
    }
    \caption{Numerical verification of Landau's DoF prediction for two parallel broadside CAPAs with $L_{{\rm{t}},\mathsf{x}}=L_{{\rm{t}},\mathsf{z}}=L_{{\rm{r}},\mathsf{x}}=L_{{\rm{r}},\mathsf{z}}=0.5~{\rm{m}}$. In (a), the numerical LoS spatial DoF is counted using the threshold $\mu_n\ge 0.5$ and compared with the Landau approximation $A_{\mathrm{t}}A_{\mathrm{r}}/(\lambda^2D^2)$ at $3.5$, $7.8$, and $15$ GHz\rev{, and the shaded area shows the range of the DoF at $15$ GHz when the threshold varies from $0.1$ to $0.9$}. In (b), the normalized eigenvalue spectra are shown at $15$ GHz for representative distances, where the vertical dotted lines mark the corresponding Landau predictions \rev{and the markers show the PSWF eigenvalue products normalized by their largest value}.}
    \label{fig:los_dof_landau_comparison}
\end{figure}

\rev{
\paragraph{Slepian's Concentration Problem and Optimal Eigenmodes}

The preceding analysis determines the number of significant modes. We next consider which current distributions realize these modes. Under the Fresnel approximation in \eqref{eq:LoS_Fresnel_kernel}, the eigenvalue problem in \eqref{eq:LoS_eigen_problem} can be separated along the $\mathsf{x}$- and $\mathsf{z}$-axes for the broadside configuration. Along the $\mathsf{x}$-axis, it can be reduced to the standard form of Slepian's \emph{concentration problem} as follows \cite{slepian1961prolate,landau1962prolate,wan2026general}:
\begin{equation}\label{eq:PSWF_concentration}
\int_{-1}^{1}
\frac{\sin\big(c_{\mathsf{x}}(t-t')\big)}{\pi(t-t')}
\Psi_n(c_{\mathsf{x}};t')
\,\mathrm{d}t'
=\Lambda_n(c_{\mathsf{x}})\Psi_n(c_{\mathsf{x}};t),
\end{equation}
where $t=2s_{\mathsf{x}}/L_{{\rm{t}},\mathsf{x}}\in[-1,1]$ is the normalized Tx coordinate and $c_{\mathsf{x}}=k_0L_{{\rm{t}},\mathsf{x}}L_{{\rm{r}},\mathsf{x}}/(4D)$ is the space-bandwidth parameter, with $c_{\mathsf{z}}$ defined analogously. The eigenfunctions $\Psi_n(c;t)$, with $c\in\{c_{\mathsf{x}},c_{\mathsf{z}}\}$, are known as prolate spheroidal wave functions (PSWFs), which are orthonormal over $[-1,1]$ and indexed such that $1>\Lambda_1(c)>\Lambda_2(c)>\cdots>0$. Accordingly, the eigenfunctions of \eqref{eq:LoS_eigen_problem} are given by \cite{miller2000communicating}
\begin{align}\label{eq:LoS_eigenmodes}
\xi_n(\mathbf{s})
&=\frac{2}{\sqrt{L_{{\rm{t}},\mathsf{x}}L_{{\rm{t}},\mathsf{z}}}}
{\rm{e}}^{{\rm{j}}\frac{k_0}{2D}\lVert\mathbf{s}_{\paral}\rVert^2}
\nonumber\\*
&\qquad\times
\Psi_p\Big(c_{\mathsf{x}};\frac{2s_{\mathsf{x}}}{L_{{\rm{t}},\mathsf{x}}}\Big)
\Psi_q\Big(c_{\mathsf{z}};\frac{2s_{\mathsf{z}}}{L_{{\rm{t}},\mathsf{z}}}\Big),
\end{align}
where $(p,q)$ indexes the $n$-th largest product $\Lambda_p(c_{\mathsf{x}})\Lambda_q(c_{\mathsf{z}})$, and the corresponding normalized eigenvalue is $\mu_n=\Lambda_p(c_{\mathsf{x}})\Lambda_q(c_{\mathsf{z}})/[\Lambda_1(c_{\mathsf{x}})\Lambda_1(c_{\mathsf{z}})]$. For a fixed current norm under \eqref{eq:simplified_power}, these orthogonal current distributions successively maximize the received-field energy. We refer the reader to~\cite{slepian1961prolate,landau1962prolate} for the properties of PSWFs and to~\cite{miller2000communicating} for the reduction to \eqref{eq:PSWF_concentration} and the resulting eigenmodes in \eqref{eq:LoS_eigenmodes}.

\paragraph{Accuracy of Landau's Approximation and Numerical Validation}
Landau's theorem provides an asymptotic characterization of the DoF. For the one-dimensional concentration operator in \eqref{eq:PSWF_concentration}, Landau and Widom \cite{landau1980eigenvalue} established that, as $c\to\infty$, the number of eigenvalues $\Lambda_n(c)$ exceeding a fixed threshold $\mu_{\rm th}\in(0,1)$ satisfies
\begin{equation}\label{eq:Landau_Widom}
\#\{n:\Lambda_n(c)>\mu_{\rm th}\}
=\frac{2c}{\pi}
+\frac{1}{\pi^2}\ln\!\Big(\frac{1-\mu_{\rm th}}{\mu_{\rm th}}\Big)\ln c
+o(\ln c).
\end{equation}
Here, the leading term $2c/\pi$ carries a clear physical interpretation. In particular, substituting $c_{\mathsf{x}}=k_0L_{{\rm{t}},\mathsf{x}}L_{{\rm{r}},\mathsf{x}}/(4D)$ yields $2c_{\mathsf{x}}/\pi=L_{{\rm{t}},\mathsf{x}}L_{{\rm{r}},\mathsf{x}}/(\lambda D)$, which corresponds to the number of significant modes along the $\mathsf{x}$-axis predicted by Landau's theorem. Multiplying this value by its $\mathsf{z}$-axis counterpart leads to the two-dimensional scaling $A_{\mathrm{t}}A_{\mathrm{r}}/(\lambda^2D^2)$ in \eqref{eq:LoS_DoF_scaling}. Furthermore, the remaining terms in \eqref{eq:Landau_Widom} describe the deviation of the one-dimensional mode count from $2c/\pi$. For sufficiently large $c$, the eigenvalues transition sharply from nearly one to nearly zero around the index $2c/\pi$. According to \eqref{eq:Landau_Widom}, the number of eigenvalues in this transition region grows logarithmically with $c$, much more slowly than $2c/\pi$. Because varying the threshold $\mu_{\rm th}$ primarily affects how many of these intermediate eigenvalues are counted, its relative impact on the DoF diminishes as $c$ grows. Conversely, when only a few significant modes are present, the mode count can become sensitive to the choice of $\mu_{\rm th}$, as illustrated by the shaded region in Fig.~\ref{fig:los_dof_vs_distance_landau}. Notably, setting $\mu_{\rm th}=0.5$ eliminates the logarithmic terms in \eqref{eq:Landau_Widom}, reducing the discrepancy between the one-dimensional count and $2c/\pi$ to $o(\ln c)$. This is why the threshold $\mu_n\geq 0.5$ is adopted in this article.

Combining these observations with the discussion following \eqref{eq:LoS_DoF_scaling}, three conditions are required for the asymptotic LoS DoF approximation to be accurate: \emph{i)} the Fresnel approximation holds, \emph{ii)} the observed wavenumber window lies entirely within the radiating disk, and \emph{iii)} the space-bandwidth parameters along both aperture dimensions are sufficiently large. When the Fresnel approximation remains valid but the space-bandwidth parameters are moderate, the DoF can be evaluated more accurately by numerically computing the PSWF eigenvalues via \eqref{eq:PSWF_concentration}. When the Fresnel approximation breaks down, the eigenvalues should instead be computed directly from the exact LoS kernel. Extensions of this concentration framework to three-dimensional fields and the joint space-time domain can be found in~\cite{wan2026general}.
}

\rev{Fig.~\ref{fig:los_dof_landau_comparison} illustrates the LoS DoF scaling for the parallel broadside configuration. In Fig.~\ref{fig:los_dof_vs_distance_landau}, the numerical DoF, obtained from the exact scalar LoS kernel using $\mu_n\ge0.5$, matches closely with Landau's prediction $A_{\mathrm{t}}A_{\mathrm{r}}/(\lambda^2D^2)$ at all three frequencies. At short distances, Landau's prediction slightly overestimates the DoF because the Fresnel approximation becomes less accurate. At long distances, only a few significant modes remain, and the DoF becomes more sensitive to the threshold, as shown by the relatively wider shaded area, which is consistent with the discussion following \eqref{eq:Landau_Widom}. In Fig.~\ref{fig:los_normalized_eigenvalues_landau}, the leading normalized eigenvalues are close to one and decrease toward zero near the mode indices predicted by Landau's theorem, marked by the dotted lines. As the Tx-Rx distance increases, this decrease occurs at smaller mode indices, indicating fewer significant spatial modes. The spectra exhibit a staircase shape, which is accurately reproduced by the PSWF prediction at $D=3$ and $6$~m. At $D=1.5$~m, the discrepancy reflects the reduced accuracy of the Fresnel approximation at short distances.}

\subsubsection{Multipath Channels}

The multipath spatial DoF is best understood from the angular-domain model in \eqref{4FPWD_Model_Final}, where $H_a(\mathbf{k},\bm{\kappa})$ tells us how strongly a plane wave transmitted toward direction $\bm{\kappa}$ is converted by the environment into a plane wave arriving from direction $\mathbf{k}$. Hence, the multipath channel can be viewed as an \emph{angular coupling map} between the directional modes seen by the Tx and Rx apertures.

We start from the number of directional modes that the two apertures can resolve. Let $\mathcal{D}_{\mathrm{t}}^{\rm{eff}}\subseteq\mathcal{D}(\bm{\kappa})$ and $\mathcal{D}_{\mathrm{r}}^{\rm{eff}}\subseteq\mathcal{D}(\mathbf{k})$ denote the effective transmit and receive angular supports, i.e., the regions where $H_a(\mathbf{k},\bm{\kappa})$ or the average power spectrum $S(\mathbf{k},\bm{\kappa})$ is non-negligible. The spatial DoF can then be understood through angular resolution. In an aperture dimension of length $L$, two plane waves whose transverse wavenumbers differ by $\Delta k$ produce a phase difference of approximately $\Delta k L$ across the aperture. If $\Delta k L\ll 2\pi$, the aperture observes them as nearly the same direction; if $\Delta k L$ is on the order of $2\pi$, they become distinguishable. Hence, the wavenumber resolution of an aperture of length $L$ is approximately $2\pi/L$. For a two-dimensional aperture $\mathcal{A}$, the corresponding resolution cell has an area of approximately $(2\pi)^2/|\mathcal{A}|$, where $|\mathcal{A}|$ is the aperture area. Therefore, given the aperture area $|\mathcal{A}|$ and the effective angular support area $|\mathcal{D}^{\rm{eff}}|$, the maximum \rev{numbers of} resolvable directional modes at the Tx and Rx are \cite{poon2005degrees,pizzo2022spatial}
\begin{align}\label{eq:multipath_DoF_scaling}
N_{\mathrm{t}}^{\rm{eff}} \approx \frac{|\mathcal{A}_{\mathrm{t}}||\mathcal{D}_{\mathrm{t}}^{\rm{eff}}|}{(2\pi)^2},
\quad
N_{\mathrm{r}}^{\rm{eff}} \approx \frac{|\mathcal{A}_{\mathrm{r}}||\mathcal{D}_{\mathrm{r}}^{\rm{eff}}|}{(2\pi)^2}.
\end{align}
The resulting spatial DoF is then upper bounded by the minimum of these two numbers, i.e.,
\begin{align}\label{eq:multipath_DoF_upper_bound}
N_{\rm{DoF}}^{\rm{NLoS}}\le \min\left\{N_{\mathrm{t}}^{\rm{eff}},N_{\mathrm{r}}^{\rm{eff}}\right\}.
\end{align}

In the most favorable rich-scattering case, the effective angular supports are as large as possible, i.e., $\mathcal{D}_{\mathrm{t}}^{\rm{eff}}=\mathcal{D}(\bm{\kappa})$ and $\mathcal{D}_{\mathrm{r}}^{\rm{eff}}=\mathcal{D}(\mathbf{k})$. For the scalar planar model, both angular supports approach the full radiating disk with area $|\mathcal{D}(\bm{\kappa})|=|\mathcal{D}(\mathbf{k})|=\pi k_0^2$, yielding
\begin{align}\label{eq:aperture_limited_DoF}
N_{\mathrm{t}}^{\rm{eff}}\approx \frac{\pi |\mathcal{A}_{\mathrm{t}}|}{\lambda^2},
\quad
N_{\mathrm{r}}^{\rm{eff}}\approx \frac{\pi |\mathcal{A}_{\mathrm{r}}|}{\lambda^2},
\end{align}
which leads to the maximum multipath spatial DoF of $N_{\rm{DoF}}^{\rm{NLoS}} \approx \pi\min\{|\mathcal{A}_{\mathrm{t}}|,|\mathcal{A}_{\mathrm{r}}|\}/\lambda^2$.

The above results show that, for CAPAs, spatial DoF is fundamentally a space-bandwidth quantity determined by the aperture geometry and the accessible wavenumber support. In LoS propagation, the field observed across the receive aperture is produced directly by the finite transmit aperture, so the available spatial-frequency window is set by the two aperture sizes, their separation distance, and their relative orientation. This window shrinks with the Tx-Rx distance and with aperture misalignment, leading to the geometry-dependent scaling in \eqref{eq:LoS_DoF_scaling} and \eqref{eq:LoS_DoF_scaling_rotated}. In multipath propagation, the limiting factor is instead the angular support created by the scattering environment. Rich scattering can spread the field over a much wider set of departure and arrival wavenumbers, allowing the continuous apertures to resolve more independent plane-wave components. However, the gain is not unbounded. Once the effective angular supports fill the radiating disk, the DoF saturates at the aperture-limited value $\pi\min\{|\mathcal{A}_{\mathrm{t}}|,|\mathcal{A}_{\mathrm{r}}|\}/\lambda^2$. Thus, CAPA spatial multiplexing is maximized when the hardware can approach the continuous EM modal density and the environment exposes a sufficiently rich angular spectrum.

\rev{\paragraph{Polarization and Evanescent Waves}
The preceding expressions count spatial modes for one polarization. For a planar aperture much larger than the wavelength that receives waves from one half-space, each propagation direction supports two independent transverse polarizations. Hence, dual polarization can at most double the DoFs in \eqref{eq:LoS_DoF_scaling} and \eqref{eq:aperture_limited_DoF}, and this doubling is approached when both polarizations are supported by the channel and the apertures. Compact EM vector antennas can access up to six local field dimensions in rich scattering \cite{poon2011degree,tian2012experimental}, as discussed in Section~\ref{sec:polarization}.

Evanescent waves can contribute additional modes outside the radiating disk when the Rx is very close to the Tx aperture. A component with transverse wavenumber $\kappa>k_0$ decays as $\mathrm{e}^{-\sqrt{\kappa^2-k_0^2}\,d}$ over a distance $d$, as discussed in Remark~\ref{remark_evanescent_wave}. Hence, such components contribute noticeably only when the separation is within about a few wavelengths. Moreover, exploiting these weakly coupled modes requires large currents, whose ohmic losses limit the achievable gain.

\paragraph{Wideband and Doubly-Dispersive Channels}
The preceding expressions describe spatial DoF at a single frequency. For a wideband signal, we instead count the independent space-time signals that can be distinguished at the Rx over an observation interval \cite{tse2005fundamentals}. Consider a signal of bandwidth $B$ centered at $f_{\mathrm{c}}$ and duration $T$, and let $N_{\rm{DoF}}(f)$ denote the spatial DoF at frequency $f$. For a channel that is approximately constant over $T$, applying Landau's theorem jointly in space and time \cite{franceschetti2017wave,franceschetti2015landau,wan2026general} gives the following approximate signal-space dimension:
\begin{align}\label{eq:wideband_dimension}
N_{\rm{DoF}}^{\rm{WB}}\approx T\int_{f_{\mathrm{c}}-B/2}^{f_{\mathrm{c}}+B/2}N_{\rm{DoF}}(f)\,\mathrm{d}f,
\end{align}
where the factor $T$ accounts for the frequency resolution $1/T$. Since both \eqref{eq:LoS_DoF_scaling} and \eqref{eq:aperture_limited_DoF} scale with $1/\lambda^2$, and thus with $f^2$, we have $N_{\rm{DoF}}(f)\approx N_{\rm{DoF}}(f_{\mathrm{c}})(f/f_{\mathrm{c}})^2$. Substituting this relation into \eqref{eq:wideband_dimension} yields
\begin{align}\label{eq:wideband_dimension_closed}
N_{\rm{DoF}}^{\rm{WB}}\approx BT\,N_{\rm{DoF}}(f_{\mathrm{c}})\Big(1+\frac{B^2}{12f_{\mathrm{c}}^2}\Big).
\end{align}
When $B\ll f_{\mathrm{c}}$, the correction term $B^2/(12f_{\mathrm{c}}^2)$ is small, e.g., about $2\%$ even at $B/f_{\mathrm{c}}=1/2$, and the dimension is approximately the spatial DoF at the carrier frequency multiplied by the number of frequency bins $BT$. Nevertheless, the spatial modes vary with frequency, so the significant modes selected at the carrier frequency may not accurately represent the entire band. In a doubly-dispersive channel, the Fourier transform of \eqref{eq:physics_based_doubly_dispersive_model} with respect to delay also depends on time. When the delay spread is much smaller than $T$ and the Doppler spread is much smaller than $B$, the delay and Doppler spreads only slightly broaden the received time-frequency support, and hence \eqref{eq:wideband_dimension} remains a good approximation of the signal-space dimension.

\paragraph{Relation to the DoF of SPDAs}
For an SPDA, the spatial DoF is defined in the same way, i.e., as the number of significant singular values of its channel matrix. Hence, it is upper bounded by the rank of this matrix, and thus by the number of antennas. For a regular square array with element spacing $\Delta$ and aperture area $|\mathcal{A}|$, the number of antennas is approximately $|\mathcal{A}|/\Delta^2$. Sampling the full radiating disk without aliasing requires $\Delta\leq\lambda/2$. At $\Delta=\lambda/2$, the array has about $4|\mathcal{A}|/\lambda^2$ antennas, exceeding the aperture-limited DoF $\pi|\mathcal{A}|/\lambda^2$ in \eqref{eq:aperture_limited_DoF}. Thus, half-wavelength sampling provides enough samples to resolve the radiating field. More generally, as a discrete current basis becomes dense over the aperture, the singular values of its channel matrix, after normalization by the element area, approach those of the CAPA operator. Denser sampling can improve field synthesis, as discussed in Section~\ref{sec:beamforming_Optimization}, but does not increase the aperture-limited DoF. In practice, element patterns \cite{yue2025directivity} and mutual coupling, as discussed in Section~\ref{sec:mutual_coupling}, affect the modal gains and can reduce the number of significant modes accessible to an SPDA. Hence, the CAPA DoF is the physical limit that a sufficiently dense SPDA approaches, and the two architectures differ mainly in the strengths of the modes rather than in their number.

\paragraph{Role of Spatial DoF in CAPA Systems}
As will be shown in Section~\ref{sec:capacity}, the spatial DoF approximately characterizes the high-SNR capacity slope. Beyond this, it plays several roles in CAPA systems. First, since the eigenvalues decrease rapidly around the DoF, as shown in Fig.~\ref{fig:los_normalized_eigenvalues_landau}, the modes beyond the DoF contribute little to the capacity at moderate SNR. Hence, the DoF approximately equals the number of streams that can be effectively multiplexed even at moderate SNR. Second, it guides hardware dimensioning. Specifically, approximately $N_{\rm{DoF}}$ independent ports can access the significant modes if their current patterns span the corresponding subspace. Third, it indicates the minimum number of modes that an accurate finite-dimensional channel model needs to retain, which affects the computational complexity of channel estimation and beamforming, as discussed in Sections~\ref{sec:beamforming_Optimization} and~\ref{sec:ce}. Finally, the scaling laws guide the choice of aperture size, carrier frequency, deployment distance, and orientation.}

\subsection{Channel Capacity} \label{sec:capacity}

The preceding subsection characterized how many spatial modes can be supported by a CAPA channel. Capacity analysis addresses the complementary question: given these modes, the transmit-power constraint, and the noise observed over the receive aperture, what spectral efficiency can be reliably achieved? In conventional MIMO, this question is answered by diagonalizing a finite-dimensional channel matrix using SVD and applying water-filling across the resulting singular modes. The same intuition carries over to CAPA systems, except that finite-dimensional vectors and matrices are replaced by continuous current and field distributions and by the integral operators that map between them. We first formulate the Shannon capacity for continuous apertures, then provide a deterministic Kolmogorov-capacity interpretation, and finally incorporate physical power coupling and colored receiver noise.

\subsubsection{Shannon Capacity}

We start from the uni-polarized frequency-flat model in \eqref{Uni_Polarized_Signal_Model_Expression}, repeated here for convenience:
\begin{align}
y(\mathbf{r})=\int_{\mathcal{S}_{\mathrm{t}}}h(\mathbf{r},\mathbf{s})x(\mathbf{s})\,{\rm{d}}\mathbf{s}+n(\mathbf{r}),
\quad \mathbf{r}\in\mathcal{S}_{\mathrm{r}}.
\end{align}
The transmit signal $x(\mathbf{s})$ is a random current distribution over the Tx aperture. Its second-order statistics are described by the covariance kernel
\begin{align}\label{eq:CAPA_input_covariance_kernel}
Q(\mathbf{s},\mathbf{s}^{\prime})
\triangleq \mathbb{E}\{x(\mathbf{s})x^*(\mathbf{s}^{\prime})\}.
\end{align}
This kernel is the continuous-aperture counterpart of the covariance matrix $\mathbf{Q}=\mathbb{E}\{\mathbf{x}\mathbf{x}^{\mathsf{H}}\}$ in conventional MIMO. \rev{Following the simplified power metric in \eqref{eq:simplified_power}}, the average transmit power \rev{constraint} can be expressed as
\begin{align}\label{eq:CAPA_trace_power_constraint}
\mathbb{E}\left\{\int_{\mathcal{S}_{\mathrm{t}}}|x(\mathbf{s})|^2\,{\rm{d}}\mathbf{s}\right\}
 = \int_{\mathcal{S}_{\mathrm{t}}}Q(\mathbf{s},\mathbf{s})\,{\rm{d}}\mathbf{s}
\le P_{\mathrm{t}}.
\end{align}
This is the continuous version of the discrete constraint $\operatorname{tr}(\mathbf{Q})\le P_{\mathrm{t}}$. \rev{As discussed in Section~\ref{sec:power_constraints}, for apertures much larger than the wavelength, this metric is approximately proportional to the radiated power when the current radiates mainly in directions not far from the aperture normal. The physical power models will be revisited in Section~\ref{sec:physical_power}.}

Given the transmit covariance kernel $Q(\mathbf{s},\mathbf{s}^{\prime})$, the noiseless received signal part has the covariance kernel
\begin{align}\label{eq:CAPA_received_signal_covariance}
K(\mathbf{r},\mathbf{r}^{\prime})
=
\int_{\mathcal{S}_{\mathrm{t}}}\int_{\mathcal{S}_{\mathrm{t}}}
h(\mathbf{r},\mathbf{s})Q(\mathbf{s},\mathbf{s}^{\prime})
h^*(\mathbf{r}^{\prime},\mathbf{s}^{\prime})
\,{\rm{d}}\mathbf{s}^{\prime}{\rm{d}}\mathbf{s}.
\end{align}
For spatially white Gaussian noise over the receive aperture, $\mathbb{E}\{n(\mathbf{r})n^*(\mathbf{r}^{\prime})\}
=N_0\delta(\mathbf{r}-\mathbf{r}^{\prime})$. 
With Gaussian signaling, the Shannon mutual information associated with $Q(\mathbf{s},\mathbf{s}^{\prime})$ is then the continuous analogue of a log-determinant:
\begin{align}\label{eq:CAPA_kernel_capacity}
I(Q)=
\log_2\det\nolimits_{\rm{F}}\left(
\delta(\mathbf{r}-\mathbf{r}^{\prime})
\,
+\,
\frac{1}{N_0}K(\mathbf{r},\mathbf{r}^{\prime})
\right),
\end{align}
where $\det_{\rm{F}}(\cdot)$ denotes the Fredholm determinant associated with the integral kernel \cite{wan2023mutual}. In this expression, $\delta(\mathbf{r}-\mathbf{r}^{\prime})$ plays the role of the identity matrix in conventional MIMO, while $\frac{1}{N_0}K(\mathbf{r},\mathbf{r}^{\prime})$ plays the role of a normalized receive-side signal covariance operator. Let $\{\lambda_m\}$ denote the eigenvalues of this normalized covariance kernel, i.e.,
\begin{align}\label{eq:CAPA_fredholm_eigenvalues}
\int_{\mathcal{S}_{\mathrm{r}}}
\frac{1}{N_0}K(\mathbf{r},\mathbf{r}^{\prime})v_m(\mathbf{r}^{\prime})
\,{\rm{d}}\mathbf{r}^{\prime}
=\lambda_m v_m(\mathbf{r}).
\end{align}
The Fredholm determinant can then be interpreted as
\begin{align}
\det\nolimits_{\rm{F}}\left(
\delta(\mathbf{r}-\mathbf{r}^{\prime})
+\frac{1}{N_0}K(\mathbf{r},\mathbf{r}^{\prime})
\right)
=\prod_{m}(1+\lambda_m).
\end{align}
Thus, \eqref{eq:CAPA_kernel_capacity} is equivalent to
\begin{align}
I(Q)=\sum_m\log_2(1+\lambda_m),
\end{align}
which has the same form as the conventional MIMO identity $\log_2\det(\mathbf{I}+\mathbf{A})=\sum_m\log_2(1+\lambda_m(\mathbf{A}))$. \rev{Unlike the finite-dimensional case, this sum contains infinitely many terms. It converges, and the Fredholm determinant in \eqref{eq:CAPA_kernel_capacity} is well defined, because $\sum_m\lambda_m=\frac{1}{N_0}\int_{\mathcal{S}_{\mathrm{r}}}K(\mathbf{r},\mathbf{r})\,{\rm{d}}\mathbf{r}$ equals the total received signal power normalized by $N_0$, which is finite whenever $h(\mathbf{r},\mathbf{s})$ is square-integrable and the transmit power is finite.} The Shannon capacity is obtained by maximizing \eqref{eq:CAPA_kernel_capacity} over all feasible covariance kernels satisfying \eqref{eq:CAPA_trace_power_constraint}.

As in conventional MIMO, this optimization can be simplified by decomposing the channel into orthogonal spatial modes. Using the Hilbert-Schmidt decomposition of the channel kernel in \eqref{eq:LoS_HS_decomposition} and the orthonormality of the singular functions gives
\begin{subequations}\label{eq:CAPA_singular_function_equations}
\begin{align}
\int_{\mathcal{S}_{\mathrm{t}}}h(\mathbf{r},\mathbf{s})\xi_n(\mathbf{s})\,{\rm{d}}\mathbf{s}
&=\sigma_n\zeta_n(\mathbf{r}),\quad \mathbf{r}\in\mathcal{S}_{\mathrm{r}},\\
\int_{\mathcal{S}_{\mathrm{r}}}h^*(\mathbf{r},\mathbf{s})\zeta_n(\mathbf{r})\,{\rm{d}}\mathbf{r}
&=\sigma_n\xi_n(\mathbf{s}),\quad \mathbf{s}\in\mathcal{S}_{\mathrm{t}}.
\end{align}
\end{subequations}
Then, the transmit signal can be expanded using the right singular functions as follows:
\begin{align}\label{eq:CAPA_current_modal_expansion}
x(\mathbf{s})=\sum_{n=1}^{\infty}\widetilde{x}_n\xi_n(\mathbf{s}),
\quad
\widetilde{x}_n=\int_{\mathcal{S}_{\mathrm{t}}}\xi_n^*(\mathbf{s})x(\mathbf{s})\,{\rm{d}}\mathbf{s}.
\end{align}
Projecting the received field onto the corresponding receive mode gives
\begin{align}
\widetilde{y}_n
=\int_{\mathcal{S}_{\mathrm{r}}}\zeta_n^*(\mathbf{r})y(\mathbf{r})\,{\rm{d}}\mathbf{r}
=\sigma_n\widetilde{x}_n+\widetilde{n}_n.
\end{align}
Thus, the continuous channel reduces to an infinite collection of parallel scalar subchannels. With entropy-maximizing independent Gaussian signaling across these modes, the transmit covariance kernel takes the diagonal form
\begin{align}
Q(\mathbf{s},\mathbf{s}^{\prime})
=\sum_{n=1}^{\infty}P_n\xi_n(\mathbf{s})\xi_n^*(\mathbf{s}^{\prime}),
\end{align}
where $P_n=\mathbb{E}\{|\widetilde{x}_n|^2\}$ is the power assigned to the $n$-th spatial mode and $\sum_n P_n\le P_{\mathrm{t}}$. The resulting channel capacity is
\begin{align}\label{eq:CAPA_water_filling_capacity}
C=\sum_{n=1}^{\infty}\log_2\left(1+\frac{\sigma_n^2P_n^\star}{N_0}\right),
\end{align}
where the optimal powers follow the same water-filling rule as in conventional MIMO:
\begin{align}\label{eq:CAPA_water_filling_power}
P_n^\star=\left(\bar{P}-\frac{N_0}{\sigma_n^2}\right)^+,\quad
\sum_{n=1}^{\infty}P_n^\star=P_{\mathrm{t}}.
\end{align}
The water level $\bar{P}$ is chosen so that the power constraint is met. Although \eqref{eq:CAPA_water_filling_capacity} contains infinitely many terms, only finitely many modes are allocated with power for any finite $P_{\mathrm{t}}$. In particular, at low SNR, power is concentrated on the strongest mode. \rev{As the SNR increases, the number of active modes quickly approaches the DoF derived above, and then grows only very slowly, since the singular values beyond the DoF decay rapidly. Hence, the DoF serves as an SNR-independent measure of the spatial multiplexing capability of the channel, and it approximately characterizes the capacity slope at least at reasonably high SNR, asymptotically, all the non-zero modes will contribute.}

\subsubsection{Kolmogorov Capacity}

In addition to Shannon capacity, the Kolmogorov $\epsilon$-capacity offers a complementary deterministic perspective for characterizing CAPA channels. Its starting point differs from Shannon theory. Shannon capacity assumes a probabilistic noise model and asks for the largest reliable communication rate. Kolmogorov capacity instead introduces a deterministic resolution level $\epsilon$ and asks how many received field patterns can be distinguished. Specifically, if two received fields $y_1(\mathbf{r})$ and $y_2(\mathbf{r})$ satisfy $\int_{\mathcal{S}_{\mathrm{r}}}
\left|y_1(\mathbf{r})-y_2(\mathbf{r})\right|^2
\,{\rm{d}}\mathbf{r}
<\epsilon^2$, 
then the receiver treats them as indistinguishable. The parameter $\epsilon$ can be interpreted as an uncertainty radius that captures finite receiver resolution, modeling errors, residual noise, or any other effect that prevents two nearby fields from being reliably separated.

\begin{figure}[t]
\centering
\includegraphics[width=\linewidth]{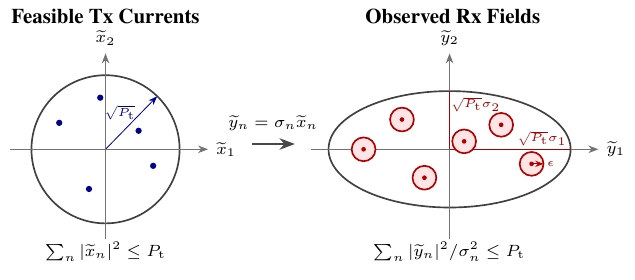}
\caption{Geometric interpretation of Kolmogorov $\epsilon$-capacity for a CAPA channel. The transmit-current ball is scaled by the singular values of the continuous channel and becomes an ellipsoid in the received-field space. The Kolmogorov capacity counts how many received fields separated by at least the resolution radius $\epsilon$ can be packed into this ellipsoid.}
\label{fig:CAPA_kolmogorov_capacity}
\end{figure}

For CAPAs, consider the set of all feasible transmit currents under the simplified power constraint
\begin{align}
\int_{\mathcal{S}_{\mathrm{t}}}|x(\mathbf{s})|^2\,{\rm{d}}\mathbf{s}\le P_{\mathrm{t}}.
\end{align}
This set forms a ball with a radius of $\sqrt{P_{\mathrm{t}}}$ in the transmit-current space.
Using the expansion in \eqref{eq:CAPA_current_modal_expansion}, the power constraint becomes $\sum_n|\widetilde{x}_n|^2\le P_{\mathrm{t}}$, while the corresponding noiseless received modal coefficients are
\begin{align}
\widetilde{y}_n=\sigma_n\widetilde{x}_n.
\end{align}
Hence, the channel maps the ball of feasible transmit modal coefficients into an ellipsoid in the received-field space, as illustrated in Fig. \ref{fig:CAPA_kolmogorov_capacity}:
\begin{align}\label{eq:CAPA_received_ellipsoid}
\sum_{n=1}^{\infty}\frac{|\widetilde{y}_n|^2}{\sigma_n^2}\le P_{\mathrm{t}}.
\end{align}
The semi-axis length of this ellipsoid along the $n$-th receive mode is therefore $\sqrt{P_{\mathrm{t}}}\sigma_n$.

The Kolmogorov $\epsilon$-capacity is the logarithm of the maximum number of $\epsilon$-separated received fields that can be packed into this ellipsoid, given by \cite{migliore2019horse}
\begin{align}
C_{\epsilon}\triangleq \log_2 (M_{\epsilon}),
\end{align}
where $M_{\epsilon}$ denotes the maximum packing number.
A useful approximation of the Kolmogorov capacity is obtained by counting the resolvable intervals along the axes of the ellipsoid. In this approximation, \rev{only the axes with $\sqrt{P_{\mathrm{t}}}\sigma_n>\epsilon$, referred to as active axes, are retained, since variations along the remaining axes fall within the resolution radius}. For the active axes, \rev{since each modal coefficient $\widetilde{y}_n$ is complex-valued and thus occupies two real dimensions, the number of $\epsilon$-sized cells along the $n$-th axis is approximately $(\sqrt{P_{\mathrm{t}}}\sigma_n/\epsilon)^2$}, which gives
\begin{align}\label{eq:CAPA_kolmogorov_capacity}
C_{\epsilon}\approx
\sum_{\sqrt{P_{\mathrm{t}}}\sigma_n>\epsilon}
\rev{\log_2\left(\frac{P_{\mathrm{t}}\sigma_n^2}{\epsilon^2}\right)}.
\end{align}
\rev{This approximation is a lower bound on $C_{\epsilon}$. To establish this bound, consider a maximal set of $\epsilon$-separated fields in the sub-ellipsoid spanned by the active axes. Balls of radius $\epsilon$ centered at these fields must cover this sub-ellipsoid; otherwise, an uncovered field could be added to the set. The number of fields is therefore at least the volume of the sub-ellipsoid divided by that of an $\epsilon$-ball. Since $M_{\epsilon}$ is at least this number, taking $\log_2$ of the volume ratio gives the right-hand side of \eqref{eq:CAPA_kolmogorov_capacity} as a lower bound on $C_{\epsilon}$. For a fixed number of active axes, this volume estimate captures the leading dependence of $C_{\epsilon}$ on $\epsilon$~\cite{dumer2004coverings}.}
\rev{Comparing \eqref{eq:CAPA_kolmogorov_capacity} with \eqref{eq:CAPA_water_filling_capacity}} shows that the deterministic packing behavior is governed by the same singular-value spectrum as the Shannon water-filling expression. The difference lies in the activation rule. \rev{Specifically}, Shannon capacity activates modes according to a noise-dependent water-filling threshold, whereas Kolmogorov capacity activates modes whose received-field axis length exceeds the prescribed resolution $\epsilon$. \rev{Notably, each active term in \eqref{eq:CAPA_kolmogorov_capacity} has the same form as the high-SNR approximation $\log_2(\sigma_n^2P_n/N_0)$ of the corresponding term in \eqref{eq:CAPA_water_filling_capacity}, with $\epsilon^2$ playing the role of the noise power $N_0$ and $P_{\mathrm{t}}$ in place of $P_n$.} \rev{Consequently, both capacities scale approximately linearly with the DoF and logarithmically with the SNR~\cite{lim2017information}.}

\subsubsection{Physical Power and Colored Noise} \label{sec:physical_power}

The simplified power constraint in \eqref{eq:CAPA_trace_power_constraint} is useful for exposing the water-filling structure, but it does not capture the power coupling induced by a physical aperture, as discussed in Sections \ref{sec:power} and \ref{sec:power_constraints}. \rev{More generally, let $R_{\mathrm{t}}(\mathbf{s},\mathbf{s}^{\prime})$ denote the transmit-side power coupling kernel that defines the power metric. For example, the radiated power in \eqref{eq:power_radiation_integral} corresponds to $R_{\mathrm{t}}(\mathbf{s},\mathbf{s}^{\prime}) = -\frac{1}{2}\Re\{\mathcal{G}_{\mathrm{p}}(\mathbf{s},\mathbf{s}')\}$.} \rev{Similarly, $R_{\mathrm{t}}(\mathbf{s},\mathbf{s}^{\prime})=\frac{1}{2}\big(-\Re\{\mathcal{G}_{\mathrm{p}}(\mathbf{s},\mathbf{s}')\}+\Re\{Z_s(\mathbf{s})\}\delta(\mathbf{s}-\mathbf{s}')\big)$ corresponds to the aperture power $P_{\mathrm{rad}}+P_{\mathrm{loss}}$. The source power $P_{\mathrm{src}}=\frac{1}{2}\mathbf{i}^{\mathsf{H}}\Re\{\mathbf{Z}\}\mathbf{i}$ is defined in the port domain and can be treated in the same way, with the matrix $\frac{1}{2}\Re\{\mathbf{Z}\}$ replacing the kernel $R_{\mathrm{t}}(\mathbf{s},\mathbf{s}^{\prime})$.} The average power constraint can then be written in the physically weighted form as follows:
\begin{align}\label{eq:CAPA_physical_power_constraint}
\int_{\mathcal{S}_{\mathrm{t}}}\int_{\mathcal{S}_{\mathrm{t}}}
R_{\mathrm{t}}(\mathbf{s},\mathbf{s}^{\prime})Q(\mathbf{s}^{\prime},\mathbf{s})
\,{\rm{d}}\mathbf{s}^{\prime}{\rm{d}}\mathbf{s}
\le P_{\mathrm{t}}.
\end{align}
Similarly, the received noise need not be spatially white. A coupled receive aperture may have a colored noise covariance kernel as follows:
\begin{align}\label{eq:CAPA_colored_noise_kernel}
K_{\mathrm{n}}(\mathbf{r},\mathbf{r}^{\prime})
=\mathbb{E}\{n(\mathbf{r})n^*(\mathbf{r}^{\prime})\},
\end{align}
which depends on the antenna structure, matching network, and thermal noise sources \cite{jeon2018capacity}. To express the corresponding mutual information in the same form as \eqref{eq:CAPA_kernel_capacity}, define the square-root and inverse-square-root kernels of $R_{\mathrm{t}}(\mathbf{s},\mathbf{s}^{\prime})$ and $K_{\mathrm{n}}(\mathbf{r},\mathbf{r}^{\prime})$ on their positive subspaces. For example, if $K_{\mathrm{n}}(\mathbf{r},\mathbf{r}^{\prime})=\sum_{l}\eta_{l}\chi_{l}(\mathbf{r})\chi_{l}^*(\mathbf{r}^{\prime})$, then $K_{\mathrm{n}}^{-1/2}(\mathbf{r},\mathbf{r}^{\prime})=\sum_{\eta_{l}>0}\eta_{l}^{-1/2}\chi_{l}(\mathbf{r})\chi_{l}^*(\mathbf{r}^{\prime})$, with the square-root operators of $R_{\mathrm{t}}(\mathbf{s},\mathbf{s}^{\prime})$ defined analogously. We then introduce the power-normalized current $\bar{x}(\mathbf{s})$ and the noise-whitened received field $\bar{y}(\mathbf{r})$ as
\begin{subequations}
\begin{align}
    \label{eq:CAPA_inverse_power_normalization}
x(\mathbf{s})
&=
\int_{\mathcal{S}_{\mathrm{t}}}
R_{\mathrm{t}}^{-1/2}(\mathbf{s},\mathbf{s}^{\prime})\bar{x}(\mathbf{s}^{\prime})
\,{\rm{d}}\mathbf{s}^{\prime},\\
\label{eq:CAPA_normalized_variables}
\bar{y}(\mathbf{r})
&=
\int_{\mathcal{S}_{\mathrm{r}}}
K_{\mathrm{n}}^{-1/2}(\mathbf{r},\mathbf{r}^{\prime})y(\mathbf{r}^{\prime})
\,{\rm{d}}\mathbf{r}^{\prime}.
\end{align}
\end{subequations}
Substituting \eqref{eq:CAPA_inverse_power_normalization} into the original channel model and applying the receive-side whitening in \eqref{eq:CAPA_normalized_variables} gives the end-to-end normalized model
\begin{align}\label{eq:CAPA_normalized_end_to_end_model}
\bar{y}(\mathbf{r})
=
\int_{\mathcal{S}_{\mathrm{t}}}
\bar h(\mathbf{r},\mathbf{s})\bar{x}(\mathbf{s})
\,{\rm{d}}\mathbf{s}
+\bar n(\mathbf{r}),
\end{align}
where $\bar n(\mathbf{r})
=
\int_{\mathcal{S}_{\mathrm{r}}}
K_{\mathrm{n}}^{-1/2}(\mathbf{r},\mathbf{r}^{\prime})n(\mathbf{r}^{\prime})
\,{\rm{d}}\mathbf{r}^{\prime}$ has the whitened covariance $\mathbb{E}\{\bar n(\mathbf{r})\bar n^*(\mathbf{r}^{\prime})\}=\delta(\mathbf{r}-\mathbf{r}^{\prime})$, and the normalized channel kernel is
\begin{align}\label{eq:CAPA_normalized_channel_kernel}
\bar h(\mathbf{r},\mathbf{s})
=
\int_{\mathcal{S}_{\mathrm{r}}}\int_{\mathcal{S}_{\mathrm{t}}}
K_{\mathrm{n}}^{-1/2}(\mathbf{r},\mathbf{r}^{\prime})
h(\mathbf{r}^{\prime},\mathbf{s}^{\prime})
R_{\mathrm{t}}^{-1/2}(\mathbf{s}^{\prime},\mathbf{s})
\,{\rm{d}}\mathbf{s}^{\prime}{\rm{d}}\mathbf{r}^{\prime}.
\end{align}
Let $\bar Q(\mathbf{s},\mathbf{s}^{\prime}) \triangleq \mathbb{E}\{\bar{x}(\mathbf{s})\bar{x}^*(\mathbf{s}^{\prime})\}$ denote the covariance kernel of the power-normalized transmit current, subject to $\int_{\mathcal{S}_{\mathrm{t}}}\bar Q(\mathbf{s},\mathbf{s})\,{\rm{d}}\mathbf{s}\le P_{\mathrm{t}}$. The normalized received signal covariance is
\begin{align}\label{eq:CAPA_normalized_received_covariance}
\bar K(\mathbf{r},\mathbf{r}^{\prime})
=
\int_{\mathcal{S}_{\mathrm{t}}}\int_{\mathcal{S}_{\mathrm{t}}}
\bar h(\mathbf{r},\mathbf{s})\bar Q(\mathbf{s},\mathbf{s}^{\prime})
\bar h^*(\mathbf{r}^{\prime},\mathbf{s}^{\prime})
\,{\rm{d}}\mathbf{s}^{\prime}{\rm{d}}\mathbf{s}.
\end{align}
The physically normalized mutual information is therefore
\begin{align}\label{eq:CAPA_physical_mutual_information}
I_{\rm{p}}(\bar Q)=
\log_2\det\nolimits_{\rm{F}}\left(
\delta(\mathbf{r}-\mathbf{r}^{\prime})
+
\bar K(\mathbf{r},\mathbf{r}^{\prime})
\right).
\end{align}
The role of the normalization is to absorb the physical transmit-power metric and the colored noise covariance into the effective channel $\bar h(\mathbf{r},\mathbf{s})$. After this transformation, the equivalent model in \eqref{eq:CAPA_normalized_end_to_end_model} has a standard trace-type power constraint on $\bar{x}(\mathbf{s})$ and unit spatially white noise $\bar n(\mathbf{r})$. Therefore, the mutual information keeps the same Fredholm log-determinant structure as \eqref{eq:CAPA_kernel_capacity}, with $\bar K(\mathbf{r},\mathbf{r}^{\prime})$ replacing the normalized received covariance kernel. In the special case where $R_{\mathrm{t}}(\mathbf{s},\mathbf{s}^{\prime})=\delta(\mathbf{s}-\mathbf{s}^{\prime})$ and $K_{\mathrm{n}}(\mathbf{r},\mathbf{r}^{\prime})=N_0\delta(\mathbf{r}-\mathbf{r}^{\prime})$, this normalization gives $\bar h(\mathbf{r},\mathbf{s})=h(\mathbf{r},\mathbf{s})/\sqrt{N_0}$, and \eqref{eq:CAPA_physical_mutual_information} reduces to \eqref{eq:CAPA_kernel_capacity}.

\rev{The inverse square-root kernels above also reveal an important physical insight. If $R_{\mathrm{t}}(\mathbf{s},\mathbf{s}^{\prime})$ accounts only for radiation, i.e., $R_{\mathrm{t}}(\mathbf{s},\mathbf{s}^{\prime})=-\frac{1}{2}\Re\{\mathcal{G}_{\mathrm{p}}(\mathbf{s},\mathbf{s}^{\prime})\}$, the associated operator is compact, and its eigenvalues decay rapidly to zero beyond approximately the aperture-limited DoF in \eqref{eq:aperture_limited_DoF}. Currents composed of these weak modes radiate almost no power, see \eqref{eq:prad_wavenumber}, yet they can still couple to a receiver located in the reactive near field. As a result, $\bar h(\mathbf{r},\mathbf{s})$ can become unbounded, and the capacity can grow without bound. Including the ohmic loss, i.e., $R_{\mathrm{t}}(\mathbf{s},\mathbf{s}^{\prime})=\frac{1}{2}\big(-\Re\{\mathcal{G}_{\mathrm{p}}(\mathbf{s},\mathbf{s}^{\prime})\}+\Re\{Z_s(\mathbf{s})\}\delta(\mathbf{s}-\mathbf{s}^{\prime})\big)$, makes $R_{\mathrm{t}}$ strictly positive and keeps the capacity finite, consistent with the regularization discussed in Section~\ref{sec:beamforming_Optimization}. Likewise, the external EM noise received by an aperture is band-limited to the radiating wavenumbers, so its covariance kernel is also compact, and the white component due to receiver circuit noise ensures that the whitening in \eqref{eq:CAPA_normalized_variables} is well defined. Material loss and circuit noise, which are often regarded as mere nonidealities, are thus essential for obtaining finite and physically meaningful capacity limits.}

\section{Summary and Conclusions} \label{sec:conclusion}

\rev{This tutorial has reviewed CAPA systems from the EM foundations of radiation to channel modeling, transceiver design, and performance limits. A continuous description of aperture currents and received fields connects these topics within the same physical model. The spatial modes of the channel provide a basis for tractable finite-dimensional approximations, while the power and noise models determine how these modes contribute to communication performance.}

\rev{The analysis shows that the available radiative DoFs are constrained by the aperture geometry, wavelength, and propagation environment. A sufficiently dense SPDA can approach the same mode count. Finer control of the aperture current can improve how the available modes are excited and combined, with benefits for spatial focusing, wavefront synthesis, and interference suppression. Realizing these benefits depends on the current distributions supported by the hardware, the accuracy of channel knowledge, and the losses and coupling within the aperture.}

\rev{These considerations also point to the next steps in CAPA research. Physical power and noise models need to be incorporated into transceiver designs with finite ports and practical excitation constraints. Wideband and time-varying operation requires beamforming and channel estimation methods that account for changes in the spatial modes across frequency and time, while large apertures call for algorithms whose complexity remains manageable as the number of significant modes grows.} \rev{In multi-user systems, aperture sharing through spatial patterns or wavenumber components~\cite{10158997,9906802}, user separation in angle and distance, and user scheduling and pilot allocation under limited spatial DoF remain important design problems~\cite{chen2025degrees}.} \rev{A central practical goal is to establish how closely realizable apertures can approach the performance predicted by continuous-current optimization under realistic power, loss, and noise conditions.}

\bibliographystyle{IEEEtran}
\bibliography{mybib}
\end{document}